\documentclass[onecolumn, amsmath,amssymb,longbibliography,aps,floatfix,superscriptaddress]{revtex4-1}

\usepackage{graphicx}

\usepackage{dblfloatfix}

\usepackage[caption=false]{subfig}
\usepackage{dcolumn}
\usepackage{bm}
\usepackage{amsmath}
\usepackage{amsfonts}
\usepackage[colorlinks=true, citecolor=blue, linkcolor=blue,urlcolor=blue ]{hyperref}

\DeclareMathOperator\erf{erf}

\DeclareMathOperator\E{E}

\usepackage{hyperref} 

\usepackage[dvipsnames]{xcolor}

\usepackage{siunitx}
\usepackage{gensymb}

\usepackage{listings}
\usepackage{color}

\usepackage[normalem]{ulem}

\hypersetup
{
	pdfauthor={Weida Liao},
	pdftitle={Theoretical model of confined thermoviscous flows for artificial cytoplasmic streaming},
}

\AtBeginDocument{%
	\newwrite\bibnotes
	\def\bibnotesext{Notes.bib}
	\immediate\openout\bibnotes=\jobname\bibnotesext
	\immediate\write\bibnotes{@CONTROL{REVTEX41Control}}
	\immediate\write\bibnotes{@CONTROL{%
			apsrev41Control,author="08",editor="1",pages="1",title="0",year="1"}}
	\if@filesw
	\immediate\write\@auxout{\string\citation{apsrev41Control}}%
	\fi
}

\begin{document}
		
	\begin{titlepage}
		
		\title{Theoretical model of confined thermoviscous flows for artificial cytoplasmic streaming}		
		
		\date{\today}
		\author{Weida Liao}
		\affiliation{Department of Applied Mathematics and Theoretical Physics, University of Cambridge, Wilberforce Road,  Cambridge, CB3 0WA, United Kingdom}
		\author{Elena Erben}
		\affiliation{Max Planck Institute of Molecular Cell Biology and Genetics, Pfotenhauerstra\ss e 108, 01307 Dresden, Germany}
		\author{Moritz Kreysing}
		\affiliation{Max Planck Institute of Molecular Cell Biology and Genetics, Pfotenhauerstra\ss e 108, 01307 Dresden, Germany}
		\affiliation{Institute of Biological and Chemical Systems, Karlsruhe Institute of Technology, Hermann-von-Helmholtz-Platz 1, 76344 Eggenstein-Leopoldshafen, Germany }
		\author{Eric Lauga}
		\email{e.lauga@damtp.cam.ac.uk}
		\affiliation{Department of Applied Mathematics and Theoretical Physics, University of Cambridge, Wilberforce Road,  Cambridge, CB3 0WA, United Kingdom}
		\begin{abstract}

Recent experiments in cell biology have probed the impact of artificially-induced intracellular flows in the spatiotemporal organisation of cells and organisms. 
In these experiments, mild dynamical heating (a few kelvins) via focused infrared light from a laser leads to long-range, thermoviscous flows of the cytoplasm inside a cell. 
To extend future use of this method in cell biology, popularised as focused-light-induced cytoplasmic streaming (FLUCS), new quantitative models are needed to link the external light forcing to the produced flows and transport. 
Here, we present a fully analytical, theoretical model describing the fluid flow induced by the dynamical laser stimulus at all length scales (both near the scan path of the laser beam and in the far field) in two-dimensional confinement.
We model the effect of the focused light as a small, local temperature change in the fluid, which causes a small change in both the density and the viscosity of the fluid locally. 
In turn, this results in a locally compressible fluid flow. 
We analytically solve for the instantaneous flow field induced by the translation of a heat spot of arbitrary time-dependent amplitude along a scan path of arbitrary length. 
We  show that the leading-order instantaneous flow field results from the thermal expansion of the fluid and is independent of the thermal viscosity coefficient. 
This leading-order velocity field is proportional to the thermal expansion coefficient and the magnitude of the temperature perturbation, with far-field behaviour typically dominated by a source or sink flow and proportional to the rate of change of the heat-spot amplitude. 	
In contrast, and in agreement with experimental measurements, the net displacement of a material point due to a full scan of the heat spot is  quadratic in the heat-spot amplitude, as it results from the interplay of thermal expansion and thermal viscosity changes. 	
The corresponding average velocity of material points over a scan is a hydrodynamic source dipole in the far field, with direction dependent on the relative importance of thermal expansion and thermal viscosity changes. 
Our quantitative findings show excellent agreement with recent experimental results and will enable the design of new controlled experiments to establish the physiological role of physical transport processes inside cells.

		\end{abstract}
		
		\maketitle
		
	\end{titlepage}

\section{Introduction}

Throughout nature, fluid flows are responsible for transport in fundamental processes that sustain life on a wide range of length scales~\cite{vogel2020life}.  
At the macroscopic scale, ocean circulation~\cite{batchelor2003perspectives,ferrari2009ocean} and coastal flows~\cite{hickey2010river,horner2015mixing} play key roles in determining the climate and shaping ecosystems, while interaction of the wind with plants disperses seeds over long distances and enhances gaseous exchange for photosynthesis~\cite{de2008effects}. 
In the human body,  both respiratory~\cite{grotberg2001respiratory} and blood flows~\cite{jensen2019blood,secomb2017blood,batchelor2003perspectives}  transport oxygen through networks of tubes.
At the length scales of micrometres lies the viscous  world of cells,  from the cilia-driven flows that determine the left--right asymmetry of developing embryos~\cite{nonaka2005novo,laugabook} to the flows inside individual cells~\cite{mogilner2018intracellular}. 
 
A notable example of intracellular fluid flow is known as cytoplasmic streaming. 
First discovered  in the 18th century~\cite{corti1774osservazioni}, cytoplasmic streaming is the bulk movement or circulation of the water-based viscous fluid, called the cytoplasm, inside a cell.
A fluid flow of this type is found in a wide variety of organisms, including plants, algae, animals, and fungi~\cite{allen1978cytoplasmic,kamiya1981physical}.
Cytoplasmic streaming has been shown to exhibit a wide range of topologies, including the fountain-like flow inside the pollen tube of a flower (\textit{Lilium longiflorum})~\cite{hepler2001polarized}, highly symmetrical streaming inside a green algal cell (\textit{Chara corallina})~\cite{van2010measurement}, and   swirls and eddies in the oocyte of the fruit fly (\textit{Drosophila})~\cite{ganguly2012cytoplasmic}.
 
Cytoplasmic streaming is driven actively: molecular motors move large cargos, such as vesicles and organelles, along polymeric filaments inside the cell, thereby entraining fluid.
The resulting flow of the whole fluid inside the cell induces transport of various substances in the fluid, including proteins, nutrients and organelles, which is important for fundamental processes such as metabolism and cell division~\cite{goldstein2015physical,goehring2011polarization}. 
This active, advective transport can be especially important in larger cells, with size on the order of a hundred micrometres~\cite{pickard1974hydrodynamic}. 
Indeed, diffusive transport by itself can be too slow on these larger length scales~\cite{goldstein2015physical}; this is further hindered by the crowded nature of the cytoplasm, with macromolecules making up 20--30\% of the volume~\cite{ellis2001macromolecular}.
Advective transport due to cytoplasmic streaming has therefore been argued to significantly impact cellular processes and the spatiotemporal organisation of cells~\cite{pickard1974hydrodynamic,hochachka1999metabolic}.

A number of theoretical and experimental studies have explored the precise role that cytoplasmic streaming plays in these processes~\cite{mogilner2018intracellular}.
For example, the green alga \textit{Chara corallina} has long, cylindrical cells.
Inside these cells, helical flows, which are driven from the cell boundary by the motion of myosin motors along actin filament tracks, have been measured experimentally~\cite{van2010measurement}.
Probing  the consequences of this rotational streaming theoretically, hydrodynamic models have suggested that the helical flow enhances mixing~\cite{goldstein2008microfluidics} and mass flux across the cell boundary~\cite{van2008nature}.
For animal cells, numerical simulations  have demonstrated that  cytoplasmic flows could result in robust positioning of organelles. 
In these simulations, the flows in the \textit{C.~elegans} embryo were driven by motors moving along microtubules~\cite{shinar2011model}, whereas in mouse oocytes the cytoplasmic flows were driven by the motion of actin filaments near the boundary~\cite{yi2011dynamic}.
Despite the contrasting molecular driving mechanisms, both led to a fountain-like flow pattern that could transport organelles to their required positions~\cite{mogilner2018intracellular}.

From an experimental standpoint, there are two significant, closely-related challenges in establishing the biological function of intracellular flows. 
The first is to create flows inside cells  reminiscent of naturally-occurring cytoplasmic streaming but driven artificially, without risking unwanted side effects that may be associated with genetic or chemical perturbations.
The second, closely-linked challenge lies in perturbing existing flows inside cells, so that the consequences for cellular processes may be observed.
Such a perturbative technique is arguably necessary to advance understanding of the physiology of intracellular flows, beyond correlation and towards causal relationships~\cite{mittasch2018non}.

To investigate in detail the causes and consequences of intracellular flows, the authors of Ref.~\cite{mittasch2018non} recently demonstrated the use of thermoviscous flows~\cite{weinert2008microscale,weinert2008optically} inside cells and developing embryos to generate and perturb cytoplasmic flows and transport, popularizing this approach in cell biology and terming it  Focused-Light-induced Cytoplasmic Streaming (FLUCS).
Using focused infrared light from a laser, localised in a small region of the cell, a thermoviscous flow is induced globally inside the cell, due to heating of the fluid. 
The laser beam produces a heat spot in the fluid and translates along a short scan path repeatedly, at a frequency of around $2~\si{\kilo\hertz}$; the repeated scanning then results in net transport of substances in the fluid,  which, near the scan path, is typically in the opposite direction to the laser motion. 
To produce net transport at physiological speeds, the  temperature changes required are only a few kelvins, avoiding  damage to the cell. 
Furthermore, this net transport, while strongest near the scan path, extends throughout the cell, sharing the long-ranged nature of cytoplasmic streaming.
This non-invasive technique enables the study of flows and transport inside cells in cellular organisation and processes.
 
Focusing on cell biology applications, the same group used FLUCS to show how fundamental processes in cell development are driven by intracellular flows~\cite{mittasch2018non}. 
For example, at the onset of development,  the  \textit{C.~elegans} zygote becomes polarized, before asymmetric cell division into two differently-sized daughter cells. 
The concentration of a particular protein (PAR2) determines which end of the zygote becomes the smaller germ cell and which becomes the larger somatic cell, thereby defining the body axis.
Flows of physiological magnitude and duration, created using FLUCS, were shown to be sufficient to localise these proteins~\cite{mittasch2018non}.
This demonstrates the biological significance of physical transport processes inside cells, as well as the potential of FLUCS for new experiments to reveal the precise role of intracellular fluid flows~\cite{kreysing2019probing}. 

The FLUCS technique has also enabled experimental perturbations to developmental processes~\cite{mittasch2018non,chartier2021hydraulic,mittasch2020regulated}. 
Guided by numerical simulations,  FLUCS was used to  redistribute PAR2 within the \textit{C.~elegans} zygote, remarkably leading to the inversion of its body axis~\cite{mittasch2018non}. 
The authors of Ref.~\cite{chartier2021hydraulic} controlled oocyte growth in \textit{C.~elegans} by artificially changing the volume of cells via FLUCS.
A combination of  FLUCS, genetics, and pharmacological intervention revealed the material properties of centrosomes in \textit{C.~elegans} embryos during development and their molecular basis~\cite{mittasch2020regulated}.
Beyond biology, thermoviscous flow and transport have been used together with closed-loop feedback control to achieve high-precision positioning of micrometre-sized particles~\cite{erben2021feedback} and high-sensitivity force measurements~\cite{stoev2021highly}.

From a hydrodynamic perspective, the flows and net transport generated in the FLUCS experiments can be explained in terms of the combined thermal expansion and thermal viscosity changes in the fluid caused by the laser, as first demonstrated in earlier  works~\cite{weinert2008microscale,weinert2008optically,yariv2004flow}. 
These studies dealt with Newtonian viscous fluid characterised by temperature-dependent density and viscosity.
This may serve as a simplified model for cytoplasm~\cite{mittasch2018non}, the complex rheology of which has been the subject of many studies~\cite{hayashi1980fluid,donaldson1972estimation,goldstein2015physical}. 
For a travelling temperature wave,  it was shown both mathematically and experimentally that thermal viscosity changes combined with flow driven by thermal expansion induced net transport of tracers along the scan path in a thin-film geometry~\cite{weinert2008microscale}.
A subsequent study instead examined localised heating by a laser, producing a heat spot that translates along a scan path repeatedly~\cite{weinert2008optically}; this is the relevant technique used in later biological (FLUCS) experiments inside cells~\cite{mittasch2018non}.
With a combination of theory, numerical simulation, and experiments, this demonstrated that the localised heating  also results in net transport of tracers, in a thin film of viscous fluid between two parallel plates~\cite{weinert2008optically}. 
The theory presented in Ref.~\cite{weinert2008optically} focused on net transport of tracers on and parallel to the scan path itself, for a heat spot of constant amplitude translating along an infinitely-long scan path.

In experiments~\cite{weinert2008optically,erben2021feedback,mittasch2018non}, both the instantaneous thermoviscous flow and the net transport induced by the laser heating are not localised to the scan path but instead extend  throughout space, as is the case for natural cytoplasmic streaming.
A recent experimental study  quantified how the average speed of tracers varies spatially, finding an inverse square law far from the scan path, in controlled microfluidic experiments on viscous fluid (glycerol-water solution) between two parallel plates~\cite{erben2021feedback}.
For the purposes of  modelling, this setup also has the advantage of separating the physical consequences of FLUCS from the biological effects.
Furthermore, in experiments~\cite{weinert2008optically,erben2021feedback,mittasch2018non}, the scan path has finite length and the amplitude of the heat spot varies with time. 
Numerical simulations, for FLUCS inside an ellipsoid representing a cell~\cite{mittasch2018non}, suggest that these two factors are crucial for understanding how the net transport of tracers in the fluid varies spatially in practical applications of FLUCS.

\begin{figure*}[t]
	\subfloat[]
	{\includegraphics[width=0.3\textwidth]{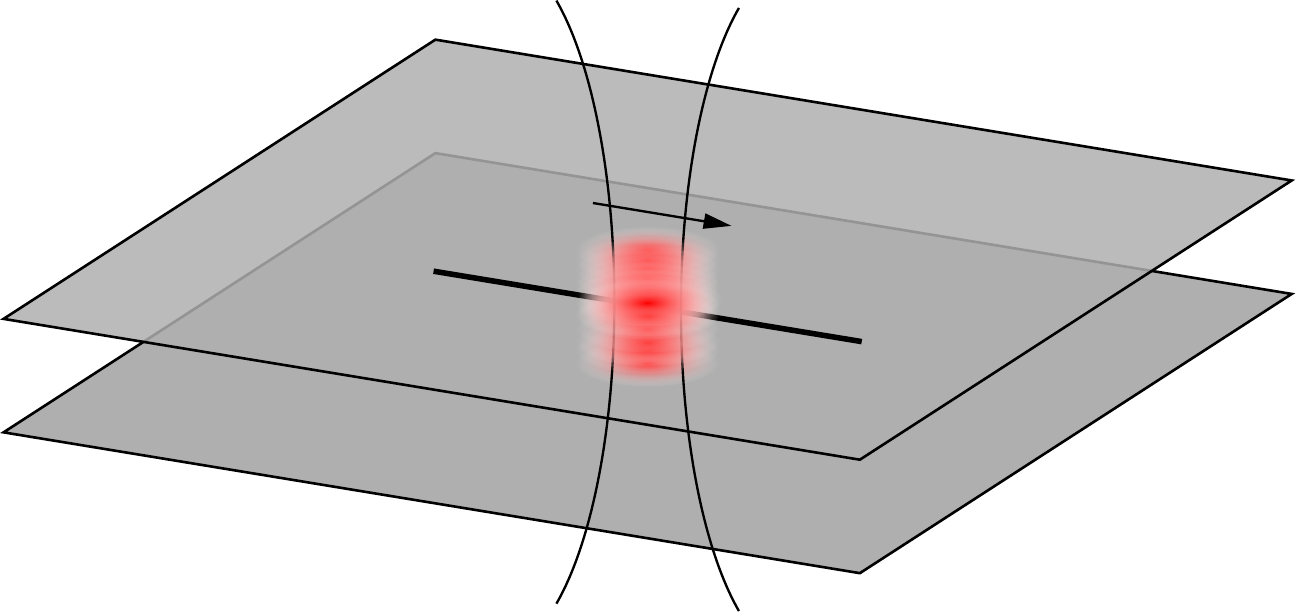}
		\label{fig:diagram_motivating_setup_sketch}}
	\subfloat[]
	{\includegraphics[width=0.3\textwidth]{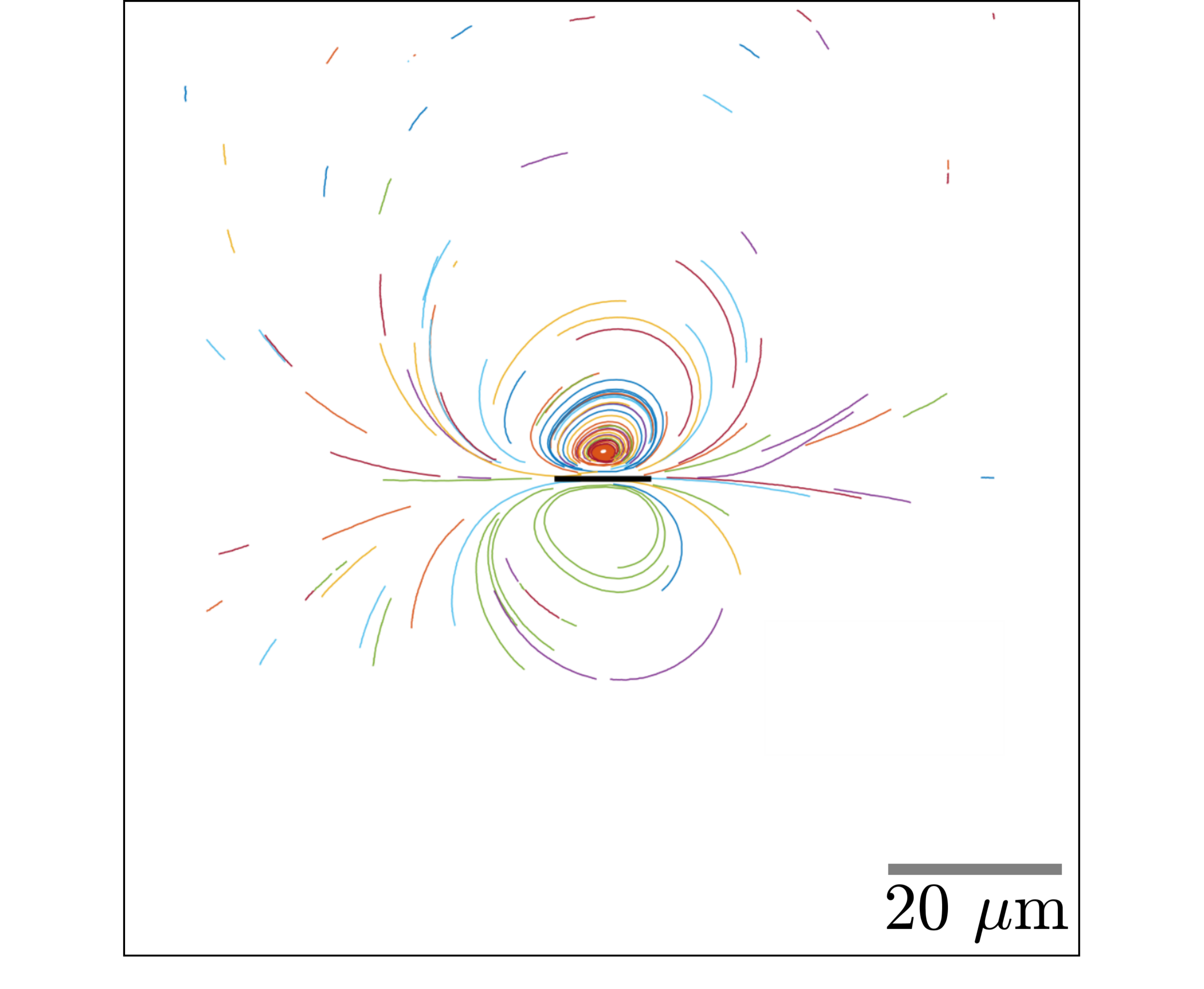}
		\label{fig:diagram_motivating_setup_expt_traj}}
	\subfloat[]
	{\includegraphics[width=0.3\textwidth]{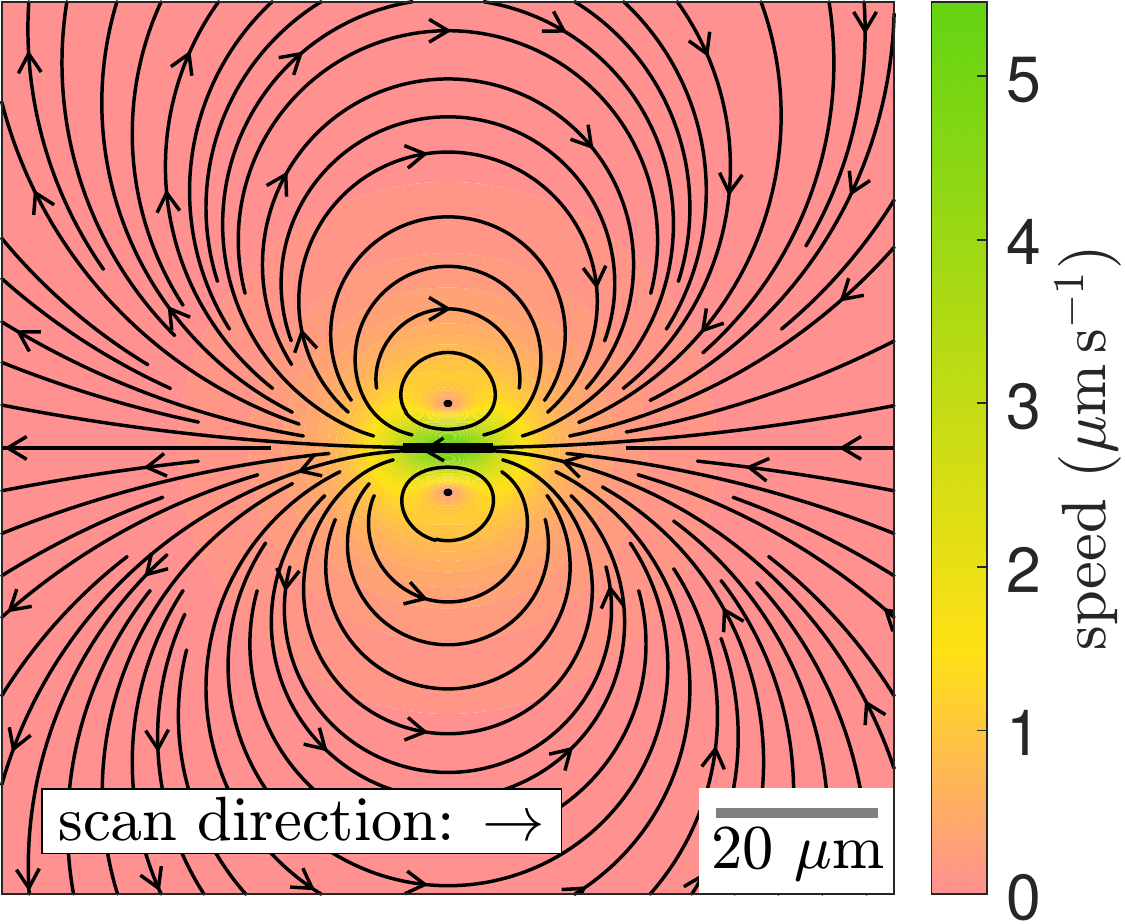}
		\label{fig:diagram_motivating_setup_theory_traj}}
	\caption{Left (Fig.~\ref{fig:diagram_motivating_setup_sketch}): sketch of setup; a laser focused on a fluid confined between two parallel plates creates a  heat spot (red), which translates along a scan path and induces fluid flow.
		Middle (Fig.~\ref{fig:diagram_motivating_setup_expt_traj}): experimentally-found trajectories of tracer beads  in the fluid, due to repeated scanning of the heat spot, viewed from above, adapted with permission from Figure 2d of Ref.~\cite{erben2021feedback} \copyright The Optical Society.
		Right (Fig.~\ref{fig:diagram_motivating_setup_theory_traj}):  theoretical time-averaged trajectories of tracers in the fluid as derived in this study (colours indicate the average speed of tracers), with parameters corresponding to experiments in Fig.~\ref{fig:diagram_motivating_setup_expt_traj}. }
	\label{fig:diagram_motivating}
\end{figure*}

In our work, summarised in Fig.~\ref{fig:diagram_motivating} and directly motivated by the controlled experiments in Ref.~\cite{erben2021feedback}, we present an analytical, theoretical model of the  flow driven in viscous fluid confined between two parallel plates by the focused light, with a scan path of arbitrary length and a fully general, time-dependent  heat-spot amplitude.  
We first solve analytically for the instantaneous fluid flow field induced by the heat spot during a scan period, valid in the entire spatial domain (i.e.,~from the near  to the far field), before analysing the trajectories of tracers in the fluid during one scan period and the net displacement of tracers due to a full scan of the heat spot.
The theory quantitatively and rigorously predicts how this net transport of tracers, induced by the repeated scanning, varies throughout space, in agreement with  data from recent microfluidic experiments~\cite{erben2021feedback}. 
Our modelling elucidates the fundamental physics of intracellular transport by FLUCS.
Our results will be useful for designing new FLUCS experiments to establish the physiological role of physical transport processes inside cells.
Our analytical descriptions will also enable the generation of more precise flow fields at even lower temperature impact on living cells, as well as the training of mathematical  models (e.g.,~machine-learning) to use flow fields for enhanced micro-manipulations.

This article is organised as follows. 
In Sec.~\ref{sec:inst_flow}, we solve analytically for the instantaneous flow field induced by a translating heat spot between two parallel plates, in the limit of small temperature change in the fluid. 
We demonstrate mathematically that thermal expansion drives the flow. 
The instantaneous flow at leading order is linear in the heat-spot amplitude and is purely due to thermal expansion, independent of thermal viscosity changes, in agreement with earlier work~\cite{weinert2008microscale,yariv2004flow}.
Our analytical solution for the instantaneous fluid velocity field shows that for a finite scan path (as in experiments), the time-dependence of the heat-spot amplitude is key, with the rate of change of the heat-spot amplitude setting the strength of the far-field source flow.
In the far field, this source flow typically dominates  over the source dipole associated with the translation of the heat spot.
We next solve  for higher-order contributions to the instantaneous flow field, quadratic in the heat-spot amplitude and including the first effect of thermal viscosity changes.
These higher-order terms are crucial for quantifying the net transport of tracers over many scan periods resulting from the instantaneous flow; the higher-order terms  arise from the amplification of the leading-order flow  by the heat spot and, in the experiments of Ref.~\cite{erben2021feedback}, are predominantly due to thermal viscosity changes. 
Building on the analytical leading-order instantaneous flow field, in Sec.~\ref{sec:one_scan}, we next solve analytically for the leading-order trajectory of an individual material point or tracer during one scan of the heat spot along a scan path. 
We then demonstrate for a general heat spot that the net displacement of  material points after a full scan period occurs at higher order.
Motivated by this, we show in Sec.~\ref{sec:net_displ} that the  leading-order net displacement of a material point due to a scan is quadratic in the heat-spot amplitude, in quantitative agreement with experiments~\cite{weinert2008optically,mittasch2018non}, and is due to the combined impact of temperature on both the density and the viscosity of the fluid, in agreement with earlier theory~\cite{weinert2008microscale}.
We then visualise the trajectories of material points due to repeated scanning of the heat spot along a finite scan path.
We characterise the average velocity of tracers (average Lagrangian velocity) over a scan period, finding that this is a hydrodynamic source dipole in the far field, in contrast with the slower-decaying far-field instantaneous fluid flow.
Finally, in Sec.~\ref{sec:comparison_expt}, we quantitatively compare the results from our theoretical model with the microfluidic experiments in Ref.~\cite{erben2021feedback}, which measured the trajectories and average speed of tracers due to repeated scanning of a heat spot. 
We conclude with a discussion of the predictions and limitations of our modelling approach, and in Sec.~\ref{sec:conclusion}, we summarise our work and propose possible further applications of our model.

\section{Theoretical model for instantaneous flow}\label{sec:inst_flow}

\subsection{Setup}\label{sec:setup}

\begin{figure}[t]
	{\includegraphics[width=0.78\textwidth]{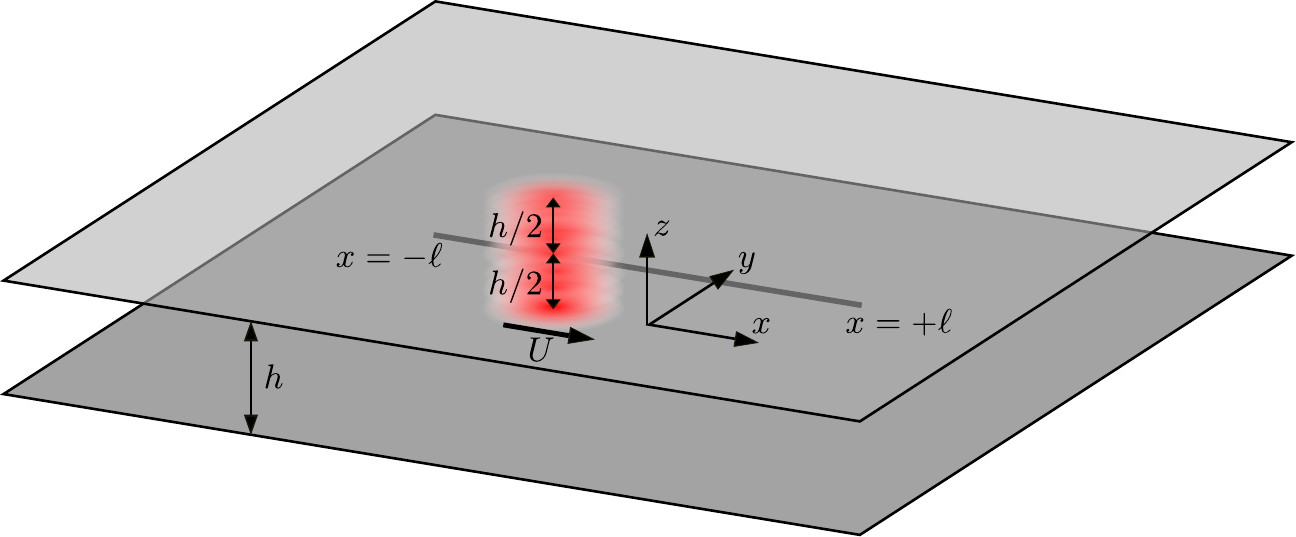}
		\caption{Setup of our theoretical model.		
		A heat spot of characteristic radius $a$ translates at speed $U$ in the $x$ direction, along a scan path  along $y=0$ from $x=-\ell$ to $x=+\ell$, causing inertialess fluid flow between the two no-slip boundaries at $z=0$ and $z=h$.}
		\label{fig:diagram_setup}}
\end{figure}

\begin{figure}[t]
	{\includegraphics[width=0.6\textwidth]{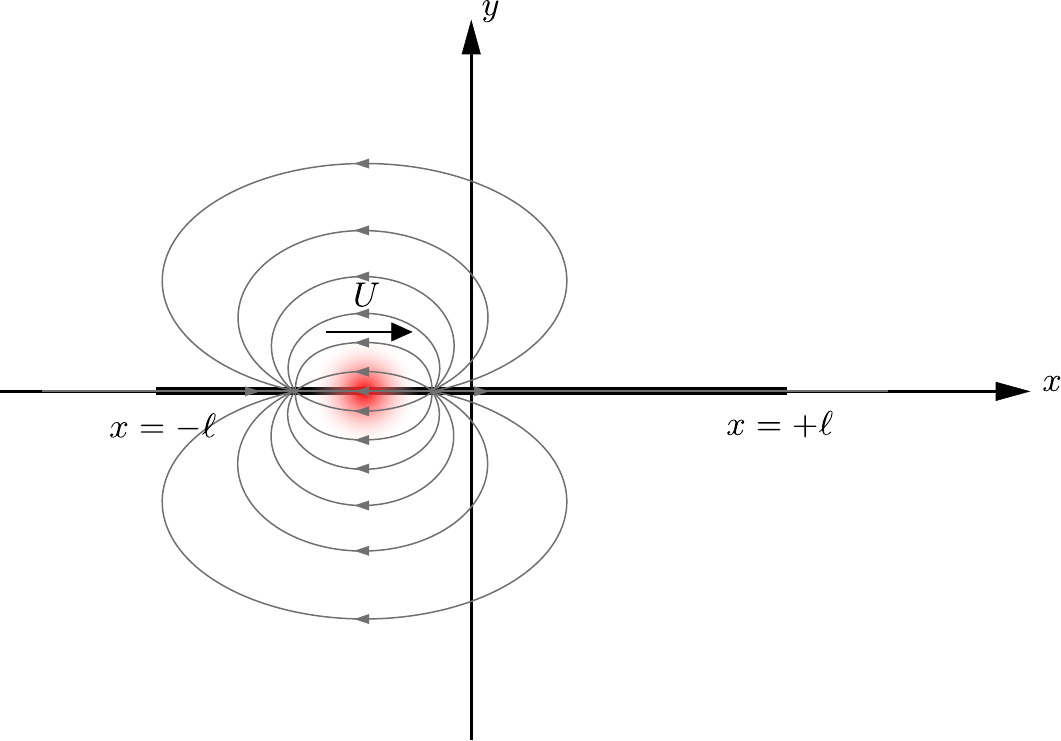}
		\caption{View from above of translating heat spot in Fig.~\ref{fig:diagram_setup}, with schematic instantaneous flow streamlines.}
		\label{fig:diagram_heat_spot}}
\end{figure}

In this first section, we introduce our theoretical model for the microfluidic experiments conducted in Ref.~\cite{erben2021feedback} as a simplified, controlled version of experiments in biological cells~\cite{mittasch2018non}. 
The setup is illustrated in Fig.~\ref{fig:diagram_setup} and a simplified view from above is given in Fig.~\ref{fig:diagram_heat_spot}, including a sketch of possible instantaneous flow streamlines. 
We consider fluid confined between parallel no-slip surfaces at $z=0$ and $z=h$ (Hele-Shaw geometry).
A heat spot of characteristic radius $a$  translates at constant speed $U$ along a scan path. 
The scan path is a line segment from $x=-\ell$ to $x=+\ell$ along $y=0$. 
At time $t$, the centre of the heat spot is thus at $(x=Ut,y=0, z=h/2)$, for $-\ell/U \leq t \leq \ell/U$ (i.e.,~during one scan). 
Initially, we focus on the  fluid flow induced instantaneously by the heat spot. 
We will examine in Sec.~\ref{sec:one_scan}   the motion of tracers or material points, with any initial position, due to the instantaneous flow during one scan period and then in Sec.~\ref{sec:net_displ} the net displacement of tracers resulting from a full scan of the heat spot.
This will allow us to understand the trajectories and average velocity of tracers due to repeated scanning as in experiments (Sec.~\ref{sec:comparison_expt}), i.e.,~where the heat spot travels from $x=-\ell$ to $x=+\ell$, then disappears and immediately reappears at $x=-\ell$, and repeats the process~\cite{weinert2008optically,erben2021feedback}; the distinction between the instantaneous velocity field and the time-averaged velocity of tracers is crucial for quantitatively reproducing experimental results.

We may write the temperature field $T(x,y,z,t)$ of the fluid as
\begin{align}
	T(x,y,z,t) = T_0 + \Delta T(x,y,z,t),
\end{align}
where $T_0$ is a constant reference temperature and $\Delta T(x,y,z,t)$ is the temperature change of the fluid due to the heat spot.
Below, we will prescribe this temperature field and discuss the conditions under which this is a good approximation.
We model the effect of the  temperature change on the fluid as follows.
Through thermal expansion (i.e.,~volume changes), an  increase in the temperature of the fluid locally decreases the density $\rho$ of the fluid (true for most fluids).
Since biological applications require small temperature changes, we may model this with a standard linear relationship
\begin{align}
	\rho &= \rho_0 (1 - \alpha \Delta T),\label{eq:density_temp_diml}
\end{align}
where $\rho_0$ is the density of the fluid at the reference temperature $T_0$ and we introduce the thermal expansion coefficient~$\alpha$.
A small, local increase in temperature also locally decreases the shear viscosity (i.e.,~dynamic viscosity)~$\eta$ of the fluid, modelled as
\begin{align}
	\eta &= \eta_0 (1 - \beta \Delta T),\label{eq:viscosity_temp_diml}
\end{align}
where $\eta_0$ is the shear viscosity of the fluid at the reference temperature $T_0$ and we introduce the thermal viscosity coefficient~$\beta$. 
For typical fluids, the coefficients $\alpha$ and $\beta$ are both usually positive~\cite{rumble2017crc}. 
In this paper, we  only consider small temperature changes, such that the relative density and viscosity changes are small, i.e.,~$\left \vert \alpha \Delta T \right \vert, \left \vert \beta \Delta T \right \vert \ll 1$. 
This is the relevant limit both for the microfluidic experiments of Ref.~\cite{erben2021feedback} (Sec.~\ref{sec:comparison_expt}) and for focused-light-induced cytoplasmic streaming in biological experiments, in order to avoid unwanted side effects of temperature changes inside cells.

Due to these effects, the heat spot induces a fluid flow, which we will solve for.
First, mass conservation is given by
\begin{align}
	\frac{\partial \rho}{\partial t} + \nabla\cdot(\rho\mathbf{u})=0,\label{eq:mass_consn_3D}
\end{align}
where $\mathbf{u} \equiv (u,v,w)$ is the velocity field.

Next, the Cauchy momentum equation is given by
\begin{align}
	\rho \frac{D \mathbf{u}}{Dt} = \nabla \cdot \boldsymbol{\Pi} + \rho \mathbf{g},\label{eq:Cauchy_momentum}
\end{align}
where $\mathbf{g}$ is the gravitational acceleration and the stress tensor $\boldsymbol{\Pi}$, under the Newtonian hypothesis, is given by 
\begin{align}
	\boldsymbol{\Pi} = - p \mathbf{1} + \kappa (\nabla \cdot \mathbf{u}) \mathbf{1} 
	+ 2 \eta \left \{ \frac{1}{2} [ \nabla \mathbf{u} + (\nabla\mathbf{u})^\text{T} ] - \frac{1}{3} \mathbf{1} (\nabla \cdot \mathbf{u} ) \right \}.\label{eq:stress_tensor_Newt}
\end{align}
Here, $p$  is the pressure field, $\mathbf{1}$ is the identity tensor, and $\kappa$ is the bulk  viscosity.
While the shear viscosity $\eta$ relates the stress to the linear deformation rate, the bulk viscosity $\kappa$ relates the stress to the volumetric deformation rate~\cite{happel} and can be important for compressible flows. 
We also note that the bulk viscosity $\kappa$, like the shear viscosity $\eta$, is in general  a function of temperature~\cite{happel,slie1966ultrasonic}.

Now, to make analytical progress, we use the lubrication limit, as in earlier work~\cite{weinert2008microscale,weinert2008optically}.
That is, we assume that the vertical separation of the plates $h$ is much smaller than the characteristic length scale over which the flow varies in the horizontal directions $x$ and $y$, given by the characteristic heat-spot diameter $2a$.
It follows that the time scale for diffusion of heat in the $z$ direction is much smaller than that in the $x$ and $y$ directions.
The laser beam intensity in experiments also decays much faster in the horizontal directions than it varies in the $z$ direction.
Thus, as a simplifying assumption, we take the temperature of the fluid to be independent of $z$, so that we model  the temperature change caused by the laser as $\Delta T = \Delta T (x,y,t)$. 

With these two assumptions, $h \ll 2a$ and $\frac{\partial \Delta T}{\partial z} = 0$, we may use a scaling argument similar to that for the classical, incompressible case~\cite{leal2007advanced} to simplify the momentum equation.  
We include the detailed derivation in Appendix~\ref{sec:lubrication_limit_scaling}.
We find that in the lubrication limit, the  momentum equations become
\begin{align}
	-\frac{\partial p}{\partial x} + \eta\frac{\partial^2 u}{\partial z^2} &=0,\label{eq:lubrication_x}\\
	-\frac{\partial p}{\partial y} + \eta \frac{\partial^2 v}{\partial z^2} &=0,\label{eq:lubrication_y}\\
	\frac{\partial p}{\partial z}&=0,\label{eq:lubrication_z}
\end{align}
the same as  the standard momentum equations for incompressible lubrication flow.
We observe  that the bulk viscosity no longer appears in these equations; from now on, for brevity, we thus refer to the shear viscosity $\eta$ simply as the viscosity.

A few comments  on the assumptions made here are in order. 
	First, we note that in Sec.~\ref{sec:limitations} we  return  to the validity of the lubrication limit in terms of the length scales in the experiments of Ref.~\cite{erben2021feedback} (which quantified the spatial variation of average tracer speed); in Sec.~\ref{sec:neglect}, we  also verify \textit{a posteriori} that the inertial terms may indeed be neglected via  a scaling argument. 
	Secondly, we remark that we have not included gravity in the equations above. 
	The primary focus of our article is the flow driven by thermal expansion. 
	With gravity, the density differences result in a horizontal gradient in hydrostatic pressure, typically driving gravity currents~\cite{simpson1982gravity}. 
	However, in Sec.~\ref{sec:neglect},  we  demonstrate  with a scaling argument that the ratio of the resulting gravity current to the thermal-expansion-driven flow is small for the experimental parameter values~\cite{yariv2004flow}.

To close the system of equations, we have the following boundary conditions. 
The velocity field is assumed to not be singular at the centre of the heat spot.
At the rigid, stationary parallel plates $z=0$ and $z=h$, the velocity field obeys the no-slip boundary condition.
The velocity is also taken to decay in the far field (unbounded fluid assumed), for a temperature change   that decays in the far field.

In this article, we   derive various equations and results that hold for a general temperature profile $\Delta T(x,y,t)$.
To illustrate these and to provide an explicit analytical solution for the flow, we will also impose in some of our results a Gaussian temperature profile for the heat spot given by
\begin{align}
	\Delta T(x,y,t) = \Delta T_0 A(t) \exp\{-[(x-Ut)^2 + y^2] / 2a^2\},\label{eq:temp_diml}
\end{align}
where $\Delta T_0 $ is the characteristic temperature change (a constant chosen to be positive) and $A(t)$ is the dimensionless amplitude of the heat spot, a function of time and kept arbitrary in our theoretical work. 
The temperature profile in Eq.~\eqref{eq:temp_diml} applies during one scan of the heat spot along the scan path.
For a heat spot, we assume that the amplitude $A(t)$ is positive, $A(t) \geq 0$.

The time-dependence of the heat-spot amplitude is a key ingredient in our theory, generalising previous theoretical modelling~\cite{weinert2008optically} that had a  constant heat-spot amplitude and an infinitely-long scan path.
In experiments, the scan path is finite. 
This is captured by the amplitude function, which must therefore be time-dependent: by definition, the amplitude is zero at each end of the  scan path, i.e.,~we always assume that~$A(\pm \ell/U)=0$. 
Furthermore, numerical work~\cite{mittasch2018non} for a different geometry (fluid inside an ellipsoid) found that time-variation of the heat-spot amplitude is necessary to reproduce experimentally-observed net transport.

The Gaussian spatial dependence of the temperature profile $\Delta T$ in Eq.~\eqref{eq:temp_diml} is motivated by measurements of the temperature field in experiments~\cite{weinert2008optically}, which showed that this is a good approximation provided that the translation of the laser beam is slow compared with the thermal equilibration of the fluid. 
In Sec.~\ref{sec:limitations}, we  discuss the validity of this assumption for the experiments in Ref.~\cite{erben2021feedback}, using a scaling argument for the advection-diffusion equation for heat. 
For experiments in cell biology, a non-invasive technique that does not damage the cells is desirable. 
The highly-localised nature of the  temperature perturbation in Eq.~\eqref{eq:temp_diml}, with small characteristic size  compared with the cell, is therefore advantageous.
We will show rigorously that this exponentially-decaying heating results in both strong net transport near the scan path and slowly-decaying flows in the far field, with algebraic instead of exponential scaling with distance.

We emphasise that while we have introduced here a Gaussian temperature field, many of the results in this article hold for any temperature profile $\Delta T(x,y,t)$. 
This generality is important because in experiments the heat spot can lose the circular symmetry assumed in Eq.~\eqref{eq:temp_diml} if its  speed of translation is too high; it may become elongated because of the thermal equilibration time of cooling the fluid~\cite{weinert2008optically,mittasch2018non}.

\subsection{Two-dimensional flow equations}\label{sec:flow}

With the setup detailed in Sec.~\ref{sec:setup}, we now  reduce the problem to two dimensions, following standard lubrication theory arguments as in Ref.~\cite{weinert2008microscale}. 
We may directly integrate the lubrication momentum equations, Eqs.~\eqref{eq:lubrication_x}--\eqref{eq:lubrication_z}, and use the no-slip boundary conditions.
The parabolic profile for the horizontal velocity field $\mathbf{u}_\text{H} \equiv (u,v) \equiv \mathbf{u}_\text{H}(x,y,z,t)$  is then given by
\begin{align}
	\mathbf{u}_\text{H} &=-\frac{\nabla_\text{H} \, p }{2\eta} z(h-z).
\end{align}
where the horizontal gradient is $\nabla_\text{H} \equiv (\partial/\partial x, \partial / \partial y)$.

We define the $z$-average as
\begin{align}
	\overline{(.)}=\frac{1}{h}\int_0^h (.) \,dz,
\end{align}
so that the two-dimensional,  $z$-averaged, horizontal velocity field $\overline{\mathbf{u}}_\text{H}(x,y,t)$ is given by
\begin{align}
	\overline{\mathbf{u}}_\text{H}&=-\frac{h^2}{12\eta} \nabla_\text{H} \, p.\label{eq:vel_press_diml}
\end{align}
We note for later comparison with experimental data in Sec.~\ref{sec:comparison_expt} that this differs from the velocity in the mid-plane, $z=h/2$, by a factor of $2/3$; in other words, we have
\begin{align}
	\left .\mathbf{u}_\text{H}\right \vert_{z=h/2} = -\frac{h^2}{8\eta} \nabla_\text{H} \, p = \frac{3}{2}\overline{\mathbf{u}}_\text{H}.
\end{align}

We also compute the $z$-average of the mass conservation equation, Eq.~\eqref{eq:mass_consn_3D}.
Combining this with the no-penetration boundary conditions on the parallel plates $z=0$ and $z=h$, we find
\begin{align}
	\frac{\partial \rho}{\partial t} + \nabla_\text{H}\cdot (\rho \overline{\mathbf{u}}_\text{H})=0.\label{eq:mass_consn_diml}
\end{align}
We note that since we allow arbitrary time-variation of the heat-spot amplitude in our theory, the solution for the flow is not in general steady in the frame of the translating heat spot. 
Hence, throughout this article, we work in the laboratory frame, in which the scan path is fixed and the fluid is at rest at infinity.

To simplify notation, from now on, we will only discuss two spatial dimensions $(x,y)$ and the corresponding two-dimensional, $z$-averaged, horizontal velocity field $\overline{\mathbf{u}}_\text{H}$.
Thus, in what follows, we drop the bar and subscript H from this velocity field and call it $\mathbf{u}$.
Similarly, we now write $\nabla$ instead of $\nabla_\text{H}$ for the horizontal gradient.

\subsection{Dimensionless problem statement}\label{sec:dimless}

We now nondimensionalise the key equations of the problem,  Eqs.~\eqref{eq:density_temp_diml},~\eqref{eq:viscosity_temp_diml},~\eqref{eq:temp_diml},~\eqref{eq:vel_press_diml}, and~\eqref{eq:mass_consn_diml}. 
We use the following characteristic scales: the heat-spot speed $U$ for velocity, the characteristic heat-spot radius $a$ for length, $a/U$ for time, the viscous lubrication stress scale $12\eta_0 a U /h^2$ for pressure (including the factor of $12$ for mathematical convenience), the characteristic temperature change $\Delta T_0$ for temperature, and the reference values $\rho_0$ for density and $\eta_0$ for viscosity.
For simplicity, we keep the same variables ($\mathbf{u}$, $x$, $y$, $\ell$, $\nabla$, $t$, $p$, $\alpha$, $\beta$, $\Delta T$, $\rho$, and $\eta$) to denote their dimensionless equivalents. 

Then, to summarise the dimensionless problem,  the  two-dimensional velocity field is determined by the pressure as 
\begin{align}
	\mathbf{u}&=-\frac{1}{\eta} \nabla p,\label{eq:vel_press}
\end{align}
and  the mass conservation equation is given by
\begin{align}
	\frac{\partial \rho}{\partial t} + \nabla \cdot (\rho \mathbf{u}) = 0.\label{eq:mass_consn}
\end{align}
We will derive results for a general prescribed temperature change $\Delta T(x,y,t)$ and also solve these equations explicitly with a Gaussian temperature profile (with arbitrary time-dependent amplitude) given by 
\begin{align}
	\Delta T(x,y,t) = A(t)\exp\{-[(x-t)^2 + y^2] /2\}, \label{eq:temp}
\end{align}
during one scan.
The temperature change determines the density $\rho$ and viscosity $\eta$ of the fluid via the equations
\begin{align}
	\rho &= 1-\alpha\Delta T,\label{eq:density_temp}\\
	\eta &= 1-\beta\Delta T,\label{eq:viscosity_temp}
\end{align}
respectively.
The boundary conditions for the two-dimensional problem are that the velocity field is assumed to not be singular at the centre of the heat spot and the fluid velocity is taken to decay at infinity.

Since the pressure field determines the velocity field, we may also derive a single equation for the pressure $p$ by substituting Eq.~\eqref{eq:vel_press} into Eq.~\eqref{eq:mass_consn} to give
\begin{align}
	\frac{\partial \rho}{\partial t} - \nabla \cdot \left ( \frac{\rho}{\eta}  \nabla p \right ) = 0.
\end{align}
Using Eqs.~\eqref{eq:density_temp} and~\eqref{eq:viscosity_temp} to write this in terms of the   general temperature field $\Delta T$ gives
\begin{align}
	\nabla \cdot \left ( \frac{1 - \alpha \Delta T}{1 - \beta \Delta T}  \nabla p \right ) = -\alpha \frac{\partial \Delta T}{\partial t} .\label{eq:pressure}
\end{align}
The pressure boundary conditions are inherited from the velocity boundary conditions, so that the pressure gradient $\nabla p$ is assumed not to be singular at the centre of the heat spot and the pressure gradient tends to zero at infinity.
(Note that the pressure itself does not necessarily decay at infinity; for example, the pressure associated with a source flow grows logarithmically at infinity.)

\subsection{Perturbation expansion}\label{sec:pert_exp}

In order to make analytical progress in solving Eq.~\eqref{eq:pressure}, we use a similar approach to that in Ref.~\cite{weinert2008microscale}.
We consider the limit of small   thermal expansion coefficient ($\alpha \ll 1$) and small thermal viscosity coefficient ($\beta \ll 1$), where we recall that these are dimensionless parameters, small because the characteristic temperature perturbation $\Delta T_0$ in experiments is sufficiently small. 
We pose a perturbation expansion in both of these parameters for the pressure $p$, given by
\begin{align}
	p &= \sum_{(m,n) = (0,0)}^{(\infty,\infty)} \alpha^m \beta^n p_{m,n} \nonumber\\
	&\equiv p_{0,0} + \alpha p_{1,0} + \beta p_{0,1} + \alpha^2 p_{2,0} + \alpha\beta p_{1,1} + \beta^2 p_{0,2} + \text{cubic and higher-order terms},\label{eq:pressure_pert_exp}
\end{align}
as $\alpha, \beta \to 0$.
In other words, the pressure at order $\alpha^m\beta^n$ is $p_{m,n}$.
We solve below for the pressure order by order. 
	This systematic approach  allows us to understand the roles of thermal expansion and thermal viscosity changes in the  flows induced.

We may then obtain the corresponding fluid velocity field   by expanding Eq.~\eqref{eq:vel_press} for small $\alpha$ and $\beta$   and collecting the terms as 
\begin{align}
	\mathbf{u} &= - \left [\sum_{k=0}^\infty (\beta \Delta T)^k \right ] \sum_{(m,n) = (0,0)}^{(\infty,\infty)}  \alpha^m \beta^n \nabla p_{m,n} \nonumber\\
	&\equiv \sum_{(m,n) = (0,0)}^{(\infty,\infty)} \alpha^m \beta^n \mathbf{u}_{m,n} \nonumber\\
	&\equiv \mathbf{u}_{0,0} + \alpha \mathbf{u}_{1,0} + \beta \mathbf{u}_{0,1} + \alpha^2 \mathbf{u}_{2,0} + \alpha\beta \mathbf{u}_{1,1} + \beta^2 \mathbf{u}_{0,2} + \text{cubic and higher-order terms}.\label{eq:vel_pert_exp}
\end{align}
We see that the velocity field $\mathbf{u}_{m,n}$ at  order $\alpha^m \beta^n$ depends explicitly on the pressure fields $p_{m,n'}$ with $0 \leq n' \leq n$, i.e.,~at lower or equal order in $\beta$.

We now expand Eq.~\eqref{eq:pressure} for the pressure $p$, using Eq.~\eqref{eq:pressure_pert_exp},  to find
\begin{align}
	&\nabla \cdot \left \{ (1 - \alpha \Delta T) \left [\sum_{k=0}^\infty (\beta \Delta T)^k \right ] \sum_{(m,n) = (0,0)}^{(\infty,\infty)}   \alpha^m \beta^n \nabla p_{m,n}  \right \}
	= -\alpha \frac{\partial \Delta T}{\partial t}\label{eq:pressure_expanded}. 
\end{align}
We emphasise that this equation for pressure holds for any temperature profile $\Delta T$ and note  that the forcing for this equation occurs at order $\alpha$.
We therefore anticipate that both the leading-order pressure and the leading-order flow it drives occur also at order $\alpha$.

We finally observe that for any temperature profile $\Delta T$, the velocity field $\mathbf{u}_{m,0}$ at order $\alpha^m$ is a potential flow (i.e.,~irrotational) for all $m$, given by 
\begin{align}
	\mathbf{u}_{m,0} = - \nabla p_{m,0}.
\end{align}
In other words, as remarked in Ref.~\cite{weinert2008microscale}, if the viscosity is constant,  the pressure acts as a velocity potential.

\subsection{Solution at order $\beta^n$}\label{sec:order_beta_n}

We now solve Eq.~\eqref{eq:pressure_expanded} order by order. 
We note that since this is a series of Poisson equations for pressure with Neumann boundary conditions,  the solution for the flow at each order is unique. 
First we consider order $\beta^n$ in Eq.~\eqref{eq:pressure_expanded} for all $n \geq 0$, i.e.,~orders $1$, $\beta$, $\beta^2$, and so on. 
We can show by induction  that the pressure $p_{0,n}$ at all these orders is zero, so that the corresponding velocity field $\mathbf{u}_{0,n}$ at each of these orders is also zero, for any prescribed temperature profile $\Delta T$ that decays at infinity. 
We include the proof in  Appendix~\ref{sec:derivation_order_beta_n}.

Hence, the leading-order instantaneous flow occurs at order $\alpha$ and  is purely due to local volume changes of the fluid upon heating. 
Furthermore, there is no flow at order $\beta$ or order $\beta^2$.
Therefore, in our perturbation expansion, the first effect of thermal viscosity changes occurs at order $\alpha\beta$.

In the microfluidic experiments of Ref.~\cite{erben2021feedback} that we discuss in Sec.~\ref{sec:comparison_expt}, the thermal viscosity coefficient~$\beta$ is much larger than the thermal expansion coefficient~$\alpha$ for the fluid used (glycerol-water solution), so that the flow at order $\alpha\beta$ is larger than that at order $\alpha^2$.
However, for a general fluid, the coefficients $\alpha$ and $\beta$ may be closer in magnitude~\cite{rumble2017crc}, so that the flow at order $\alpha^2$ could potentially be as important as the flow at order $\alpha\beta$.
Therefore, in this section, we  include the solution for the instantaneous flow up to quadratic order.
Both quadratic terms (order $\alpha\beta$ and order $\alpha^2$)  are crucial for understanding the trajectories of material points over many scans, as we will find that the  leading-order instantaneous flow (order $\alpha$) in fact gives rise to zero net displacement of material points after one full scan of the heat spot, at order $\alpha$.

In summary, the perturbation expansion for the velocity field $\mathbf{u}$ simplifies to
\begin{align}
	\mathbf{u} = \alpha \mathbf{u}_{1,0} + \alpha^2 \mathbf{u}_{2,0} + \alpha\beta \mathbf{u}_{1,1} + \text{cubic and higher-order terms}.\label{eq:vel_short_pert_exp}
\end{align}
Since there is no flow at order $\beta^n$, all terms in this updated perturbation expansion include the thermal expansion coefficient~$\alpha$. 
Physically, this means that the instantaneous flow is driven by thermal expansion. 
Thermal expansion can produce flow without thermal viscosity changes, but thermal viscosity changes can only lead to flow through interaction with thermal expansion.

\subsection{Solution at order $\alpha$}\label{sec:order_alpha}

We now solve for the leading-order instantaneous flow induced by the translating heat spot with arbitrary time-dependent amplitude.
This occurs at order $\alpha$, matching the forcing of the mass conservation equation by the temperature changes; in other words, this flow is associated purely with thermal expansion.

We begin by deriving the result that for a heat spot of arbitrary shape, the instantaneous fluid velocity field at order $\alpha$ is the time-derivative of a function proportional to the amplitude $A(t)$ of the heat spot.
This is important for understanding the net displacement of material points after a full scan period in Sec.~\ref{sec:one_scan}.
We next discuss 
	the physical mechanism for the flow induced by a general heat spot.
	Then we specialise to a Gaussian heat spot and  obtain an explicit analytical formula for the leading-order flow field $\mathbf{u}_{1,0}$ at order $\alpha$, 
	given by Eqs.~\eqref{eq:u10_decomposn},~\eqref{eq:u_switch}, and~\eqref{eq:u_translate}. 
To visualise this key result, we plot the streamlines of the two separate contributions to the flow in Fig.~\ref{fig:plot_u10_switch_streamlines} and Fig.~\ref{fig:plot_u_translate_streamlines};
we then illustrate the full velocity field during one scan in Fig.~\ref{fig:plot_u10_streamlines_snapshots} and Fig.~\ref{fig:plot_u10_streamlines_snapshots_trapezium_amplitude}.

\subsubsection{General heat spot}\label{sec:general_heat_spot_order_alpha}

At order $\alpha$, the equation for the pressure [Eq.~\eqref{eq:pressure_expanded}] reads
\begin{align}
	\nabla^2 p_{1,0} = -\frac{\partial \Delta T}{\partial t},\label{eq:order_alpha_original}
\end{align}
and holds for any temperature profile $\Delta T$; this is the fully-linearised, leading-order problem. 
We observe that the forcing in Eq.~\eqref{eq:order_alpha_original} is a time-derivative.
If we find a function $F(x,y,t)$ such that
\begin{align}
	\nabla^2 F = - \Delta T,\label{eq:order_alpha_time-integrated}
\end{align}
then $p_{1,0} = \partial F / \partial t$ is a solution of Eq.~\eqref{eq:order_alpha_original}.

We now assume that the temperature profile $\Delta T(x,y,t)$ during a scan is set by some arbitrary shape function $\Theta(x-t,y)$, steady in the frame moving with the heat spot, multiplied by the (arbitrary) time-dependent amplitude function $A(t)$, i.e.,
\begin{align}
	\Delta T(x,y,t) = A(t) \Theta(x-t, y)\label{eq:temp_general}.
\end{align}
The amplitude function $A(t)$ is taken to be positive if $\Delta T$ is a heat spot and negative if $\Delta T$ is a cool spot.
The shape function $\Theta(x-t,y)$ has the interpretation of a heat spot with constant amplitude that translates at unit speed in the positive $x$ direction. 
For example, the Gaussian heat spot in Eq.~\eqref{eq:temp} corresponds to shape function $\Theta(x,y) = \exp[-(x^2+y^2)/2]$; in general, the shape function need not have circular symmetry.

To solve Eq.~\eqref{eq:order_alpha_time-integrated}, with the general heat spot in Eq.~\eqref{eq:temp_general}, we pose the ansatz
\begin{align}
	F(x,y,t) = A(t) p_{1,0}^\text{(S)}(x-t,y),
\end{align}
where we will see later that $p_{1,0}^\text{(S)}(x-t,y)$ is the pressure field associated with the time-variation of the amplitude of the heat spot, as the heat spot switches on or off gradually (S stands for ``switch" in the superscript).
The pressure field $p_{1,0}^\text{(S)}(x-t,y)$ satisfies the Poisson equation
\begin{align}
	\nabla^2 p_{1,0}^\text{(S)}(x-t,y) = -\Theta(x-t,y).\label{eq:p_switch_Poisson}
\end{align}

For any given shape function $\Theta$, we can solve Eq.~\eqref{eq:p_switch_Poisson} for $p_{1,0}^\text{(S)}(x-t,y)$ (non-singular at the centre of the heat spot and with decaying gradient at infinity).
We then deduce that the full pressure field $p_{1,0}$ at order $\alpha$ is given by
\begin{align}
	p_{1,0}(x,y,t) &= \frac{\partial}{\partial t} [A(t) p_{1,0}^\text{(S)}(x-t,y)] \nonumber\\
	&\equiv A'(t) p_{1,0}^\text{(S)}(x-t,y) + A(t) p_{1,0}^\text{(T)}(x-t,y),\label{eq:p10_general_decomposn}
\end{align}
where prime ($'$) denotes differentiation with respect to the argument (here, time $t$) and we define the pressure field $p_{1,0}^\text{(T)}(x-t,y)$ as
\begin{align}
	p_{1,0}^\text{(T)}(x-t,y) \equiv \frac{\partial}{\partial t} p_{1,0}^\text{(S)}(x-t,y).
\end{align}
We will see later that  the pressure field $p_{1,0}^\text{(T)}(x-t,y)$ is associated with translation of the heat spot (T stands for ``translate" in the superscript).

In Eq.~\eqref{eq:p10_general_decomposn}, we have decomposed the pressure field at order $\alpha$ into two contributions.
The corresponding velocity field is given by
\begin{align}
	\mathbf{u}_{1,0}(x,y,t) &= -\frac{\partial}{\partial t} \nabla [A(t) p_{1,0}^\text{(S)}(x-t,y)] \nonumber\\
	&= \frac{\partial}{\partial t} [A(t)\mathbf{u}_{1,0}^\text{(S)}(x-t,y)]\nonumber\\
	&\equiv A'(t)\mathbf{u}_{1,0}^\text{(S)}(x-t,y) + A(t) \mathbf{u}_{1,0}^\text{(T)}(x-t,y),\label{eq:u10_general_decomposn}
\end{align}
where the two separate velocity fields $\mathbf{u}_{1,0}^\text{(S)}$ (associated with the switching-on of the heat spot) and $\mathbf{u}_{1,0}^\text{(T)}$ (associated with the translation of the heat spot) are given by
\begin{align}
	\mathbf{u}_{1,0}^\text{(S)} &= - \nabla p_{1,0}^\text{(S)},\label{eq:u_switch_general}\\
	\mathbf{u}_{1,0}^\text{(T)} &= - \nabla p_{1,0}^\text{(T)} \label{eq:u_translate_general}.
\end{align}
Importantly, we observe that this velocity field $\mathbf{u}_{1,0}$ is the time-derivative of a function proportional to $A(t)$. 
We will see in Sec.~\ref{sec:net_displ_zero_alpha} that this implies that the net displacement at order $\alpha$ of a material point due to one full scan of the heat spot is precisely zero;  for this reason, to understand the time-averaged trajectories of tracers seen in experiments~\cite{erben2021feedback} due to repeated scanning of the heat spot, we   solve for higher-order contributions to the instantaneous flow field in Sec.~\ref{sec:order_alpha_beta} and Sec.~\ref{sec:order_alpha_sq}.

\subsubsection{Physical mechanism for general heat spot}\label{sec:physical_mechanism_order_alpha}

Although we have not yet fully solved explicitly for the leading-order instantaneous flow (at order $\alpha$) induced by a heat spot, it is already possible to gain physical understanding from the general results of Sec.~\ref{sec:general_heat_spot_order_alpha}, in terms of thermal expansion. 

We start by interpreting physically the two terms contributing to the general velocity field $\mathbf{u}_{1,0}$ at order $\alpha$ in Eq.~\eqref{eq:u10_general_decomposn}.
In experimental  temperature-field measurements~\cite{mittasch2018non}, the heat spot appears to switch on gradually at the start of the scan path, i.e.,~the amplitude of the temperature perturbation increases; similarly the temperature perturbation appears to switch off gradually at the end of the scan path. 
In those experiments, the reason for this was reduced efficiency of the laser deflector for large angles; however, more generally, for any finite scan path, the amplitude of the heat spot must increase from zero at the start of the scan path and decrease to zero at the end, i.e.,~the amplitude must vary with time.

The first contribution to the velocity field, $A'(t) \mathbf{u}_{1,0}^\text{(S)}(x-t,y)$, is proportional to the rate of change of the heat-spot amplitude.
Thus, the velocity field $\mathbf{u}_{1,0}^\text{(S)}(x-t,y)$ is associated with the switching-on of the heat spot.
In more detail, Eq.~\eqref{eq:p_switch_Poisson} states that the divergence of the flow contribution $\mathbf{u}_{1,0}^\text{(S)}$ is given by the shape of the heat spot $\Theta$, a heat source. 
Physically, the heat spot causes a local increase in the volume of the fluid as it switches on, i.e.,~thermal expansion.
Mass is conserved, so there must be a fluid flux outwards from the heat spot.
We therefore expect  the instantaneous flow $\mathbf{u}_{1,0}^\text{(S)}$ to be a (2D) source flow in the far field, decaying spatially as $1/r$, where $r$ is the distance from the centre of the heat spot.

Note that experimental measurements show that the average speed of tracers over many scans of the heat spot  instead decays as $1/r^2$ in the far field~\cite{erben2021feedback}.
It is therefore important to distinguish between the instantaneous fluid velocity field induced by the heat spot during one scan (which we focus on in this section)  and the time-averaged velocity of  tracers (or material points) in the fluid.
We emphasise that existing experimental data~\cite{weinert2008optically,mittasch2018non,erben2021feedback} deal with the average velocity of tracers over many scans, not with the instantaneous fluid flow.

\begin{figure}[t]
	{\includegraphics[width=0.25\textwidth]{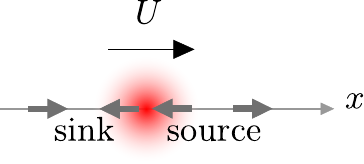} 
		\caption{Cartoon of the physical mechanism for the instantaneous flow field contribution $\mathbf{u}_{1,0}^\text{(T)}$ associated with translation of the heat spot (i.e.,~temperature perturbation) at order $\alpha$, based on the case of a constant-amplitude heat spot in Ref.~\cite{weinert2008optically}. 
			The heat-spot shape (shown in red), which is specified by the function $\Theta(x-t,y)$ in our mathematical model, translates in the positive $x$ direction.
			At the front of the heat spot, the heat spot  is arriving, so the temperature is locally increasing, $\partial \Theta/\partial t>0$, producing a hydrodynamic source.
			At the back, the heat spot is leaving, so instead $\partial \Theta/\partial t<0$, resulting in a sink. }
		\label{fig:diagram_alpha_mechanism_cartoon_translate}}
\end{figure}

The second  contribution to the velocity field, $A(t) \mathbf{u}_{1,0}^\text{(T)}(x-t,y)$, is instead proportional to the amplitude itself of the heat spot; it is nonzero even if the heat spot has constant amplitude.  
We  explain here why the flow $\mathbf{u}_{1,0}^\text{(T)}$ is associated with the translation of the heat spot, adapting for this term the physical mechanism from Ref.~\cite{weinert2008optically}. 
The velocity field $\mathbf{u}_{1,0}^\text{(T)}$  satisfies the equation
\begin{align}
	\nabla \cdot \mathbf{u}_{1,0}^\text{(T)}(x-t,y) = \frac{\partial }{\partial t} \Theta(x-t,y),\label{eq:p_translate_Poisson}
\end{align}
where we recall that the heat-spot shape $\Theta$ has the interpretation of a heat spot of constant amplitude. 
We illustrate the physical mechanism in Fig.~\ref{fig:diagram_alpha_mechanism_cartoon_translate}.
At the front of the heat spot, the forcing $\partial \Theta/\partial t$ is positive, a source term, because the translating heat spot is arriving, i.e.,~the fluid is heating up locally.
At the back of the heat spot, the forcing $\partial \Theta/\partial t$ is negative, a sink term, because the heat spot is leaving, i.e.,~the fluid is cooling down locally. 
Therefore, based on this mechanism, we expect the  velocity field $\mathbf{u}_{1,0}^\text{(T)}$ to be a hydrodynamic source dipole in the far field, decaying as $1/r^2$.

The physical mechanism presented here, for the full flow $\mathbf{u}_{1,0}$, builds on that presented in earlier work~\cite{weinert2008optically} that focused on a constant-amplitude heat spot that translates along an infinite scan path.
For that case, from the mechanism in Ref.~\cite{weinert2008optically}, the source at the front and sink at the back suggest that the far-field, leading-order instantaneous flow is a source dipole, decaying as $1/r^2$. 
This is reflected in the translation contribution in our theory.
However, for a general, time-varying heat-spot amplitude, there is an important difference between the prediction by our theory and by the constant-amplitude mechanism. 
Our theory predicts that the time-variation of the amplitude generically produces the far field of the full leading-order instantaneous flow, decaying as $1/r$, instead of the $1/r^2$ scaling purely due to heat-spot translation for the special, constant-amplitude case.
We  show these results explicitly for a Gaussian heat spot in Sec.~\ref{sec:velocity_order_alpha}, confirming the far-field behaviour in Sec.~\ref{sec:alpha_far}.

We note that the physical mechanism for the leading-order instantaneous flow does not rely on thermal viscosity changes, which are quantified by the thermal viscosity coefficient~$\beta$; this reflects the fact that thermal expansion alone is responsible for the leading-order flow.

Finally, we observe that the velocity field $\mathbf{u}_{1,0}$, given by Eq.~\eqref{eq:u10_general_decomposn}, is linear in the heat-spot amplitude $A(t)$.
To provide physical interpretation of this mathematical symmetry, we consider the case of localised cooling instead of heating.
	This could potentially be achieved in future experiments by uniformly heating a large domain except a localised spot, the ``cool" spot (in a relative sense). 
	From our theory, if we have a cool spot [$A(t) \leq 0$] instead of a heat spot, then the instantaneous flow at order $\alpha$ is reversed.
This is consistent with the mechanism described above.

\subsubsection{Pressure field}\label{sec:pressure_alpha}

We discussed above the physical mechanism for the leading-order instantaneous flow induced by a general heat spot.
To confirm this intuition mathematically, we now focus on the specific Gaussian heat spot in Eq.~\eqref{eq:temp}, i.e.,~Eq.~\eqref{eq:temp_general} with shape function $\Theta(x,y) = \exp[-(x^2+y^2)/2]$, and solve Eq.~\eqref{eq:order_alpha_original} explicitly for the pressure $p_{1,0}$ at order $\alpha$.

\begin{figure}[t]
	{\includegraphics[width=0.3\textwidth]{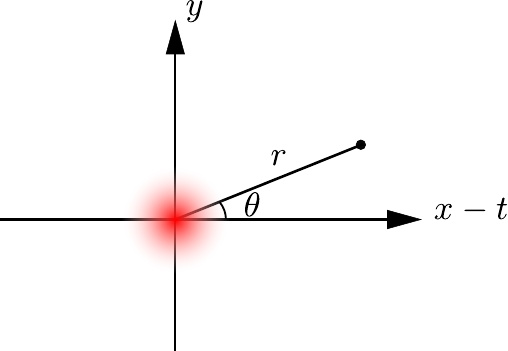}
		\caption{Polar coordinates with origin coinciding with the centre of the heat spot at time $t$.}
		\label{fig:diagram_polar_coord}}
\end{figure}
The Gaussian heat spot has circular symmetry, so we introduce plane polar coordinates $(r,\theta)$ with origin at the centre of the heat spot at time $t$ (Fig.~\ref{fig:diagram_polar_coord}), given by
\begin{align}
	x-t &= r\cos\theta,\\
	y &= r\sin\theta.
\end{align}

As explained in Sec.~\ref{sec:general_heat_spot_order_alpha}, we first solve Eq.~\eqref{eq:p_switch_Poisson} for the pressure field $p_{1,0}^\text{(S)}$ associated with the time-variation of the amplitude of the heat spot.
Since the forcing has circular symmetry, we choose an ansatz with the same symmetry, $p_{1,0}^\text{(S)} = p_{1,0}^\text{(S)}(r)$.
The Poisson equation then simplifies to the ordinary differential equation
\begin{align}
	\frac{1}{r} \frac{\partial }{\partial r} \left ( r \frac{\partial p_{1,0}^\text{(S)} }{\partial r} \right ) = - \exp(-r^2/2).
\end{align}
We integrate this, imposing the boundary condition that the solution is non-singular at the centre of the heat spot.
This gives the pressure $p_{1,0}^\text{(S)}$ as
\begin{align}
	p_{1,0}^\text{(S)}(r) = -\frac{1}{2}\E_1(r^2/2) - \ln r,\label{eq:p10_switch}
\end{align}
where the exponential integral $\E_1$ is defined as
\begin{align}
	\E_1(z) \equiv \int_z^\infty \frac{\exp(-s)}{s} \, ds.\label{eq:E1}
\end{align}

From Eq.~\eqref{eq:p10_general_decomposn}, the full pressure field $p_{1,0}$ at order $\alpha$ is thus given by
\begin{align}
	p_{1,0}(x,y,t) &= A'(t) p_{1,0}^\text{(S)}(x-t,y) + A(t) p_{1,0}^\text{(T)}(x-t,y),
\end{align}
where the pressure $p_{1,0}^\text{(T)}(x-t,y)$ is given by
\begin{align}
	p_{1,0}^\text{(T)}(x-t,y) = \frac{(x-t)[1-\exp(-r^2/2)]}{r^2}.
\end{align}

\subsubsection{Velocity field: $\mathbf{u}_{1,0}^\text{(S)}$ term associated with time-variation of heat-spot amplitude}\label{sec:velocity_order_alpha}

The corresponding full instantaneous velocity field at order $\alpha$ [repeating Eq.~\eqref{eq:u10_general_decomposn} for convenience] is given by
\begin{align}
	\mathbf{u}_{1,0}(x,y,t)
	&= \frac{\partial}{\partial t} [A(t)\mathbf{u}_{1,0}^\text{(S)}(x-t,y)]\nonumber\\
	&\equiv A'(t)\mathbf{u}_{1,0}^\text{(S)}(x-t,y) + A(t) \mathbf{u}_{1,0}^\text{(T)}(x-t,y),\label{eq:u10_decomposn}
\end{align}
where, from Eq.~\eqref{eq:u_switch_general} and Eq.~\eqref{eq:u_translate_general}, the two contributing, irrotational velocity fields are given by
\begin{align}
	\mathbf{u}_{1,0}^\text{(S)}(x-t,y) 
	=&  \frac{(x-t)[1-\exp(-r^2/2)]}{r^2} \mathbf{e}_x + \frac{y[1-\exp(-r^2/2)]}{r^2} \mathbf{e}_y\nonumber\\
	\equiv&  \frac{1-\exp(-r^2/2)}{r} \mathbf{e}_r,\label{eq:u_switch}\\
	\mathbf{u}_{1,0}^\text{(T)}(x-t,y) =& -\mathbf{e}_x \left \{\frac{1}{r^2} - \frac{2(x-t)^2}{r^4} + \left [-\frac{1}{r^2} + \frac{2(x-t)^2}{r^4} + \frac{(x-t)^2}{r^2} \right ] \exp(-r^2/2) \right \} \nonumber\\
	&- \mathbf{e}_y \left \{-\frac{2(x-t)y}{r^4} + \left [\frac{2(x-t)y}{r^4} + \frac{(x-t)y}{r^2}\right ] \exp(-r^2/2) \right \}.\label{eq:u_translate}
\end{align}
In the above, $\mathbf{e}_x$ and $\mathbf{e}_y$ are unit vectors in the $x$ and $y$ directions, respectively; $\mathbf{e}_r\equiv \frac{x-t}{r} \mathbf{e}_x + \frac{y}{r} \mathbf{e}_y$ is the radial unit vector (from the centre of the heat spot). 
This is a linear superposition of two separate flows, as interpreted physically in Sec.~\ref{sec:physical_mechanism_order_alpha}: $A'(t)\mathbf{u}_{1,0}^\text{(S)}(x-t,y)$ associated with the time-variation of the amplitude of the heat spot (``switching on" and ``switching off") and $A(t)\mathbf{u}_{1,0}^\text{(T)}(x-t,y)$ associated with the translation of the heat spot.
We now examine the flow fields $\mathbf{u}_{1,0}^\text{(S)}$ and $\mathbf{u}_{1,0}^\text{(T)}$.

\begin{figure*}[t]
	\subfloat[]
	{\includegraphics[width=0.5\textwidth]{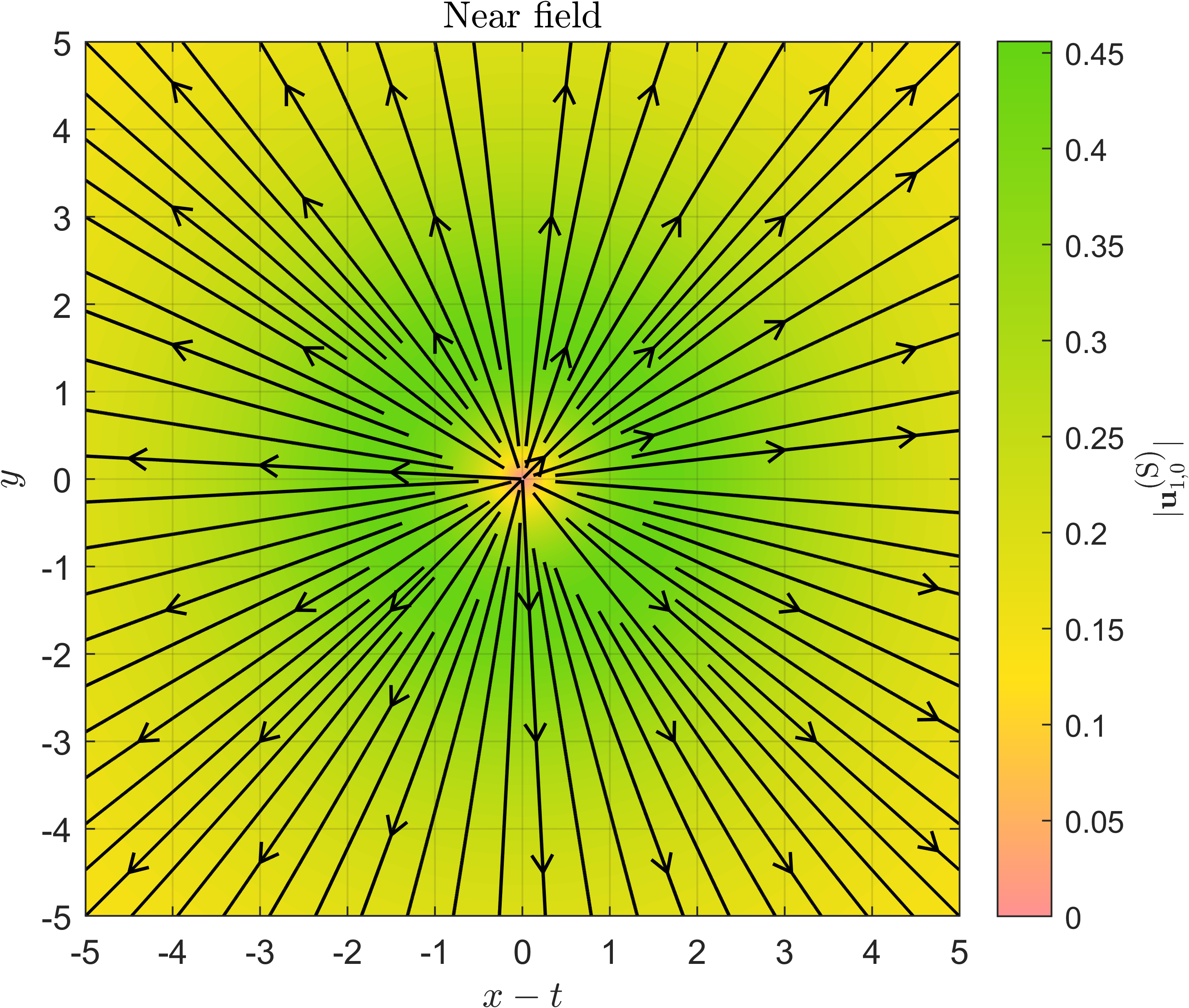}
		\label{fig:plot_u10_switch_streamlines_near}}
	\subfloat[]
	{\includegraphics[width=0.5\textwidth]{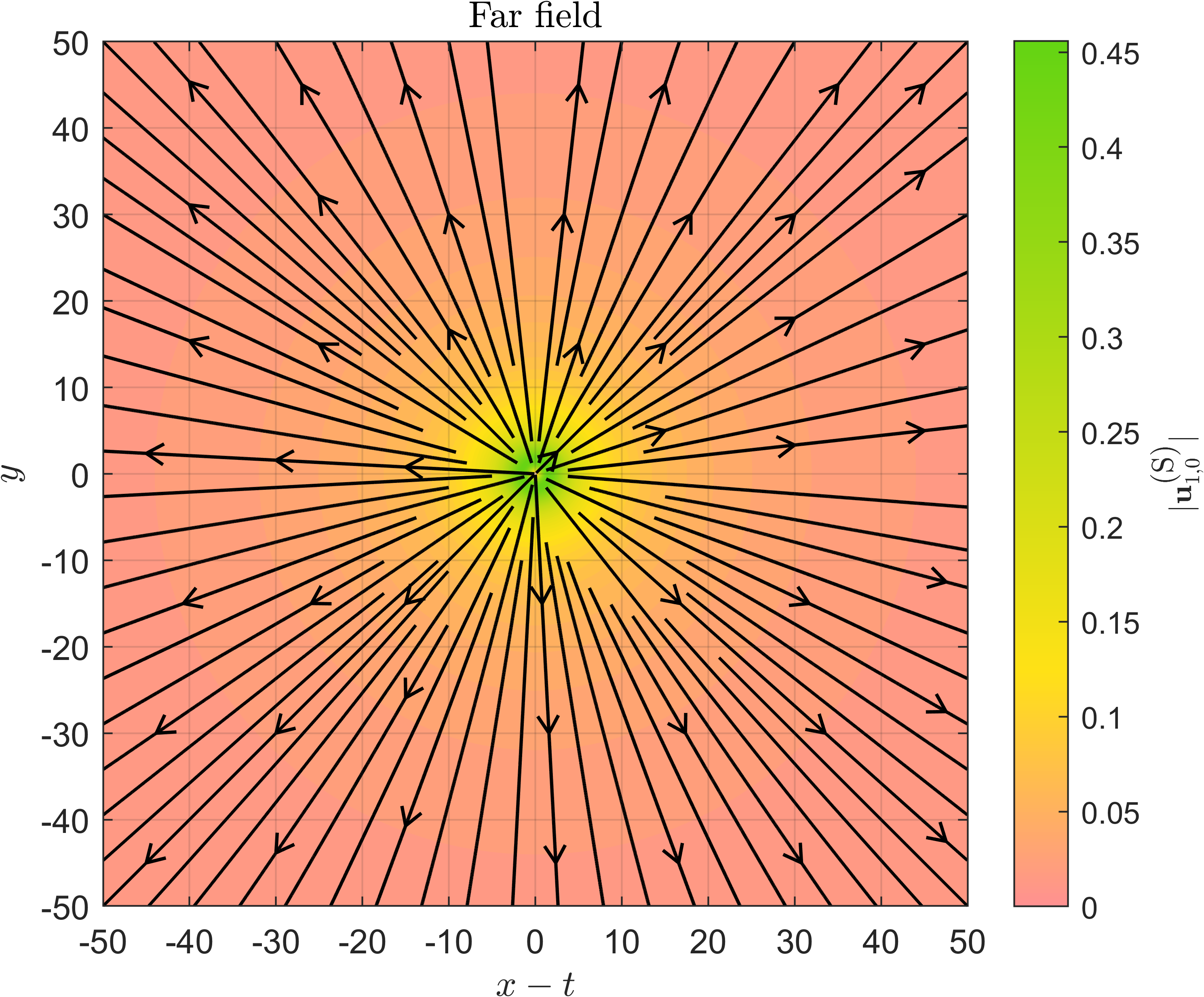}
		\label{fig:plot_u10_switch_streamlines_far}} 
	\caption{Streamlines for the flow field $\mathbf{u}_{1,0}^\text{(S)}(x-t,y)$ [as in Eq.~\eqref{eq:u10_general_decomposn} for the instantaneous flow at order $\alpha$] associated with the time-variation of the heat-spot amplitude.  Left (Fig.~\ref{fig:plot_u10_switch_streamlines_near}): streamlines for $-5 \leq x-t, y \leq 5$, close to the heat spot (near field), with magnitude of the velocity field $ \vert \mathbf{u}_{1,0}^\text{(S)}  \vert$ indicated by colour. Right (Fig.~\ref{fig:plot_u10_switch_streamlines_far}): streamlines for $-50 \leq x-t,y \leq 50$ to illustrate far-field behaviour.}
	\label{fig:plot_u10_switch_streamlines}
\end{figure*}
\begin{figure}[t]
	{\includegraphics[width=0.6\textwidth]{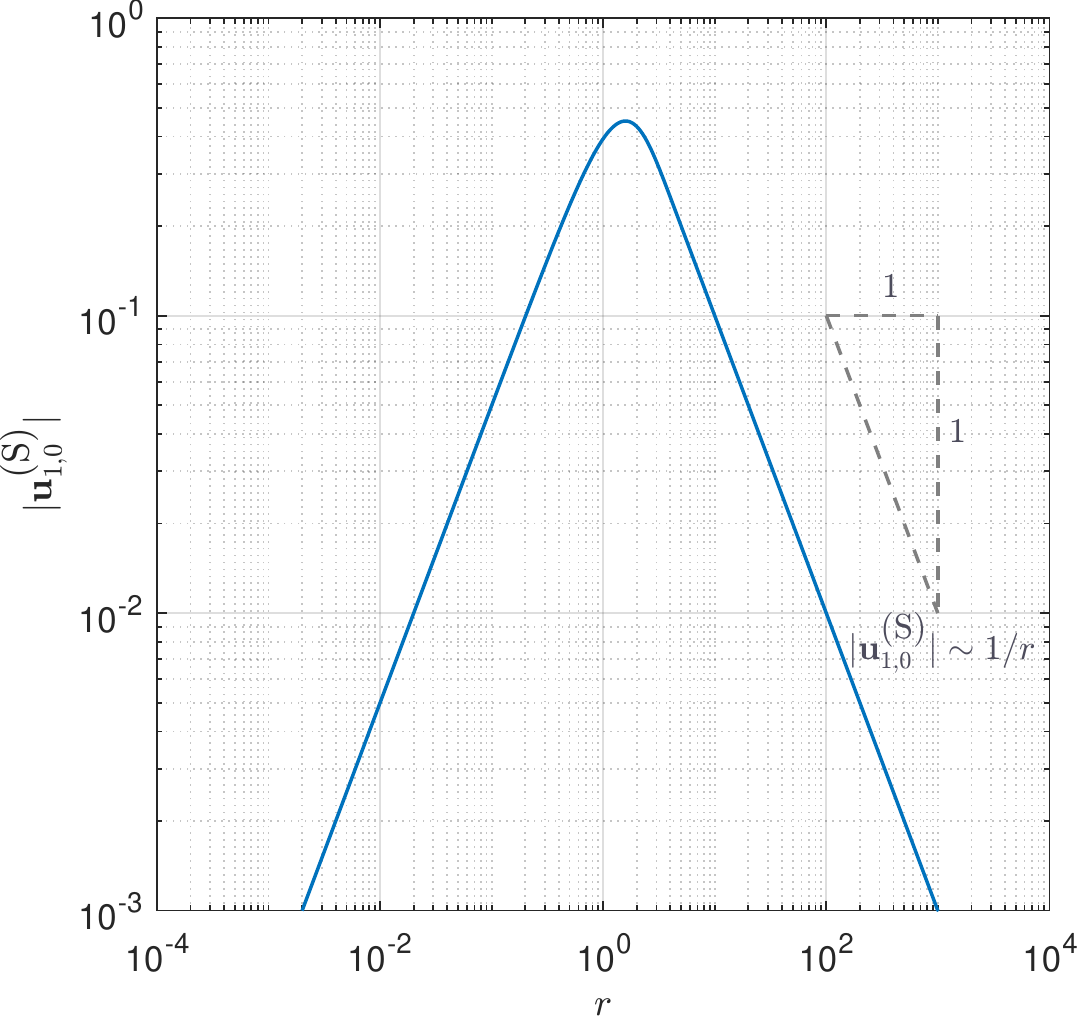}
		\caption{Magnitude of the flow field at order $\alpha$ associated with the time-variation of the heat-spot amplitude, $ \vert \mathbf{u}_{1,0}^\text{(S)} \vert$ [Eq.~\eqref{eq:magnitude_u10S}], against the radial distance $r$ to the centre of the heat spot, plotted on a log--log scale.
		The flow is a regularised source: it is zero at the centre of the heat spot, while its  far-field behaviour is given by $\vert \mathbf{u}_{1,0}^\text{(S)} \vert \sim 1/r$.}
		\label{fig:plot_u_switch_speed_radius}}
\end{figure}
We illustrate the flow field $\mathbf{u}_{1,0}^\text{(S)}$ in Fig.~\ref{fig:plot_u10_switch_streamlines}, with the near field in Fig.~\ref{fig:plot_u10_switch_streamlines_near} and the far field in Fig.~\ref{fig:plot_u10_switch_streamlines_far}. (Recall that the length scales are in units of $a$, the characteristic radius of the Gaussian temperature profile.)
The flow is purely radial, inheriting the symmetry of the heat spot.
Its magnitude, illustrated on a log--log scale in Fig.~\ref{fig:plot_u_switch_speed_radius}, is given by
\begin{align}
	\vert \mathbf{u}_{1,0}^\text{(S)} \vert = \frac{1-\exp(-r^2/2)}{r}.\label{eq:magnitude_u10S}
\end{align}
At the centre of the heat spot, this is zero, so we have a regularised version of the flow due to a point source.
In the far field, we have
\begin{align}
	\mathbf{u}_{1,0}^\text{(S)}(x-t,y) 
	\sim  \frac{(x-t)}{r^2} \mathbf{e}_x + \frac{y}{r^2} \mathbf{e}_y \equiv \frac{1}{r} \mathbf{e}_r.\label{eq:u_switch_far-field}
\end{align}
This is a source flow, centred on the heat spot, as we expect from the physical mechanism (Sec.~\ref{sec:physical_mechanism_order_alpha}).
The magnitude of this flow correspondingly decays as $\vert \mathbf{u}_{1,0}^\text{(S)} \vert \sim 1/r$.

\subsubsection{Velocity field: $\mathbf{u}_{1,0}^\text{(T)}$ term associated with translation of the heat spot}

\begin{figure*}[t]
	\subfloat[]
	{\includegraphics[width=0.544\textwidth]{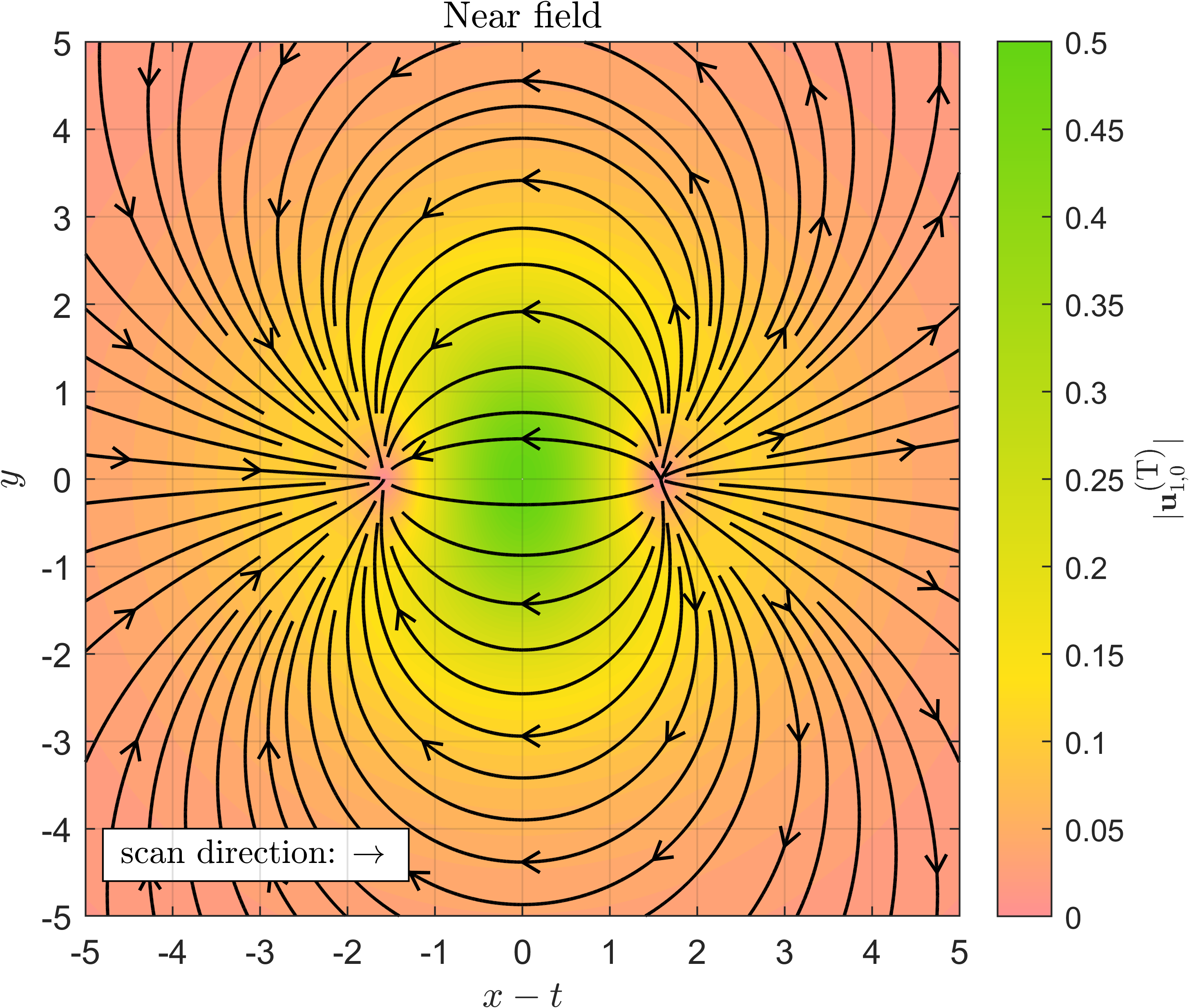}
		\label{fig:plot_u_translate_streamlines_near}}
	\subfloat[]
	{\includegraphics[width=0.456\textwidth]{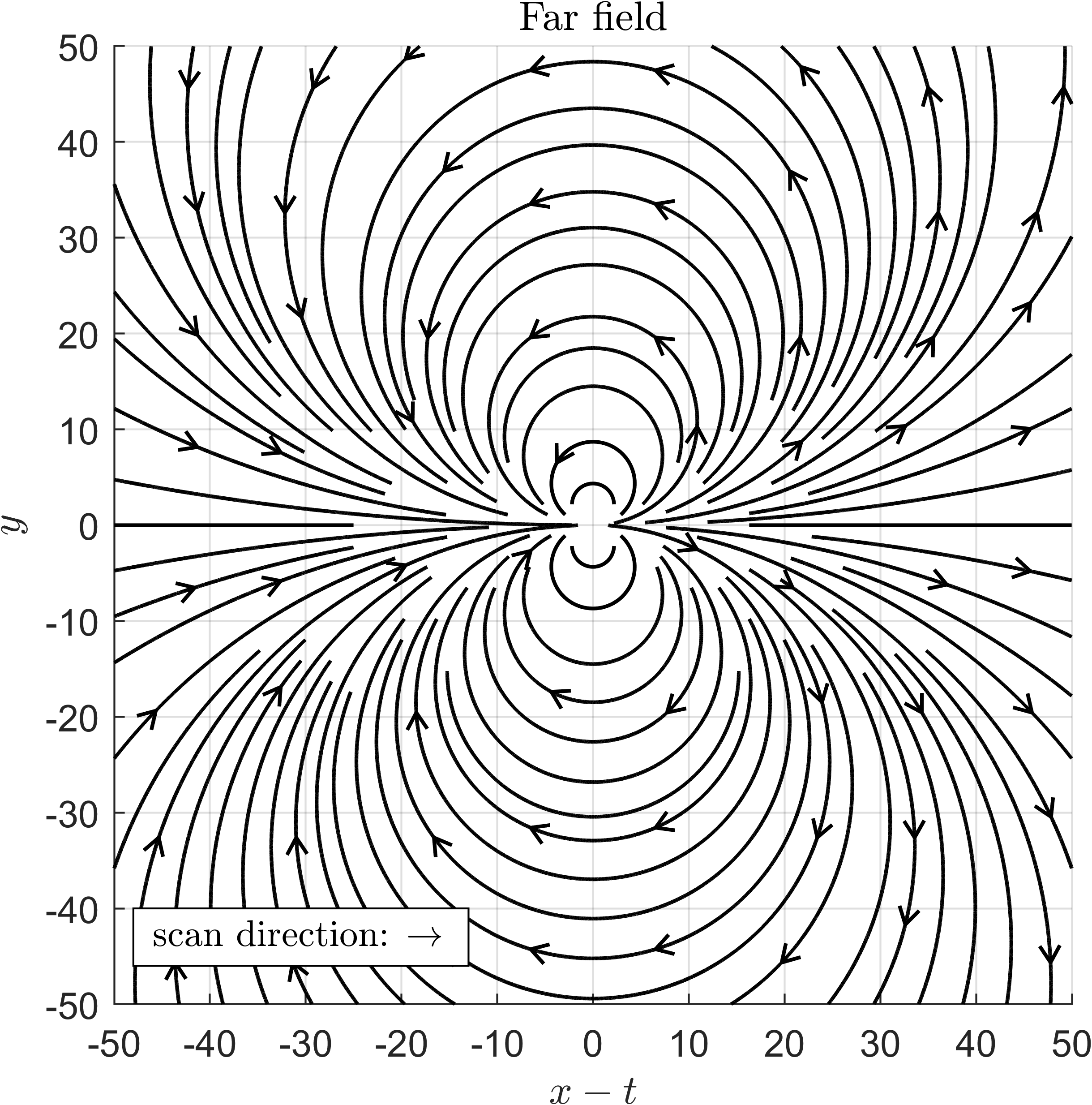}
		\label{fig:plot_u_translate_streamlines_far}} 
	\caption{Streamlines for the velocity field $\mathbf{u}_{1,0}^\text{(T)}(x-t,y)$ [as in Eq.~\eqref{eq:u10_general_decomposn} for the instantaneous flow at order $\alpha$] associated with translation of the heat spot in the positive $x$ direction (scan direction).  Left (Fig.~\ref{fig:plot_u_translate_streamlines_near}): streamlines for $-5 \leq x-t, y \leq 5$, close to the heat spot (near field), with magnitude of the velocity field $ \vert \mathbf{u}_{1,0}^\text{(T)}  \vert$ indicated by colour. Right (Fig.~\ref{fig:plot_u_translate_streamlines_far}): streamlines for $-50 \leq x-t,y \leq 50$ to illustrate far-field behaviour.}
	\label{fig:plot_u_translate_streamlines}
\end{figure*}
\begin{figure}[t]
	{\includegraphics[width=0.6\textwidth]{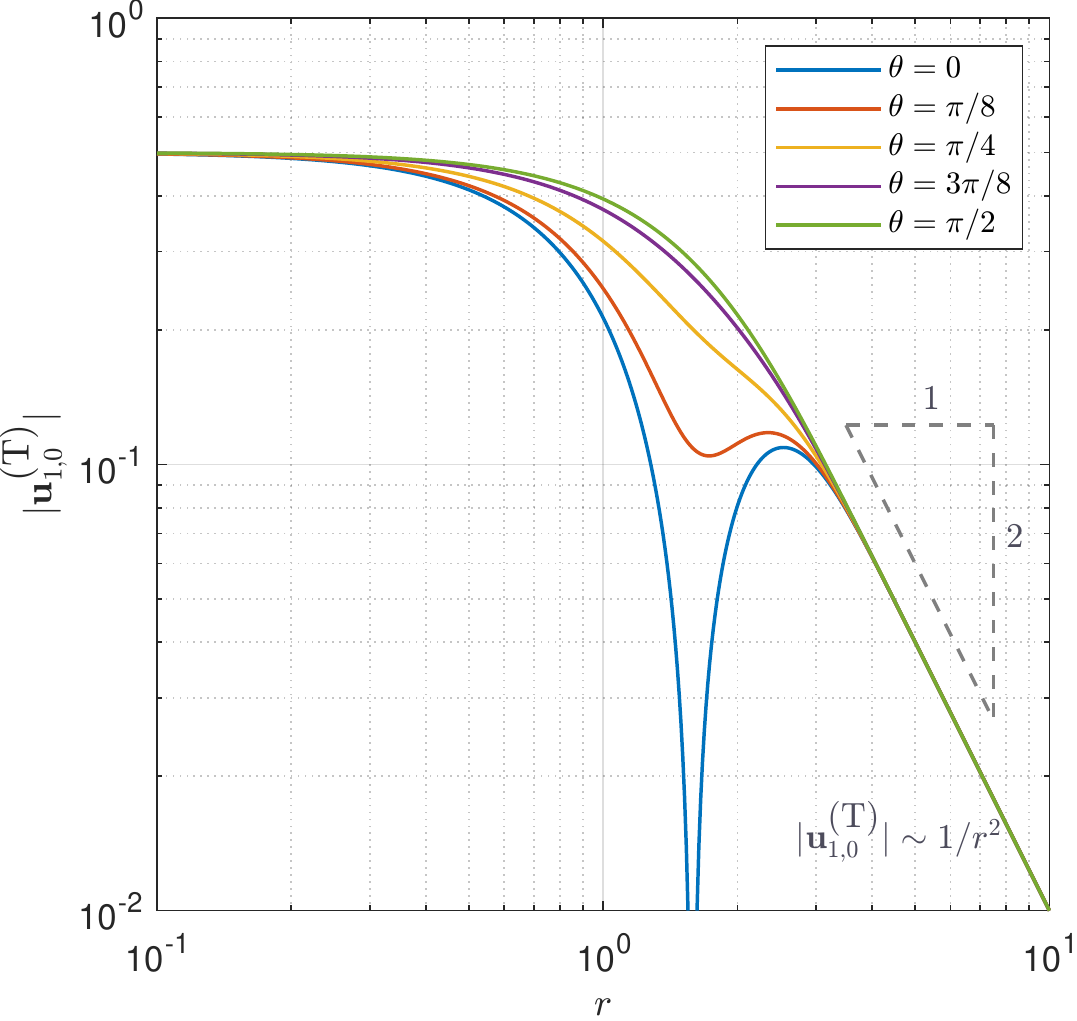}
		\caption{Magnitude of the flow field at order $\alpha$ associated with translation of the heat spot, $\vert \mathbf{u}_{1,0}^\text{(T)} \vert$ [Eq.~\eqref{eq:magnitude_u10T}], against the radial distance $r$ to the centre of the heat spot, plotted on a log--log scale, for $\theta = 0$, $\pi/8$, $\pi/4$, $3\pi/8$, and $\pi/2$.
		The far-field behaviour is given by $\vert \mathbf{u}_{1,0}^\text{(T)}   \vert \sim 1/r^2$.}
		\label{fig:plot_u_translate_speed_radius}}
\end{figure}
We illustrate in Fig.~\ref{fig:plot_u_translate_streamlines} the flow field $\mathbf{u}_{1,0}^\text{(T)}$ associated with translation of the heat spot, with a plot of the streamlines in the near field in Fig.~\ref{fig:plot_u_translate_streamlines_near} and the far field in Fig.~\ref{fig:plot_u_translate_streamlines_far}.
The magnitude, $\vert \mathbf{u}_{1,0}^\text{(T)} \vert$, is given by
\begin{align}
	\vert \mathbf{u}_{1,0}^\text{(T)}  \vert = \frac{1}{r^2} \{ 
	[(1+r^2)\exp(-r^2/2)-1]^2 + r^2 \exp(-r^2)  [2 \exp(r^2/2) - (2+r^2)]\sin^2\theta\}^{1/2},\label{eq:magnitude_u10T}
\end{align}
which is an increasing function of $\sin^2\theta$ at any fixed radius $r$.
The magnitude $\vert \mathbf{u}_{1,0}^\text{(T)} \vert$ is visualised in Fig.~\ref{fig:plot_u_translate_speed_radius}, a log--log plot of the speed $ \vert \mathbf{u}_{1,0}^\text{(T)}    \vert$ against the radius $r$,  along radial lines $\theta = 0$, $\pi/8$, $\pi/4$, $3\pi/8$, and $\pi/2$.

These plots support the physical mechanism for this flow, $\mathbf{u}_{1,0}^\text{(T)}$, proposed in Sec.~\ref{sec:physical_mechanism_order_alpha}.
We see in Fig.~\ref{fig:plot_u_translate_streamlines_near} a source on the right, where the heat spot is arriving so that the fluid is heating up, and a sink on the left, where the heat spot is leaving so that the fluid is cooling down.
In agreement with this mechanism, we have shown mathematically that the far field of the flow  $\mathbf{u}_{1,0}^\text{(T)}(x-t,y)$ is a hydrodynamic source dipole, given by
\begin{align}
	\mathbf{u}_{1,0}^\text{(T)}(x-t,y) \sim -  \left \{  \mathbf{e}_x \left [\frac{1}{r^2} - \frac{2(x-t)^2}{r^4}  \right]   + \mathbf{e}_y \left [-\frac{2(x-t)y}{r^4}\right]  \right \},\label{eq:u_translate_far-field}
\end{align}
with magnitude decaying as $\vert \mathbf{u}_{1,0}^\text{(T)} \vert \sim 1/r^2$.

\subsubsection{Far-field behaviour}\label{sec:alpha_far}

As a reminder, the full instantaneous velocity field $\mathbf{u}_{1,0}$ at order $\alpha$ is a linear superposition [given by Eq.~\eqref{eq:u10_decomposn} for general amplitude function $A(t)$] of the flows $\mathbf{u}_{1,0}^\text{(S)}$ [Eq.~\eqref{eq:u_switch}] and $\mathbf{u}_{1,0}^\text{(T)}$ [Eq.~\eqref{eq:u_translate}]. 
We have seen that the far field of $\mathbf{u}_{1,0}^\text{(S)}$ is a source flow [Eq.~\eqref{eq:u_switch_far-field}], which decays as $1/r$, and the far field of $\mathbf{u}_{1,0}^\text{(T)}$ is a source dipole [Eq.~\eqref{eq:u_translate_far-field}], which decays as $1/r^2$.
Therefore, provided $A'(t) \neq 0$, 
the far field of the full instantaneous flow $\mathbf{u}_{1,0}$ at order $\alpha$ is given by
\begin{align}
	\mathbf{u}_{1,0}(x,y,t) 
	\sim \frac{A'(t)}{r}\mathbf{e}_r,
\end{align} 
i.e.,~a source flow, proportional to the rate of change of the amplitude of the heat spot and decaying spatially as $1/r$.
This dominates over the contribution due to the translation of the heat spot, thus illustrating the importance of the time-dependence of the heat-spot amplitude in our model.
For example, this is the relevant behaviour for the case of a sinusoidal amplitude function, which we will illustrate later in this section.

At times $t$ such that the  heat-spot amplitude function  is stationary with respect to time [i.e.,~at a local maximum or minimum, $A'(t)=0$], the far-field behaviour of the instantaneous flow $\mathbf{u}_{1,0}$ is instead given by
\begin{align}
	\mathbf{u}_{1,0}(x,y,t) \sim -A(t) \left \{  \mathbf{e}_x \left [\frac{1}{r^2} - \frac{2(x-t)^2}{r^4}  \right ]  + \mathbf{e}_y \left [-\frac{2(x-t)y}{r^4}\right ] \right \},\label{eq:u10_far_A_const}
\end{align}
which is a hydrodynamic source dipole, proportional to the amplitude of the heat spot, and decaying spatially as $1/r^2$ (thus slower than the previous case).
This is relevant if, for example,  the amplitude of the heat spot remains constant at some maximum value for most of the scan period (when the  heat spot is away from the ends of the scan path where it switches on and off) or alternatively, as in previous theoretical modelling~\cite{weinert2008optically}, the heat-spot amplitude is identically a constant and the scan path is infinitely long.

\subsubsection{Full velocity field}

In our work so far, we have allowed the amplitude function $A(t)$ to be fully general.
To visualise the full instantaneous velocity field $\mathbf{u}_{1,0}(x,y,t)$ at order $\alpha$ at a given time $t$ [Eqs.~\eqref{eq:u10_decomposn},~\eqref{eq:u_switch}, and~\eqref{eq:u_translate}] in this section, we need to choose a particular amplitude function $A(t)$.

\begin{figure}[t]
	{\includegraphics[width=\textwidth]{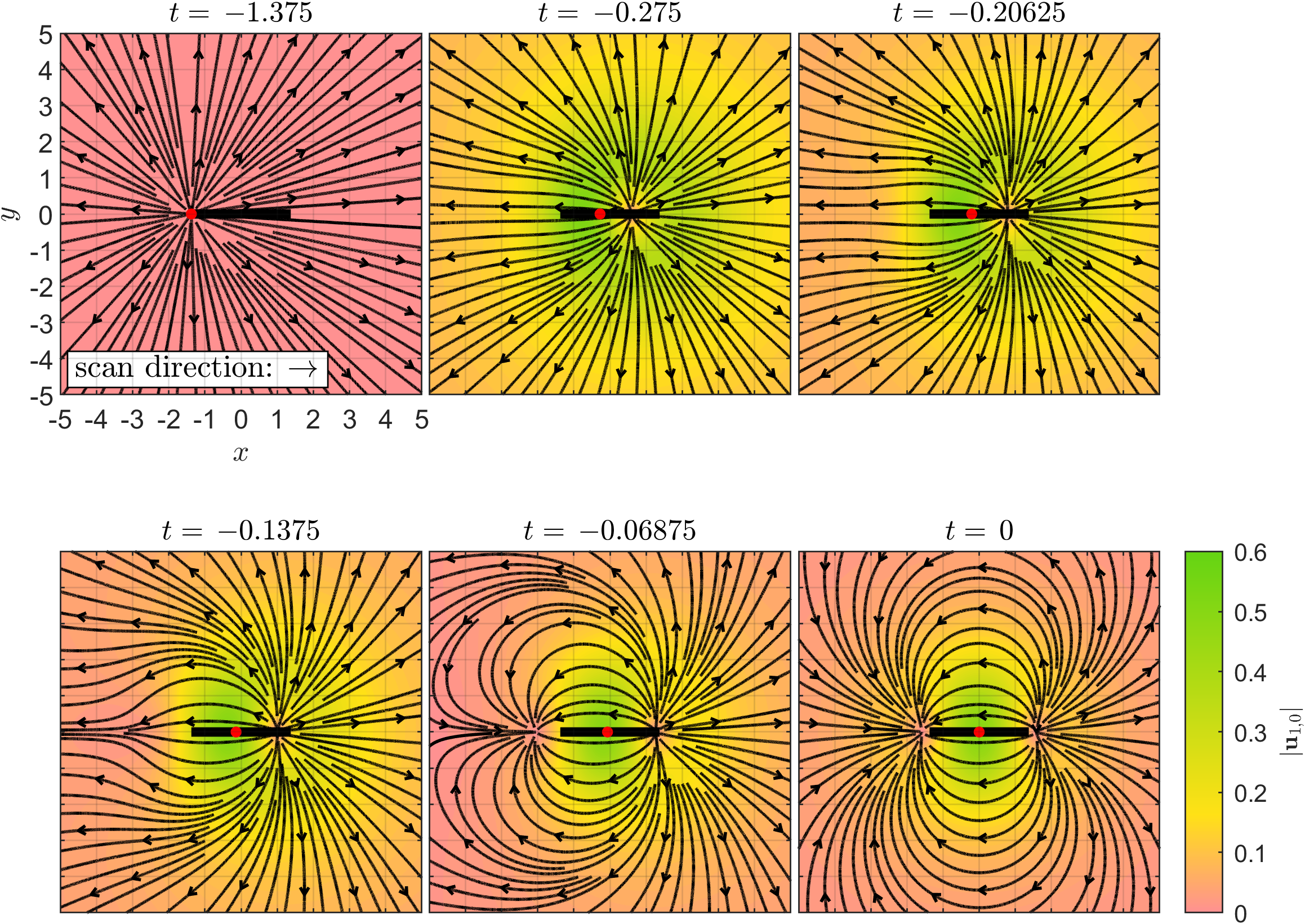}
		\caption{Snapshots of the streamlines of the instantaneous fluid velocity field $\mathbf{u}_{1,0}$ at order $\alpha$, i.e.,~leading order, over the course of half a scan period, $-t_0 \leq t \leq 0$, with $t_0=1.375$ and sinusoidal amplitude function given by Eq.~\eqref{eq:A_sinusoidal}. 
		The heat spot translates in the positive $x$ direction. 
		The centre of the heat spot is indicated with a red dot, while the scan path is shown as a thick black line segment.}
		\label{fig:plot_u10_streamlines_snapshots}}
\end{figure}
Following  Ref.~\cite{mittasch2018non}, we choose a sinusoidal amplitude function $A(t)$ given by
\begin{equation}
	A(t) = \cos^2 \left ( \frac{\pi t}{2 t_0} \right ),\label{eq:A_sinusoidal}
\end{equation}
valid for $-t_0 \leq t \leq t_0$, 
where  $t_0=\ell$ is half the scan period in dimensionless terms.
For this amplitude function [Eq.~\eqref{eq:A_sinusoidal}] and for $t_0 = 1.375$ (to match experiments~\cite{erben2021feedback}), we plot the instantaneous streamlines of the velocity field $\mathbf{u}_{1,0}$ at order $\alpha$ in Fig.~\ref{fig:plot_u10_streamlines_snapshots} over the course of the first half of a scan period.
The centre of the heat spot is indicated with a red dot.
The streamlines for the second half of the scan period may be obtained by symmetry.

As  expected, near the start of the scan period (e.g.,~panels for $t=-1.375$ and $t=-0.275$), when the amplitude of the heat spot is small and increasing, the instantaneous flow is dominated by the contribution $A'(t)\mathbf{u}_{1,0}^\text{(S)}(x-t,y)$, associated with the time-variation of the amplitude of the heat spot.
The heat spot is switching on, giving a source flow.

At the instant $t=0$ (halfway through the scan period) when  the amplitude of the heat spot reaches its maximum value, there is no switching-on contribution to the flow.
The instantaneous flow is then simply given by the contribution $A(t) \mathbf{u}_{1,0}^\text{(T)}(x-t,y)$, associated with the translation of the heat spot, with a source at the front and a sink at the back.

For the sinusoidal amplitude function chosen in Eq.~\eqref{eq:A_sinusoidal}, the characteristic time scale over which the heat spot switches on and off is the same as the scan period, as in the experiments and numerical simulations of Ref.~\cite{mittasch2018non}. 
Except at $t=-t_0$, $t=0$, and $t=t_0$, the rate of change of the amplitude is nonzero.
For this sinusoidal amplitude function, we therefore predict that the instantaneous flow field is generically dominated by a source or sink flow in the far field, decaying as $1/r$, not by a hydrodynamic source dipole (decaying as $1/r^2$).

\begin{figure}[t]
	{\includegraphics[width=\textwidth]{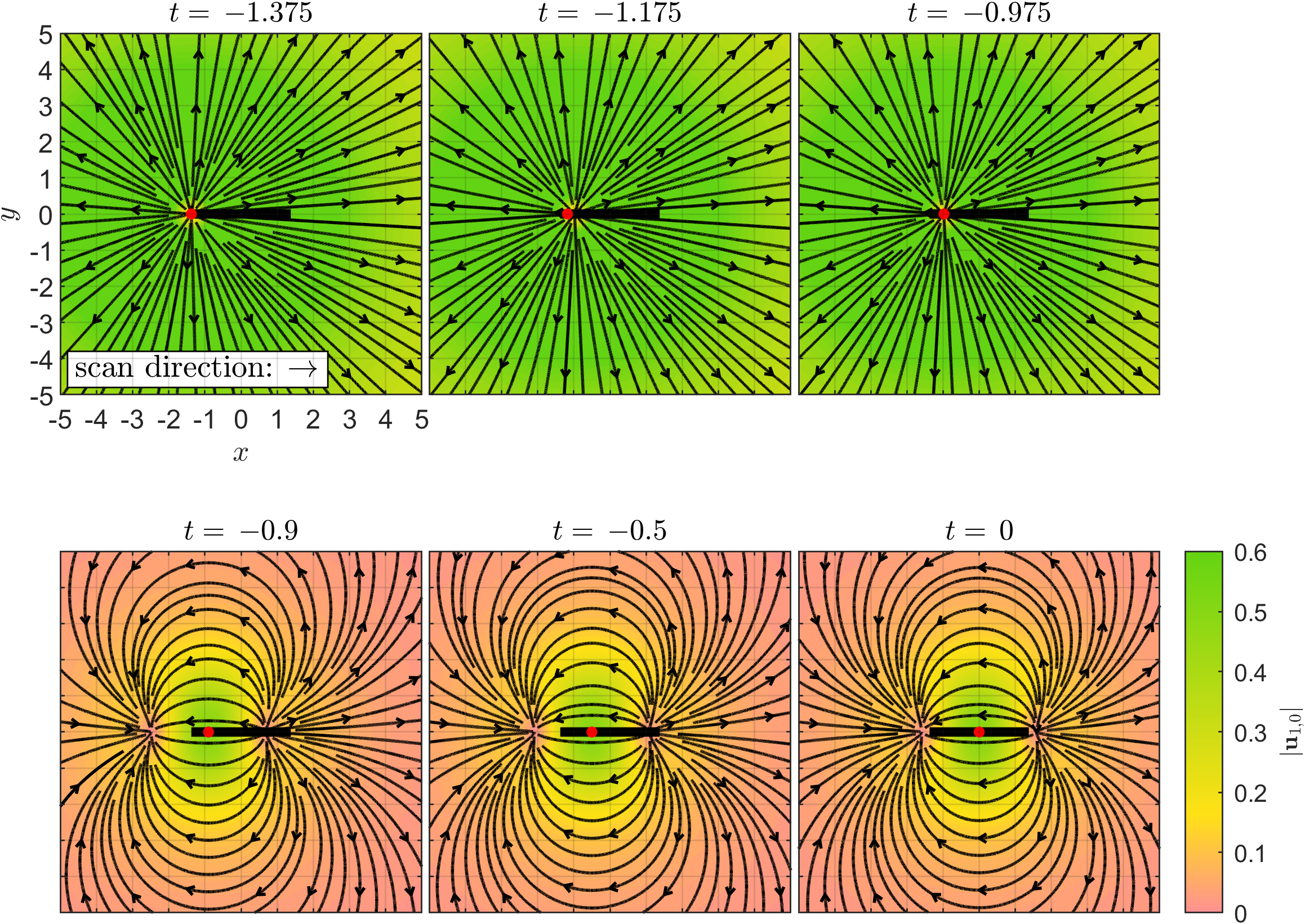} 
		\caption{Same as Fig.~\ref{fig:plot_u10_streamlines_snapshots} but with a trapezoidal heat-spot amplitude function instead of sinusoidal, illustrating the case where the switching-on and switching-off of the heat spot are rapid and the amplitude is constant for most of the scan period. 
		Snapshots of the streamlines of the instantaneous fluid velocity field $\mathbf{u}_{1,0}$ at order $\alpha$, i.e.,~leading order, over the course of half a scan period, $-t_0 \leq t \leq 0$, with $t_0=1.375$ and trapezoidal amplitude function given by Eq.~\eqref{eq:A_trapezoidal} with switching-on time $t_\text{s}=0.4$. 
		The heat spot translates in the positive $x$ direction. 
		The centre of the heat spot is indicated with a red dot, while the scan path is shown as a thick black line segment.}
		\label{fig:plot_u10_streamlines_snapshots_trapezium_amplitude}}
\end{figure}

Clearly, alternative amplitude functions are possible. 
In other  applications of thermoviscous flows, the heat spot may instead switch on quickly, then translate at constant amplitude for most of the scan period, before switching off quickly. 
An example of such a ``trapezoidal" amplitude function is given by
\begin{align}
	A(t) = 
	\begin{cases}
		\frac{1}{t_\text{s}}(t+t_0), & \text{ if } -t_0 \leq t \leq -t_0 + t_\text{s},\\
		1, & \text{ if } -t_0 + t_\text{s} \leq t \leq t_0 - t_\text{s},\\
		-\frac{1}{t_\text{s}}(t-t_0), & \text{ if } t_0 - t_\text{s} \leq t \leq t_0,
	\end{cases}\label{eq:A_trapezoidal}
\end{align}
where $t_\text{s}$ is the ``switching-on" time, over which the amplitude changes from $0$ to $1$ or vice versa.
The limit of $t_\text{s}\to 0$ corresponds to a heat spot that remains approximately stationary while it switches on quickly at the start of the scan path, then translates at constant amplitude to the end of the scan path, where it switches off.
For this amplitude function, we illustrate the instantaneous streamlines of the fluid velocity $\mathbf{u}_{1,0}$ at order $\alpha$ during the first half of a scan period in Fig.~\ref{fig:plot_u10_streamlines_snapshots_trapezium_amplitude} (for the case $t_\text{s}=0.4$). 
In this case, the instantaneous flow field is  a source dipole in the far field for most of the scan period. 
We remark that even though the amplitude is constant for most of the scan period, the fact that the heat spot switches on and off at the ends of the scan path is crucial for explaining the resulting net displacement of tracers.

\subsection{Solution at order $\alpha\beta$}\label{sec:order_alpha_beta}

In Sec.~\ref{sec:order_alpha}, we solved for the leading-order instantaneous flow (order $\alpha$), for a heat spot of arbitrary amplitude and Gaussian shape.
We now proceed to the quadratic terms in the perturbation expansion for the flow field [Eq.~\eqref{eq:vel_pert_exp}], as these give rise to the leading-order net displacement of tracers (Sec.~\ref{sec:net_displ}), as in experiments.

As discussed in Sec.~\ref{sec:order_beta_n}, order $\alpha\beta$ provides the first effect of thermal viscosity changes, since the flow at order $\beta$ and order $\beta^2$ is zero. 
Furthermore, for the particular fluid (glycerol-water solution) in experiments in Ref.~\cite{erben2021feedback} that we  compare our theory with in Sec.~\ref{sec:comparison_expt}, the thermal viscosity coefficient~$\beta$ is much larger than the thermal expansion coefficient~$\alpha$.
Hence, we begin  in this section with order $\alpha\beta$;  we study the flow at order $\alpha^2$ (which could be important for other liquids) in the next section.
The key result in Eqs.~\eqref{eq:u11_translate} and~\eqref{eq:u11} is an explicit, analytical formula for the instantaneous flow at order $\alpha\beta$ induced by the heat spot; we plot the streamlines in Fig.~\ref{fig:plot_u11_streamlines} and explain the physical mechanism for the flow in Sec.~\ref{sec:physical_mechanism_order_alphabeta}.

\subsubsection{General heat spot}

The Poisson equation for the pressure at order $\alpha\beta$, from Eq.~\eqref{eq:pressure_expanded}, is  given by
\begin{align}
	\nabla^2 p_{1,1} &= - \nabla \cdot (\Delta T \, \nabla p_{1,0})\nonumber\\
	&= \nabla \cdot (\Delta T \, \mathbf{u}_{1,0}),\label{eq:order_alphabeta_original}
\end{align}
where we have used the result  that the pressure at order $\beta$ is zero  to simplify the equation and we have used Eq.~\eqref{eq:vel_press} at order $\alpha$ to write the forcing in terms of the velocity field at order $\alpha$.
The result in Eq.~\eqref{eq:order_alphabeta_original} holds for any temperature profile $\Delta T(x,y,t)$ and states that the flow at order $\alpha\beta$ is incompressible.

As in Sec.~\ref{sec:general_heat_spot_order_alpha}, we now consider the general heat spot given by Eq.~\eqref{eq:temp_general} and reproduced below for convenience,
\begin{align}
	\Delta T(x,y,t) = A(t) \Theta(x-t,y),
\end{align}
where $A(t)$ is the amplitude function and $\Theta(x-t,y)$ is  the shape function, steady in the frame translating with the heat spot.
Recall that at order $\alpha$ we decomposed the velocity field $\mathbf{u}_{1,0}$ [Eq.~\eqref{eq:u10_general_decomposn}] into two contributions, due to the time-variation of the heat-spot amplitude and due to the translation of the heat spot. 
Since Eq.~\eqref{eq:order_alphabeta_original} is linear in the pressure $p_{1,1}$ at order $\alpha\beta$, we can similarly decompose $p_{1,1}$ as
\begin{align}
	p_{1,1}(x,y,t) = A(t) A'(t) p_{1,1}^\text{(S)}(x-t,y) + A(t)^2 p_{1,1}^\text{(T)}(x-t,y),\label{eq:p11_decomposn}
\end{align}
where the two contributing pressure fields $ p_{1,1}^\text{(S)}(x-t,y)$ and $ p_{1,1}^\text{(T)}(x-t,y)$ satisfy the Poisson equations
\begin{align}
	\nabla^2 p_{1,1}^\text{(S)}(x-t,y) &= \nabla \cdot [\Theta(x-t,y) \mathbf{u}_{1,0}^\text{(S)}(x-t,y)],\label{eq:Poisson_p11_switch} \\
	\nabla^2 p_{1,1}^\text{(T)}(x-t,y) &= \nabla \cdot [\Theta(x-t,y) \mathbf{u}_{1,0}^\text{(T)}(x-t,y)],\label{eq:Poisson_p11_translate}
\end{align}
respectively. 
The  velocity field at order $\alpha\beta$, using Eq.~\eqref{eq:vel_press}, is given by
\begin{align}
	\mathbf{u}_{1,1} = -\nabla p_{1,1} + \Delta T \, \mathbf{u}_{1,0},\label{eq:u11_p11_Theta_u10}
\end{align}
which can be decomposed as
\begin{align}
	\mathbf{u}_{1,1}(x,y,t) = A(t) A'(t) \mathbf{u}_{1,1}^\text{(S)}(x-t,y) + A(t)^2 \mathbf{u}_{1,1}^\text{(T)}(x-t,y),\label{eq:u11_decomposn}
\end{align}
where
\begin{align}
	\mathbf{u}_{1,1}^\text{(S)}(x-t,y) &= -\nabla p_{1,1}^\text{(S)}(x-t,y) + \Theta(x-t,y)\mathbf{u}_{1,0}^\text{(S)}(x-t,y),\label{eq:u11_switch_general} \\
	\mathbf{u}_{1,1}^\text{(T)}(x-t,y) &= -\nabla p_{1,1}^\text{(T)}(x-t,y) + \Theta(x-t,y)\mathbf{u}_{1,0}^\text{(T)}(x-t,y).\label{eq:u11_translate_general}
\end{align}

\subsubsection{Pressure field $p_{1,1}^\text{(S)}$, velocity field $\mathbf{u}_{1,1}^\text{(S)}$, and physical mechanism associated with time-variation of heat-spot amplitude for heat spot with circular symmetry}\label{sec:u11_switch}

We now specialise to the case of a heat spot with circular symmetry, $\Theta(x-t,y) = \Theta(r)$. 
In this section, we  solve Eq.~\eqref{eq:Poisson_p11_switch} for the pressure $p_{1,1}^\text{(S)}$ at order $\alpha\beta$ associated with the time-variation of the heat-spot amplitude and then deduce that the corresponding velocity field $\mathbf{u}_{1,1}^\text{(S)}$ is zero.

For a heat spot with circular symmetry, both the forcing and the boundary conditions for the Poisson equation for the pressure field $p_{1,0}^\text{(S)}$ at order $\alpha$  [Eq.~\eqref{eq:p_switch_Poisson}]  have circular symmetry, so that the solution $p_{1,0}^\text{(S)}$ and hence the velocity field $\mathbf{u}_{1,0}^\text{(S)}$  inherit the same symmetry. 
Consequently, at order $\alpha\beta$,  the forcing in Eq.~\eqref{eq:Poisson_p11_switch} has circular symmetry, which is inherited by  the solution for the pressure, $p_{1,1}^\text{(S)} = p_{1,1}^\text{(S)}(r)$. 
The Poisson equation  [Eq.~\eqref{eq:Poisson_p11_switch}] is therefore an ordinary differential equation in $r$ given by
\begin{align}
	\frac{1}{r} \frac{\partial}{\partial r} \left (r \frac{\partial p_{1,1}^\text{(S)}}{\partial r} \right ) = -\frac{1}{r} \frac{\partial}{\partial r} \left ( r \Theta \frac{\partial p_{1,0}^\text{(S)}}{\partial r} \right ).
\end{align}
Integrating this, we find
\begin{align}
	\frac{\partial p_{1,1}^\text{(S)}}{\partial r} = - \Theta \frac{\partial p_{1,0}^\text{(S)}}{\partial r},
\end{align}
where we have set the constant of integration to be zero because the velocity is not singular at the centre of the heat spot.
We note that in vector form, this is simply
\begin{align}
	\nabla p_{1,1}^\text{(S)} =  \Theta\mathbf{u}_{1,0}^\text{(S)}.\label{eq:nabla_p11_switch_circular}
\end{align}

Integrating again gives the pressure field $p_{1,1}^\text{(S)}$ at order $\alpha\beta$ associated with the time-variation of the amplitude of the heat spot as
\begin{align}
	 p_{1,1}^\text{(S)}(r) = \int_r^\infty \Theta(\tilde{r}) \frac{\partial p_{1,0}^\text{(S)}(\tilde{r})}{\partial \tilde{r}} \, d\tilde{r},\label{eq:p11_switch_circular}
\end{align}
where we have chosen the constant of integration such that this pressure decays at infinity.

From this and Eq.~\eqref{eq:u11_switch_general}, we deduce that the corresponding velocity field $\mathbf{u}_{1,1}^\text{(S)}$ at order $\alpha\beta$ associated with the time-variation of the heat-spot amplitude is given by 
\begin{align}
	\mathbf{u}_{1,1}^\text{(S)} = \mathbf{0}.
\end{align}

With this result, we may now simplify Eq.~\eqref{eq:u11_decomposn} to find
\begin{align}
	\mathbf{u}_{1,1}(x,y,t) = A(t)^2\mathbf{u}_{1,1}^\text{(T)}(x-t,y).\label{eq:u11_decomposn_circular}
\end{align}
In other words, the flow at order $\alpha\beta$ is proportional to the flow associated with the translation of the heat spot; the rate of change of the amplitude $A'(t)$ does not feature in this expression.
This holds for any heat spot with circular symmetry (the relevant case for sufficiently slow scanning).

We can now generalise and explain physically this result that for any heat spot with circular symmetry, there is no ``switching-on" contribution to the velocity field at order $\alpha\beta$. 
Specifically, under the same assumption of a heat spot with circular symmetry, we show  that the rate of change of the heat-spot amplitude $A'(t)$ in fact only contributes to the flow at orders $\alpha^m$ for $m \geq 0$; the flow associated with the switching-on of the heat spot originates solely from thermal expansion. 
Any flows due to the coupling of thermal expansion and thermal viscosity changes are therefore independent of the rate of change of the heat-spot amplitude; instead, they depend on the instantaneous value of the heat-spot amplitude itself.

Substituting the relationship between density and temperature [Eq.~\eqref{eq:density_temp}] into mass conservation [Eq.~\eqref{eq:mass_consn}], we find
\begin{align}
	-\alpha \left [A'(t) \Theta(r) -   \frac{A(t) \Theta '(r)(x-t)}{r} \right ] + \nabla \cdot[\rho(r,t) \mathbf{u}] = 0,
\end{align}
where prime ($'$) indicates differentiation with respect to the argument.
We introduce the velocity field $\mathbf{u}^\text{(S)}$ associated with the switching-on of the heat spot, which satisfies
\begin{align}
	- \alpha A'(t) \Theta(r) + \nabla \cdot [\rho(r,t) \mathbf{u}^\text{(S)}] = 0,
\end{align}
with the boundary conditions that $\mathbf{u}^\text{(S)}$ is non-singular at the centre of the heat spot and decays at infinity.
In other words, by linearity, any contributions to the full velocity field $\mathbf{u}$ that involve the rate of change of the amplitude, $A'(t)$, are contained in the switching-on velocity.

At this stage, we note the circular symmetry of the problem and expect that the solution inherits this. 
With an ansatz of $\mathbf{u}^\text{(S)} = u^\text{(S)} (r,t) \mathbf{e}_r$, the equation becomes 
\begin{align}\label{eq:81}
	-\alpha A'(t) \Theta(r) + \frac{1}{r} \frac{\partial}{\partial r} [r \rho(r,t) u^\text{(S)}(r,t)] = 0.
\end{align}
Integrating Eq.~\eqref{eq:81} and using the boundary conditions, we find the switching-on velocity field as
\begin{align}
	\mathbf{u}^\text{(S)}(r,t) = \frac{\alpha A'(t)}{r \rho(r,t)} \int_0^r \tilde{r} \Theta (\tilde{r}) \, d \tilde{r} \, \mathbf{e}_r.
\end{align}
This is linear in the rate of change of the heat-spot amplitude $A'(t)$.

Importantly, this depends only on the thermal expansion coefficient~$\alpha$, not the thermal viscosity coefficient~$\beta$.
This is because to obtain the flow $\mathbf{u}^\text{(S)}$ here, we  have not needed to use the momentum equations. 
The circular symmetry of the temperature profile results in circular symmetry of the density and viscosity fields. 
This allows us to find the flow $\mathbf{u}^\text{(S)}$ purely from mass conservation; the flow obtained in this way also solves the  momentum equations, together with a pressure field $p^\text{(S)}$ that has circular symmetry too. 

In addition to this, note that the solution $\mathbf{u}^\text{(S)}$ is valid for all $\alpha$ and $\beta$; it does not rely on the limit $\alpha$, $\beta \ll 1$. 
To enable comparison with our results  from perturbation expansions, we expand $\mathbf{u}^\text{(S)}$ for small $\alpha$ to give
\begin{align}
	\mathbf{u}^\text{(S)} &= \frac{\alpha A'(t)}{r } \sum_{k=0}^\infty [\alpha A(t) \Theta(r)]^k \int_0^r \tilde{r} \Theta (\tilde{r}) \, d \tilde{r} \, \mathbf{e}_r \nonumber\\
	&\equiv \alpha A'(t) \mathbf{u}^\text{(S)}_{1,0} + \alpha^2 A'(t) A(t) \mathbf{u}^\text{(S)}_{2,0} + O(\alpha^3),
\end{align}
where we have included the factors of $A(t)$ and $A'(t)$ so that the notation is consistent with that introduced previously.
In particular, at order $\alpha$ we obtain 
\begin{align}
	\mathbf{u}^\text{(S)}_{1,0} 
	&= \frac{1}{r }  \int_0^r \tilde{r} \Theta (\tilde{r}) \, d \tilde{r} \, \mathbf{e}_r,
\end{align}
and at order $\alpha^2$ we find
\begin{align}
	\mathbf{u}^\text{(S)}_{2,0} 
	&= \frac{\Theta(r)}{r }  \int_0^r \tilde{r} \Theta (\tilde{r}) \, d \tilde{r} \, \mathbf{e}_r.
\end{align}
For the Gaussian temperature profile $\Theta (r) = \exp(-r^2/2)$, this recovers the result in Eq.~\eqref{eq:u_switch} 
for order $\alpha$ and  agrees with the result in Eq.~\eqref{eq:u20_switch_circular} when we  consider order $\alpha^2$ in Sec.~\ref{sec:order_alpha_sq}.

\subsubsection{Pressure field}

Building on our results for more general temperature profiles, we now focus on  the specific Gaussian temperature profile in Eq.~\eqref{eq:temp}.
From Eq.~\eqref{eq:u11_decomposn_circular} for the flow $\mathbf{u}_{1,1}$ at order $\alpha\beta$, we  only need to find the velocity field $\mathbf{u}_{1,1}^\text{(T)}$, associated with the translation of the heat spot.
To do this, we solve in this section Eq.~\eqref{eq:Poisson_p11_translate} for the corresponding pressure field~$p_{1,1}^\text{(T)}$.

Substituting  Eq.~\eqref{eq:u_translate} into Eq.~\eqref{eq:Poisson_p11_translate}, the Poisson equation becomes, in polar coordinates,
\begin{align}
	\frac{1}{r}\frac{\partial}{\partial r} \left (r \frac{\partial p_{1,1}^\text{(T)}}{\partial r}\right ) + \frac{1}{r^2} \frac{\partial^2 p_{1,1}^\text{(T)}}{\partial \theta^2} = \cos\theta \left [  2 r \exp(-r^2) - \frac{\exp(-r^2/2)}{r} + \frac{\exp(-r^2)}{r} \right ].\label{eq:Poisson_p11_translate_polar}
\end{align}
This may be solved, for example, by separation of variables and reduction of order.
The pressure $p_{1,1}^\text{(T)}$ at order $\alpha\beta$ associated with translation of the heat spot is then given by
\begin{align}
	p_{1,1}^\text{(T)}(x-t,y) = (x-t) \left [  - \frac{1}{4r^2} - \frac{1}{2r^2} \exp(-r^2/2) + \frac{3}{4r^2} \exp(-r^2) + \frac{1}{4} \E_1(r^2/2) - \frac{1}{4} \E_1(r^2) \right ],\label{eq:p11_translate}
\end{align}
where we recall that $\E_1$ is the exponential integral [defined in Eq.~\eqref{eq:E1}].

\subsubsection{Velocity field}

From Eqs.~\eqref{eq:u11_translate_general},~\eqref{eq:p11_translate}, and~\eqref{eq:u_translate}, we deduce that the velocity field $\mathbf{u}_{1,1}^\text{(T)}$ associated with the translation of the heat spot is given by
\begin{align}
	\mathbf{u}_{1,1}^\text{(T)}(x-t,y)
	= \mathbf{e}_x  \bigg \{&\frac{1}{4r^2} - \frac{(x-t)^2}{2r^4} + \left [-\frac{1}{2r^2} + \frac{(x-t)^2}{r^4} \right ] \exp(-r^2/2) \nonumber\\
	&+ \left [\frac{1}{4r^2} - \frac{(x-t)^2}{2r^4} \right ] \exp(-r^2) - \frac{1}{4} \E_1(r^2/2) + \frac{1}{4} \E_1(r^2)  \bigg \} \nonumber\\
	+ \mathbf{e}_y& \left [- \frac{(x-t)y}{2r^4} + \frac{(x-t)y}{r^4} \exp(-r^2/2) - \frac{(x-t)y}{2r^4}  \exp(-r^2) \right ].\label{eq:u11_translate}
\end{align}
The full instantaneous flow $\mathbf{u}_{1,1}$ at order $\alpha\beta$ is then given by Eq.~\eqref{eq:u11_decomposn_circular}, reproduced here for convenience,
\begin{align}
	\mathbf{u}_{1,1}(x,y,t) = A(t)^2\mathbf{u}_{1,1}^\text{(T)}(x-t,y).\label{eq:u11}
\end{align}

\begin{figure*}[t]
	\subfloat[]
	{\includegraphics[width=0.54\textwidth]{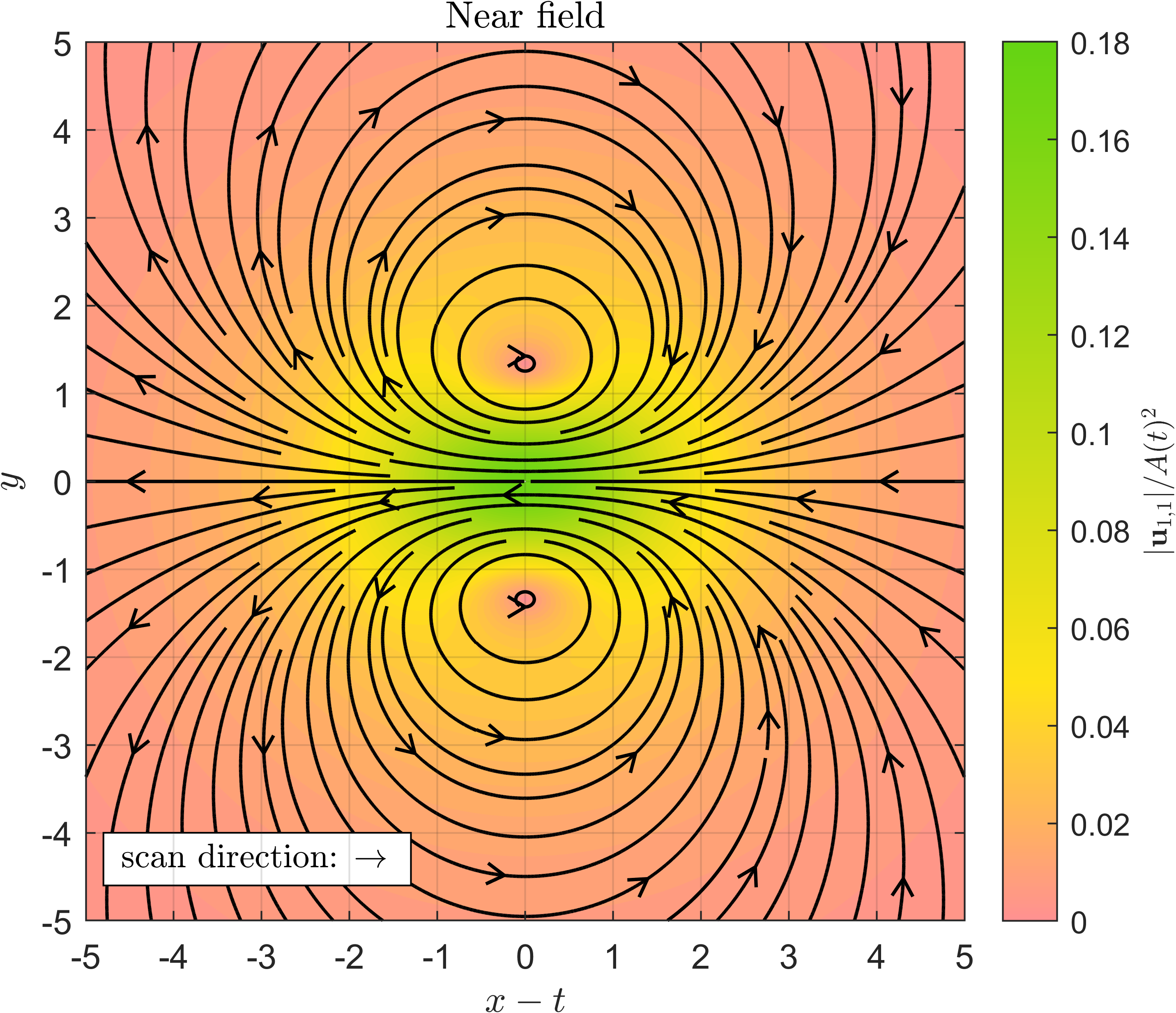}
		\label{fig:plot_u11_streamlines_near}}
	\subfloat[]
	{\includegraphics[width=0.46\textwidth]{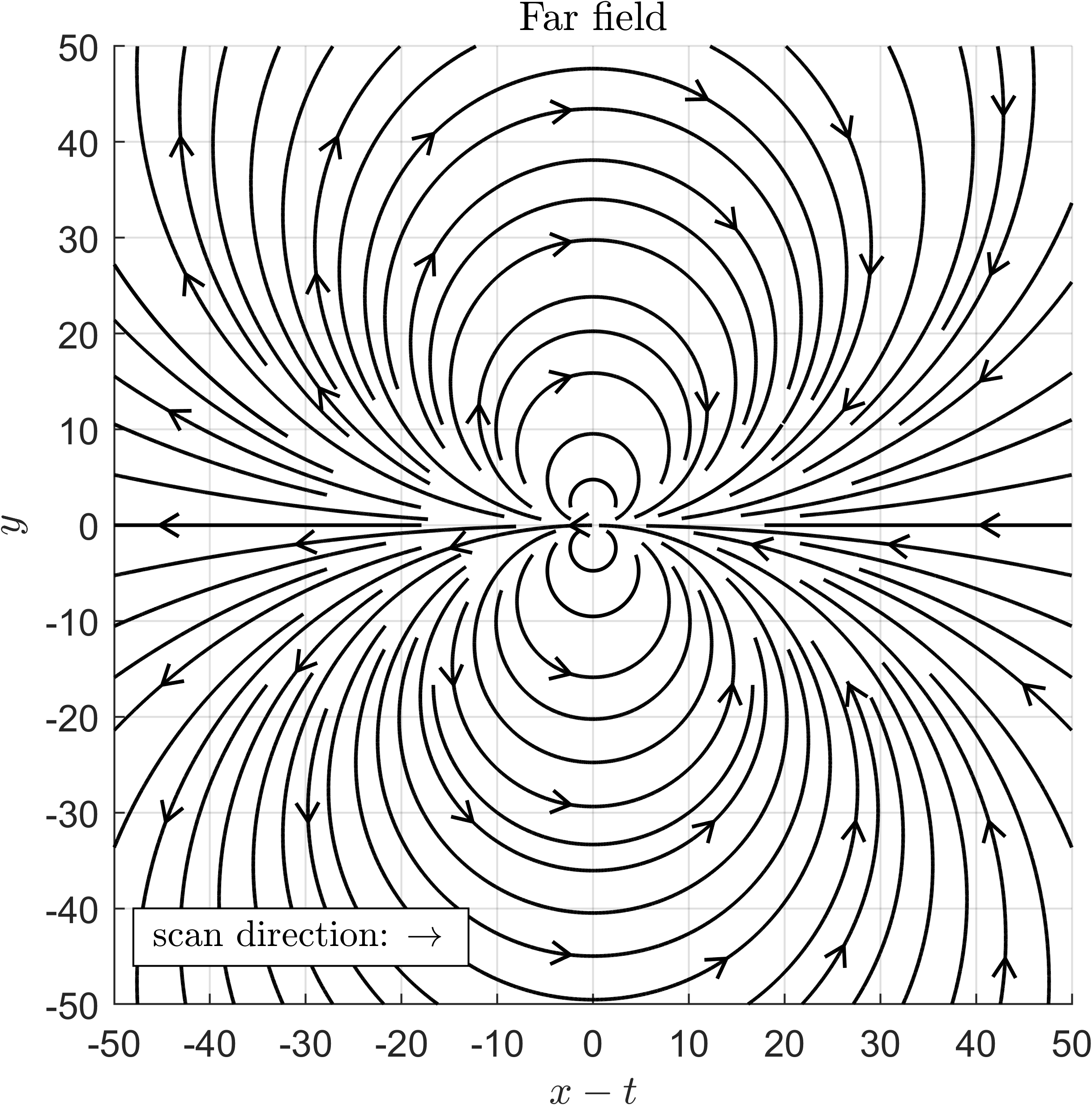}
		\label{fig:plot_u11_streamlines_far}} 
	\caption{Streamlines for the flow  $\mathbf{u}_{1,1}$ at order $\alpha\beta$, which includes the first effect of viscosity variation with temperature. 
	The heat spot translates in the positive $x$ direction (scan direction).  
	Left (Fig.~\ref{fig:plot_u11_streamlines_near}): streamlines for $-5 \leq x-t, y \leq 5$, close to the heat spot (near field). The magnitude of the velocity field divided by the square of the amplitude, $ \vert \mathbf{u}_{1,1}  \vert / A(t)^2 = \vert \mathbf{u}_{1,1}^\text{(T)} \vert $, is indicated by colour. 
		Right (Fig.~\ref{fig:plot_u11_streamlines_far}): streamlines for $-50 \leq x-t, y \leq 50$ to illustrate far-field behaviour.}
	\label{fig:plot_u11_streamlines}
\end{figure*}
\begin{figure}[t]
	{\includegraphics[width=0.6\textwidth]{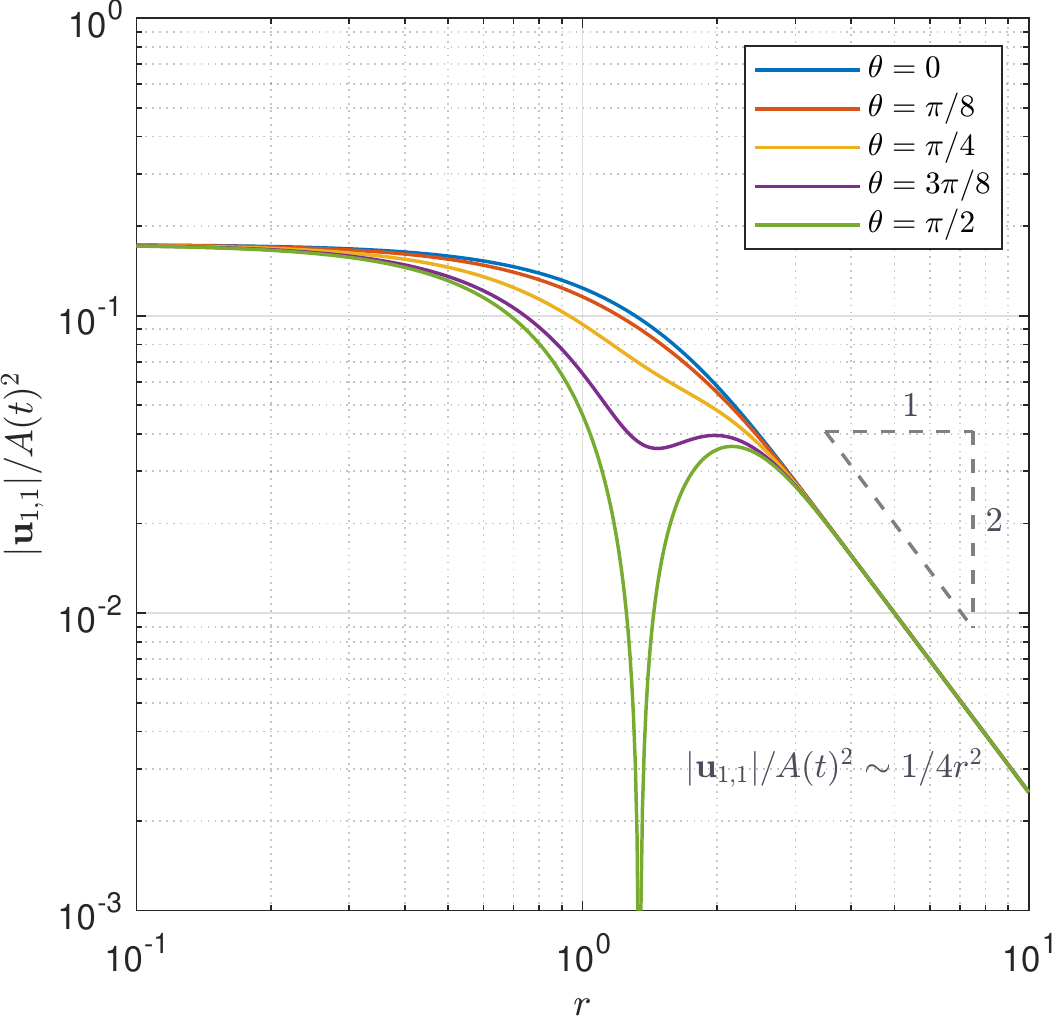}
		\caption{Scaled magnitude of the order-$\alpha\beta$ velocity field, $ \vert \mathbf{u}_{1,1}  \vert / A(t)^2 = \vert \mathbf{u}_{1,1}^\text{(T)} \vert$ [Eq.~\eqref{eq:magnitude_u11}], against the radial distance $r$ to the centre of the heat spot, plotted on a log--log scale, for $\theta = 0$, $\pi/8$, $\pi/4$, $3\pi/8$, and $\pi/2$.
			The far-field behaviour is given by $ \vert \mathbf{u}_{1,1}   \vert /A(t)^2 \sim 1/4r^2$. 
		}
		\label{fig:plot_u11_speed_radius}}
\end{figure}
We plot the streamlines of the flow $\mathbf{u}_{1,1}$ at order $\alpha\beta$ in Fig.~\ref{fig:plot_u11_streamlines}.
To quantify the spatial variation, we plot on a log--log scale the scaled magnitude $ \vert \mathbf{u}_{1,1}  \vert / A(t)^2 = \vert \mathbf{u}_{1,1}^\text{(T)} \vert $ as a function of the radius $r$, at fixed angles $\theta = 0$, $\pi/8$, $\pi/4$, $3\pi/8$, and $\pi/2$,  in Fig.~\ref{fig:plot_u11_speed_radius}.
This magnitude is given by
\begin{align}
	\frac{ \vert \mathbf{u}_{1,1}  \vert}{A(t)^2} = \vert \mathbf{u}_{1,1}^\text{(T)} \vert = \frac{1}{4r^2} \{& [1 - 2 \exp(-r^2/2) + \exp(-r^2) -r^2 \E_1(r^2/2) + r^2 \E_1(r^2) ]^2 \nonumber\\
	&+ 4 r^2 [1-\exp(-r^2/2)]^2 [\E_1(r^2/2) - \E_1(r^2)] \cos^2\theta
	\}^{1/2},\label{eq:magnitude_u11}
\end{align}
which decreases as the angle $\theta$ increases from $0$ to $\pi/2$.

Some properties of this instantaneous flow   will be important when we compare our theory with experiments in Sec.~\ref{sec:comparison_expt}.
First, the flow at order $\alpha\beta$ is quadratic in the amplitude, in contrast with the instantaneous flow at order $\alpha$, which is instead linear in the heat-spot amplitude.
Secondly, unlike at order $\alpha$, this flow is incompressible.
There are no sources or sinks in the flow.
The streamlines instead form closed loops with leftward velocities on the $x$ axis (i.e.,~in the opposite direction to the translation of the heat spot) and rightward velocities far from the $x$ axis.

\subsubsection{Far-field behaviour}

In the far field, the instantaneous flow $\mathbf{u}_{1,1}$ at order $\alpha\beta$ is a source dipole, given by
\begin{align}
	\mathbf{u}_{1,1} \sim \frac{1}{4} A(t)^2 \left \{\mathbf{e}_x  \left  [\frac{1}{r^2} - \frac{2(x-t)^2}{r^4} \right  ]  + \mathbf{e}_y \left [ -\frac{2(x-t)y}{r^4} \right ]\right \},\label{eq:u11_far-field}
\end{align}
with magnitude decaying as $ \vert \mathbf{u}_{1,1}   \vert  \sim A(t)^2/4r^2$.

The $1/r^2$ decay of this flow at order $\alpha\beta$ is faster than the $1/r$ decay of the source flow that typically [for $A'(t)\neq 0$] dominates the far field of the instantaneous flow at leading order, i.e.,~order $\alpha$.
However, in Sec.~\ref{sec:one_scan}, Sec.~\ref{sec:net_displ}, and Sec.~\ref{sec:comparison_expt}, we show that the leading-order average velocity of tracers inherits key features from the instantaneous flow at order $\alpha\beta$.

 \subsubsection{Physical mechanism}\label{sec:physical_mechanism_order_alphabeta}

We now propose an explanation for the physical mechanism behind our analytical result for the instantaneous flow field $\mathbf{u}_{1,1}$ at order $\alpha\beta$, i.e.,~for the first effect of thermal viscosity changes.
As a reminder, we have already considered the contribution $A'(t)A(t)\mathbf{u}_{1,1}^\text{(S)}$, associated with the time-variation of the heat-spot amplitude at order $\alpha\beta$, in Sec.~\ref{sec:u11_switch}.
We found that this contribution is in fact zero, $\mathbf{u}_{1,1}^\text{(S)} = \mathbf{0}$, and justified this physically: for a heat spot with circular symmetry, switching on produces a flow purely due to thermal expansion and mass conservation, independent of viscosity.
The flow $\mathbf{u}_{1,1}$ is therefore proportional to the contribution $\mathbf{u}_{1,1}^\text{(T)}$ associated with the translation of the heat spot [Eq.~\eqref{eq:u11_decomposn_circular}].

\begin{figure*}[t]
	\subfloat[]
	{\includegraphics[width=0.5\textwidth]{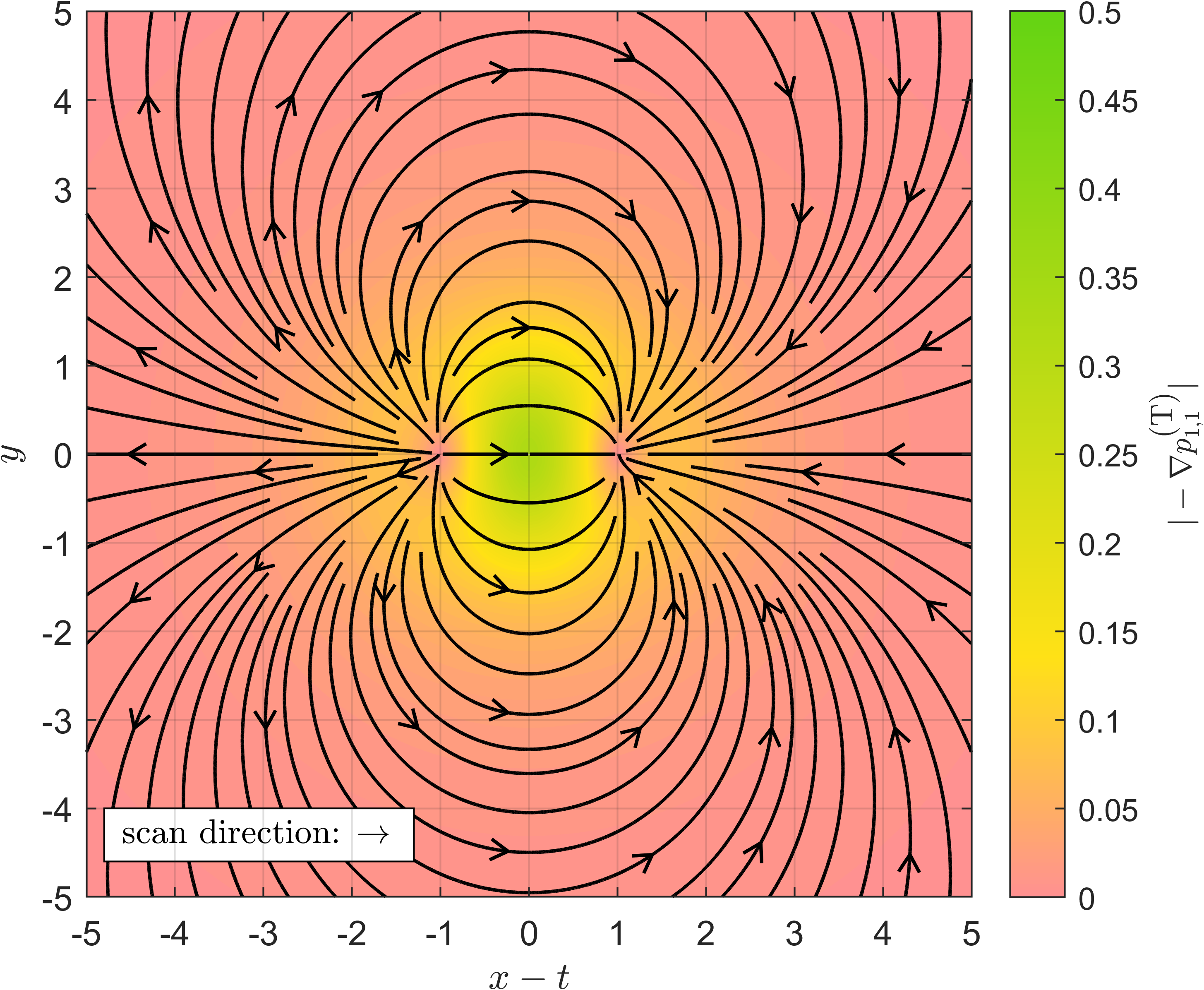}
		\label{fig:plot_minusgradp11translate_streamlines}} 
	\subfloat[]
	{\includegraphics[width=0.5\textwidth]{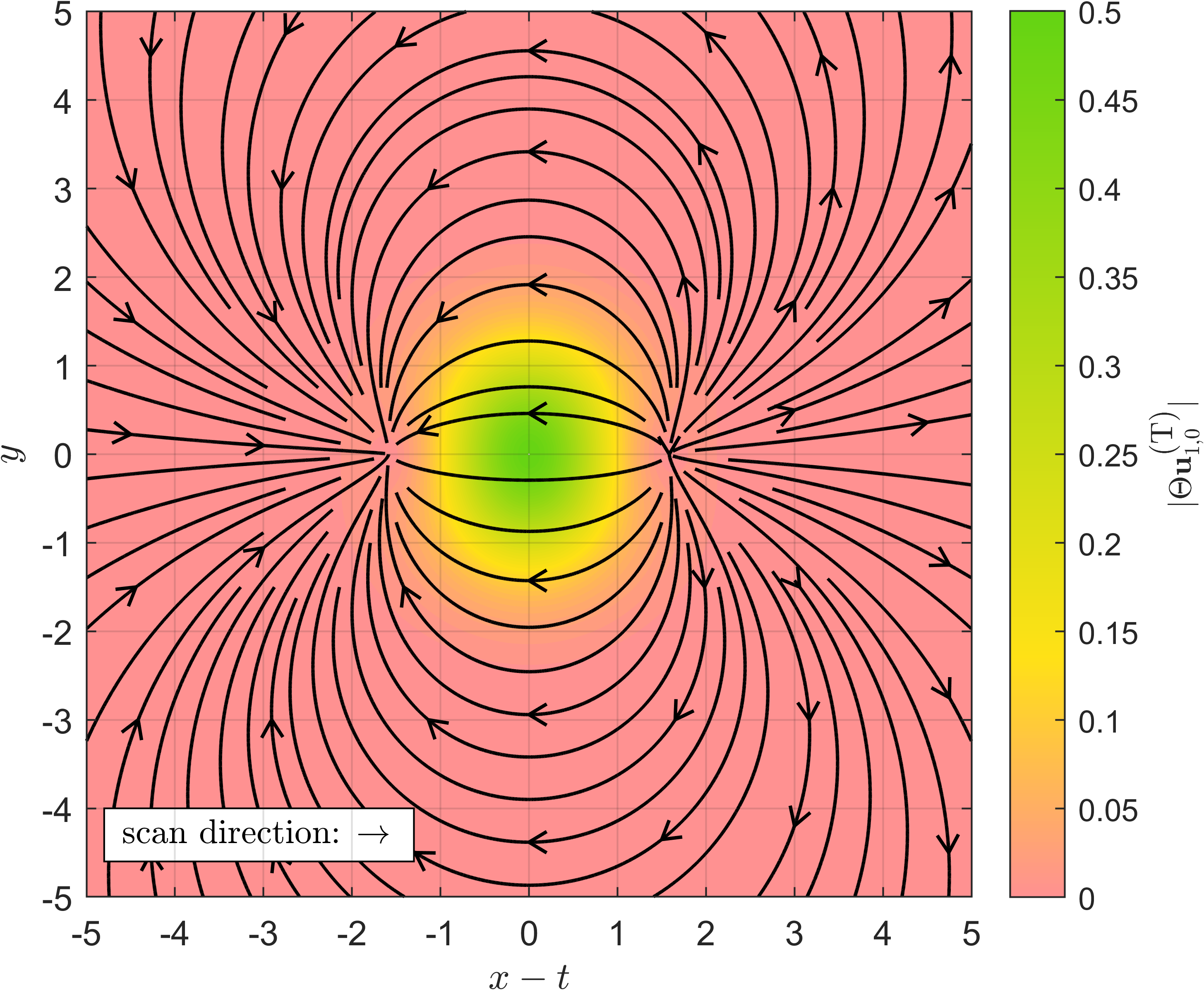}
		\label{fig:plot_Su10translate_streamlines}}
	\caption{Streamlines for the two terms that make up the velocity field $\mathbf{u}_{1,1}^\text{(T)} = -\nabla p_{1,1}^\text{(T)} + \Theta \mathbf{u}_{1,0}^\text{(T)}$ at order $\alpha\beta$, associated with the translation of the heat spot in the positive $x$ direction (scan direction).  The magnitude of each velocity field is indicated by colour. 
		Left (Fig.~\ref{fig:plot_minusgradp11translate_streamlines}): streamlines of the potential flow $-\nabla p_{1,1}^\text{(T)}(x-t,y)$ for $-5 \leq x-t,y \leq 5$. 
		The potential flow decays algebraically in the far field, whereas the flow $\Theta\mathbf{u}_{1,0}^\text{(T)}$ decays exponentially, so the potential flow dominates the far field of the flow $\mathbf{u}_{1,1}^\text{(T)}$ at order $\alpha\beta$.
		Right (Fig.~\ref{fig:plot_Su10translate_streamlines}): streamlines of the flow at leading order associated with translation of the heat spot, modulated by the heat-spot shape, $\Theta(x-t,y)\mathbf{u}_{1,0}^\text{(T)}(x-t,y)$, for $-5 \leq x-t, y \leq 5$. 
		The  envelope $\Theta$  suppresses the magnitude away from the heat spot exponentially.
		The contribution of the flow $\Theta\mathbf{u}_{1,0}^\text{(T)}$ to the velocity field $\mathbf{u}_{1,1}^\text{(T)}$ at order $\alpha\beta$ is thus mostly confined to the near field.
		In the near field, the flow $\Theta\mathbf{u}_{1,0}^\text{(T)}$ has larger magnitude than the potential flow $-\nabla p_{1,1}^\text{(T)}$, producing the leftward flow seen in the near field of the flow $\mathbf{u}_{1,1}^\text{(T)}$.}
	\label{fig:plots_minusgradp11translate_Su10translate_alphabeta_streamlines}
\end{figure*}
Recall from Eq.~\eqref{eq:u11_translate_general} that the  velocity field $\mathbf{u}_{1,1}^\text{(T)}$ is the sum of two terms, a potential flow, $-\nabla p_{1,1}^\text{(T)}$, and the velocity field at order $\alpha$ associated with the translation of the heat spot, modulated by the heat-spot shape, $\Theta\mathbf{u}_{1,0}^\text{(T)}$. 
The streamlines for these two separate flows are plotted in Fig.~\ref{fig:plots_minusgradp11translate_Su10translate_alphabeta_streamlines}, with $-\nabla p_{1,1}^\text{(T)}$ in Fig.~\ref{fig:plot_minusgradp11translate_streamlines} and $\Theta\mathbf{u}_{1,0}^\text{(T)}$ in Fig.~\ref{fig:plot_Su10translate_streamlines}. 
We can interpret  the physical origin of  each contribution to the flow field $\mathbf{u}_{1,1}^\text{(T)}$ and how they give rise to the hydrodynamic source dipole in the far field of $\mathbf{u}_{1,1}^\text{(T)}$, the direction of circulation of fluid flow, and finally the quadratic scaling of the full flow $\mathbf{u}_{1,1}$ at order $\alpha\beta$ with the heat-spot amplitude.

\begin{figure}[t]
	{\includegraphics[width=0.65\textwidth]{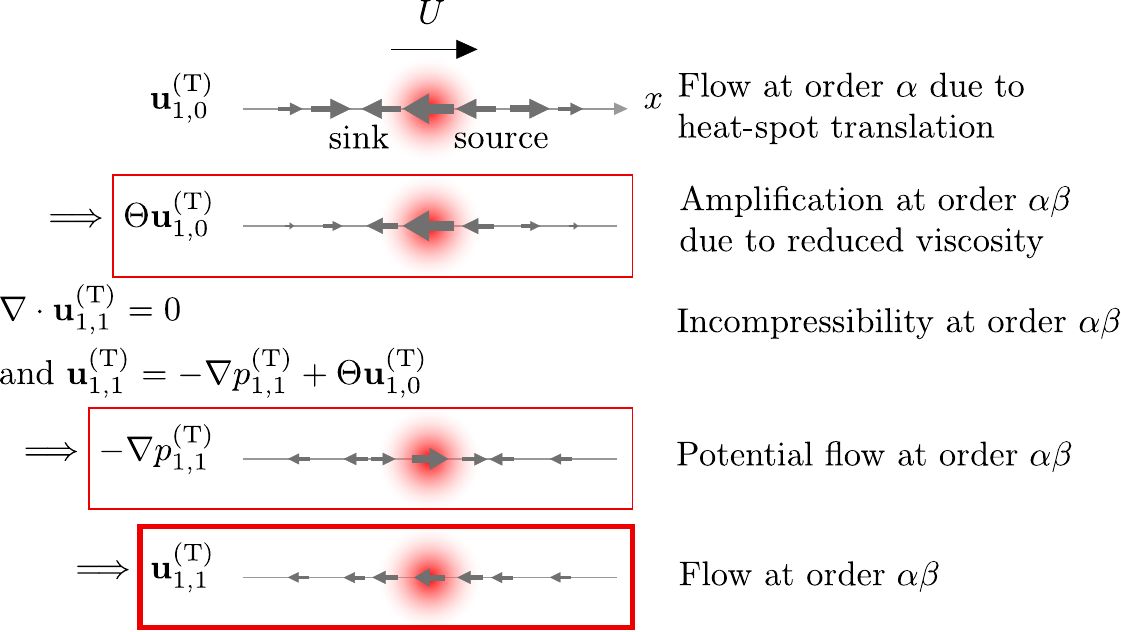}
		\caption{Schematic explanation of the physical mechanism for the  flow $\mathbf{u}_{1,1}^\text{(T)}$ at order $\alpha\beta$, the leading-order effect of viscosity variation, on the $x$ axis. We begin with the flow $\mathbf{u}_{1,0}^\text{(T)}$ associated with the translation of the heat spot at leading order (i.e.,~order $\alpha$), explained in the cartoon in Fig.~\ref{fig:diagram_alpha_mechanism_cartoon_translate}. 
			Then, we consider the first effect of thermal viscosity changes, at order $\alpha\beta$. 
			Viscosity decreases locally due to heating. 
			Intuitively, this  amplifies the leading-order flow  in the heat spot (compared with if the viscosity were instead its reference value), producing the compressible contribution $\Theta\mathbf{u}_{1,0}^\text{(T)}$ to the order-$\alpha\beta$ flow. 
			However, the order-$\alpha\beta$ flow is  incompressible, which is enforced by the pressure $p_{1,1}^\text{(T)}$ associated with heat-spot translation at order $\alpha\beta$.
			This adds a potential flow $-\nabla p_{1,1}^\text{(T)}$; mathematically, this arises from the parallel-plates geometry. 
			The  flow $\mathbf{u}_{1,1}^\text{(T)}$ at order $\alpha\beta$ is the combination of these two effects, amplification and incompressibility. 
		}
		\label{fig:diagram_alphabeta_mechanism_cartoon}}
\end{figure}
We illustrate in Fig.~\ref{fig:diagram_alphabeta_mechanism_cartoon} the interaction of  thermal viscosity changes with thermal expansion.
We may summarise the physical mechanism for the flow at order $\alpha\beta$  briefly as follows. 
The heat spot amplifies the leading-order flow (order $\alpha$) associated with the translation of the heat spot.
However, to enforce incompressibility of the full flow at order $\alpha\beta$, a second flow must be added to this, which compensates for the compressibility of the amplification effect.

In more detail, first, we focus on  the modified leading-order flow term, $\Theta\mathbf{u}_{1,0}^\text{(T)}$ (Fig.~\ref{fig:plot_Su10translate_streamlines}).
We build on previous work~\cite{weinert2008optically,weinert2008microscale} that focused on flow on the $x$ axis. 
First, recall from  Sec.~\ref{sec:order_alpha}  the physical mechanism behind the leading-order flow (order $\alpha$) associated with translation of the heat spot, which arises purely due to thermal expansion.
This is  illustrated in Fig.~\ref{fig:diagram_alpha_mechanism_cartoon_translate}.
Considering flow on the $x$ axis for a heat spot  translating rightwards, there is a source on the right, due to heating as the heat spot arrives, and a sink on the left, due to cooling as the heat spot leaves.
Thus, the flow near the heat spot is leftwards, i.e.,~from the source to the sink, and the flow on the $x$ axis outside these two stagnation points is rightwards.
This flow at order $\alpha$ does not take into account any thermal viscosity changes; it treats the viscosity as constant at its reference value.

We can now use this flow $\mathbf{u}_{1,0}^\text{(T)}$ to explain the contribution $\Theta \mathbf{u}_{1,0}^\text{(T)}$ to the flow $\mathbf{u}_{1,1}^\text{(T)}$.
Heating decreases the viscosity of the fluid locally [Eq.~\eqref{eq:viscosity_temp}] from the reference value and is quantified by the heat-spot shape function $\Theta(x-t,y)$ [Eq.~\eqref{eq:temp_general}].
Therefore, a given pressure gradient is able to drive a larger flow [Eq.~\eqref{eq:vel_press}]; in other words, the reduced viscosity locally amplifies the leading-order flow  $\mathbf{u}_{1,0}^\text{(T)}$ associated with heat-spot translation (at order $\alpha$).
This amplification is the leading-order effect of thermal viscosity changes, captured by the thermal viscosity coefficient~$\beta$ and occurring at order $\alpha\beta$.
From our theory, this amplification correction at order $\alpha\beta$  is given by $\Theta \mathbf{u}_{1,0}^\text{(T)}$, which is highly localised to the heat spot due to its exponential decay, inherited from the localised temperature perturbation.
It is therefore mainly the  flow near the heat spot, which is leftward, that is amplified, instead of the rightward flow further away, where the temperature and viscosity perturbations are exponentially small.

\begin{figure*}[t]
	\subfloat[]
	{\includegraphics[width=0.5\textwidth]{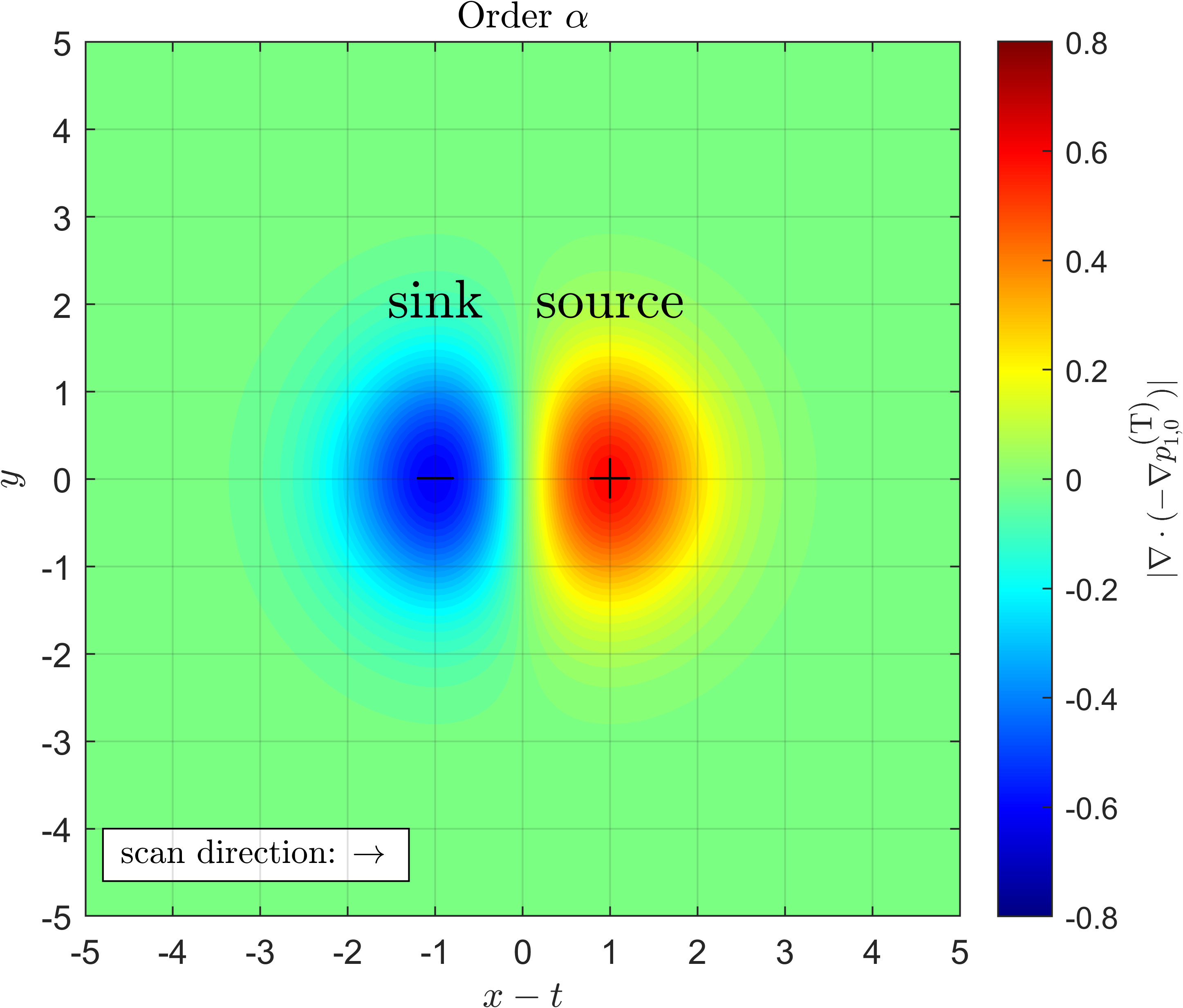}
		\label{fig:plot_div_minusgradp10translate_contours}} 
	\subfloat[]
	{\includegraphics[width=0.5\textwidth]{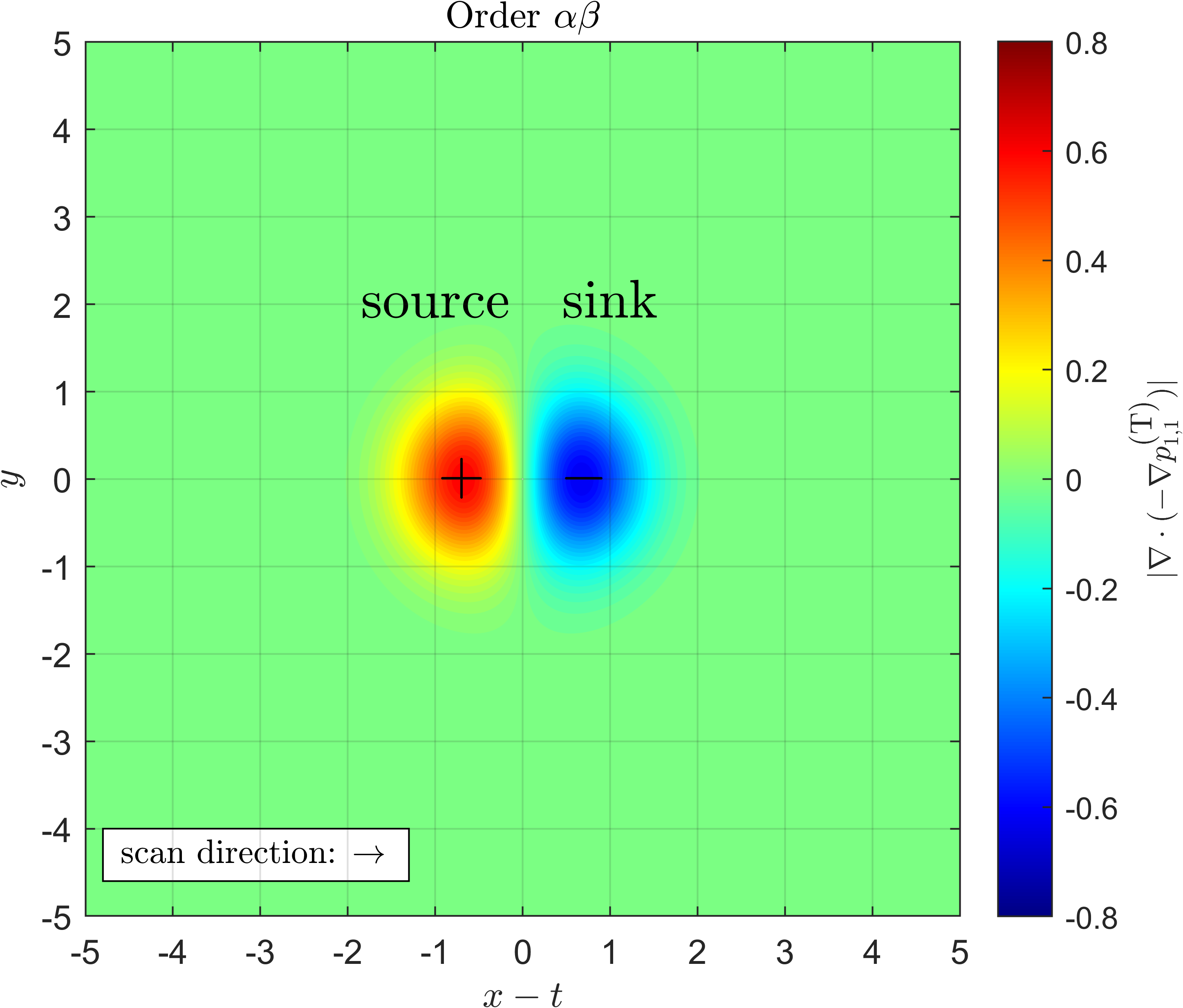}
		\label{fig:plot_div_minusgradp11translate_contours}}
	\caption{Divergence of potential flow fields $-\nabla p_{1,0}^\text{(T)}$ at order $\alpha$ (left) and $-\nabla p_{1,1}^\text{(T)}$ at order $\alpha\beta$ (right) associated with translation of the heat spot in the positive $x$ direction (scan direction).  Red indicates a source (positive divergence); blue indicates a sink (negative divergence); green indicates that the magnitude of the divergence is small. 
		Left (Fig.~\ref{fig:plot_div_minusgradp10translate_contours}): filled contour plot for $\nabla \cdot (-\nabla p_{1,0}^\text{(T)})$.
		Right (Fig.~\ref{fig:plot_div_minusgradp11translate_contours}): filled contour plot for $\nabla \cdot (-\nabla p_{1,1}^\text{(T)})$.
		}
	\label{fig:plots_div_minusgradp10translate_p11translate_alphabeta_contours}
\end{figure*}

However, the term $\Theta\mathbf{u}_{1,0}^\text{(T)}$ explained above cannot be the full  velocity field $\mathbf{u}_{1,1}^\text{(T)}$ associated with heat-spot translation at order $\alpha\beta$.
This is because the amplification correction flow $\Theta\mathbf{u}_{1,0}^\text{(T)}$ is compressible; the streamlines in Fig.~\ref{fig:plot_Su10translate_streamlines} clearly show a source on the right and a sink on the left.
On the other hand,  the total flow $\mathbf{u}_{1,1}^\text{(T)}$ at order $\alpha\beta$ is incompressible. 
We found analytically that the flow $\mathbf{u}_{1,1}^\text{(T)}$ is leftwards everywhere on the $x$ axis (Fig.~\ref{fig:plot_u11_streamlines}), whereas the amplification correction flow $\Theta\mathbf{u}_{1,0}^\text{(T)}$ is only leftwards in the near field, inherited from $\mathbf{u}_{1,0}^\text{(T)}$.  
These two physical features of $\Theta\mathbf{u}_{1,0}^\text{(T)}$ indicate that there must be another contribution to the flow $\mathbf{u}_{1,1}^\text{(T)}$ at order $\alpha\beta$.
This is the potential flow $-\nabla p_{1,1}^\text{(T)}$, which arises mathematically from the parallel-plates geometry.

This  potential flow $-\nabla p_{1,1}^\text{(T)}$ (Fig.~\ref{fig:plot_minusgradp11translate_streamlines})  enforces the incompressibility of the velocity field $\mathbf{u}_{1,1}^\text{(T)}$ at order $\alpha\beta$.
The divergence of the potential flow therefore must exactly cancel that of the amplification correction flow $\Theta\mathbf{u}_{1,0}^\text{(T)}$, which is expressed mathematically as
\begin{align}
	\nabla \cdot (-\nabla p_{1,1}^\text{(T)}) = - \nabla \cdot (\Theta \mathbf{u}_{1,0}^\text{(T)}).
\end{align}
This is simply the Poisson equation we solved for the pressure field earlier [Eq.~\eqref{eq:Poisson_p11_translate}].
We plot the left-hand side, $\nabla \cdot (-\nabla p_{1,1}^\text{(T)})$, in Fig.~\ref{fig:plot_div_minusgradp11translate_contours}.
Red indicates a source [$\nabla \cdot (-\nabla p_{1,1}^\text{(T)})>0$]; blue indicates a sink [$\nabla \cdot (-\nabla p_{1,1}^\text{(T)})<0$].
We see a source on the left, to compensate for the sink on the left in the amplification correction flow $\Theta\mathbf{u}_{1,0}^\text{(T)}$, and a sink on the right, to compensate for the source in $\Theta\mathbf{u}_{1,0}^\text{(T)}$.
Away from these red and blue regions in the near field in Fig.~\ref{fig:plot_div_minusgradp11translate_contours}, the divergence is exponentially small due to  decay of the temperature profile.
For comparison, we also plot the equivalent at order $\alpha$, $\nabla \cdot (-\nabla p_{1,0}^\text{(T)})$, in Fig.~\ref{fig:plot_div_minusgradp10translate_contours}.

At order $\alpha\beta$, the source on the left and sink on the right give rise to the far field of the potential flow $-\nabla p_{1,1}^\text{(T)}$, a  hydrodynamic source dipole, which decays algebraically.
This is also precisely the far field of the full velocity field $\mathbf{u}_{1,1}^\text{(T)}$ at order $\alpha\beta$ due to heat-spot translation, because  the amplification contribution $\Theta \mathbf{u}_{1,0}^\text{(T)}$ decays exponentially.

We return now to the  incompressible velocity field $\mathbf{u}_{1,1}^\text{(T)}=-\nabla p_{1,1}^\text{(T)} + \Theta \mathbf{u}_{1,0}^\text{(T)}$ at order $\alpha\beta$, shown in Fig.~\ref{fig:plot_u11_streamlines}.
Near the heat spot (translating rightwards), the leftward flow $\Theta \mathbf{u}_{1,0}^\text{(T)}$ due to the amplification effect dominates. 
In the far field, the hydrodynamic source dipole from the potential flow $-\nabla p_{1,1}^\text{(T)}$ dominates.
Together, these contributions give rise to a circulatory flow on each side of the $x$ axis. 
We  revisit this explanation in Sec.~\ref{sec:net_displ} when we consider the net displacement of material points due to the scanning of the heat spot.

Finally, we note that one may adapt the physical explanation of the flow $\mathbf{u}_{1,1}(x,y,t) = A(t)^2 \mathbf{u}_{1,1}^\text{(T)}(x-t,y)$ at order $\alpha\beta$ to the case of a cool spot [$A(t) \mapsto -A(t)$]; the flow is the same  as for a heat spot, consistent with the quadratic scaling with heat-spot amplitude.

\subsection{Solution at order $\alpha^2$}\label{sec:order_alpha_sq}

Having computed the leading-order instantaneous flow (order $\alpha$ in Sec.~\ref{sec:order_alpha}) and the leading-order effect of thermal viscosity changes (order $\alpha\beta$ in Sec.~\ref{sec:order_alpha_beta}), we now turn our attention to order $\alpha^2$, i.e.,~the other quadratic order. 
Previous work~\cite{weinert2008optically} has focused  on the interaction between thermal expansion and thermal viscosity changes  at order $\alpha\beta$, and neglected flow at order $\alpha^2$.
However, for some fluids, the thermal expansion coefficient~$\alpha$ and thermal viscosity coefficient~$\beta$ may be of comparable magnitude~\cite{rumble2017crc}; hence, the flow at order $\alpha^2$ could be of comparable magnitude to that at order $\alpha\beta$.
In this section, to complete our understanding of quadratic effects for an arbitrary fluid, we  therefore solve for the instantaneous flow at order $\alpha^2$. 
	The analytical formula for this is given by Eqs.~\eqref{eq:u20_decomposn},~\eqref{eq:u20_switch}, and~\eqref{eq:u20_translate}. The streamlines of the two separate contributions to the flow at order $\alpha^2$ are plotted in Fig.~\ref{fig:plot_u20_switch_streamlines} and Fig.~\ref{fig:plot_u20_translate_streamlines}, while in Sec.~\ref{sec:physical_mechanism_order_alphasq}, we provide a physical explanation for the flow.

\subsubsection{General heat spot}

The Poisson equation for pressure at order $\alpha^2$ [from Eq.~\eqref{eq:pressure_expanded}] is given by
\begin{align}
	\nabla^2 p_{2,0} &= \nabla\cdot (\Delta T \, \nabla p_{1,0}) \nonumber\\
	&= - \nabla \cdot (\Delta T \, \mathbf{u}_{1,0}),
\end{align}
where we recall that $\Delta T$ is the temperature profile (general) and $\mathbf{u}_{1,0}$ is the instantaneous flow at order $\alpha$.
Importantly, this Poisson equation is almost identical to the equation at order $\alpha\beta$, Eq.~\eqref{eq:order_alphabeta_original}, differing only by a factor of $-1$.
By linearity, the solution for the pressure $p_{2,0}$ at order $\alpha^2$ is therefore given by
\begin{align}
	p_{2,0}(x,y,t) &= -p_{1,1}(x,y,t).\label{eq:p20_p11}
\end{align}

For a general heat spot as in Eq.~\eqref{eq:temp_general}, we decompose this as
\begin{align}
	p_{2,0}(x,y,t) = A(t)A'(t) p_{2,0}^\text{(S)}(x-t,y) + A(t)^2 p_{2,0}^\text{(T)}(x-t,y),\label{eq:p20_decomposn}
\end{align}
just as in Eq.~\eqref{eq:p11_decomposn}.
The pressure field $p_{2,0}^\text{(S)}$ is associated with the time-variation of the amplitude of the heat spot (``switching on"). 
The pressure field $p_{2,0}^\text{(T)}$ is associated with the translation of the heat spot. 
By Eq.~\eqref{eq:p20_p11}, we can relate these pressures at order $\alpha^2$ to those at order $\alpha\beta$ as
\begin{align}
	p_{2,0}^\text{(S)}(x-t,y) &= - p_{1,1}^\text{(S)}(x-t,y),\label{eq:p20_switch_general}\\
	p_{2,0}^\text{(T)}(x-t,y) &= - p_{1,1}^\text{(T)}(x-t,y).\label{eq:p20_translate_general}
\end{align}

The velocity field at order $\alpha^2$, from Eq.~\eqref{eq:vel_pert_exp}, is a potential flow  given by
\begin{align}
	\mathbf{u}_{2,0} = - \nabla p_{2,0}.\label{eq:u20_p20}
\end{align}
We decompose this as
\begin{align}
	\mathbf{u}_{2,0}(x,y,t) = A(t)A'(t) \mathbf{u}_{2,0}^\text{(S)}(x-t,y) + A(t)^2 \mathbf{u}_{2,0}^\text{(T)}(x-t,y),\label{eq:u20_decomposn_general}
\end{align}
where
\begin{align}
	\mathbf{u}_{2,0}^\text{(S)}(x-t,y) &= - \nabla p_{2,0}^\text{(S)}(x-t,y),\label{eq:u20_switch_general}\\
	\mathbf{u}_{2,0}^\text{(T)}(x-t,y) &= - \nabla p_{2,0}^\text{(T)}(x-t,y).\label{eq:u20_translate_general}
\end{align}

While the pressure fields at order $\alpha^2$ and order $\alpha\beta$ differ only by a factor of $-1$, the velocity fields will be qualitatively different in structure.
This is because the instantaneous flow at order $\alpha^2$ is purely a potential flow and is compressible, whereas the flow at order $\alpha\beta$ instead  has two separate contributions (a potential flow and the heat-spot-modulated leading-order flow) and is incompressible. 

\subsubsection{Pressure field $p_{2,0}^\text{(S)}$, velocity field $\mathbf{u}_{2,0}^\text{(S)}$, and physical mechanism associated with time-variation of heat-spot amplitude for heat spot with circular symmetry}\label{sec:u20_switch}

As in Sec.~\ref{sec:u11_switch} for order $\alpha\beta$, we now specialise to a heat spot with circular symmetry, i.e.,~with shape function $\Theta(x-t,y) = \Theta(r)$.
By Eqs.~\eqref{eq:p20_switch_general} and~\eqref{eq:p11_switch_circular}, the pressure field $p_{2,0}^\text{(S)}$ at order $\alpha^2$ associated with the time-variation of the heat-spot amplitude is given by
\begin{align}
	p_{2,0}^\text{(S)}(r) = -\int_r^\infty \Theta(\tilde{r}) \frac{\partial p_{1,0}^\text{(S)}(\tilde{r})}{\partial \tilde{r}} \, d\tilde{r}.\label{eq:p20_switch_circular}
\end{align}

In agreement with results in Sec.~\ref{sec:u11_switch}, the corresponding velocity field, from Eqs.~\eqref{eq:u20_switch_general} and~\eqref{eq:nabla_p11_switch_circular}, is given by
\begin{align}
	\mathbf{u}_{2,0}^\text{(S)}(r) = \Theta(r)\mathbf{u}_{1,0}^\text{(S)}(r).\label{eq:u20_switch_circular}
\end{align}
This is the source-like flow $\mathbf{u}_{1,0}^\text{(S)}(r)$ at leading order,  modulated by the heat-spot shape.
The flow $A(t)A'(t)\mathbf{u}_{2,0}^\text{(S)}$ at order $\alpha^2$  therefore locally amplifies the leading-order source-like flow associated with an increasing heat-spot amplitude, $A'(t)\mathbf{u}_{1,0}^\text{(S)}$. 
Physically, heating decreases the density of the fluid locally from its reference value, so that the flow speed must increase to compensate for this, to satisfy mass conservation.

\subsubsection{Pressure field}

We now build on our general results and  write down the pressure field for the case of the Gaussian temperature profile in Eq.~\eqref{eq:temp}.
This is possible because, as we saw in the previous sections, much of the mathematics is shared with Sec.~\ref{sec:order_alpha_beta}.

Recall the decomposition of the pressure $p_{2,0}$ at order $\alpha^2$ into two contributions, given by Eq.~\eqref{eq:p20_decomposn}.
We already have the pressure field $p_{2,0}^\text{(S)}$ associated with the time-variation of the heat-spot amplitude in Eq.~\eqref{eq:p20_switch_circular}.
From Eqs.~\eqref{eq:p11_translate} and~\eqref{eq:p20_translate_general}, we can write down the pressure $p_{2,0}^\text{(T)}$ associated with the translation of the heat spot as
\begin{align}
	p_{2,0}^\text{(T)}(x-t,y) = (x-t) \left [ \frac{1}{4r^2} + \frac{1}{2r^2} \exp(-r^2/2) - \frac{3}{4r^2} \exp(-r^2) - \frac{1}{4} \E_1(r^2/2) + \frac{1}{4} \E_1(r^2) \right ].\label{eq:p20_translate}
\end{align}

\subsubsection{Velocity field}

The instantaneous flow at order $\alpha^2$ may be written as
\begin{align}
	\mathbf{u}_{2,0}(x,y,t) = A(t)A'(t) \mathbf{u}_{2,0}^\text{(S)}(x-t,y) + A(t)^2 \mathbf{u}_{2,0}^\text{(T)}(x-t,y),\label{eq:u20_decomposn}
\end{align}
by Eq.~\eqref{eq:u20_decomposn_general}, reproduced here for convenience.

\begin{figure}[t]
	{\includegraphics[width=0.54\textwidth]{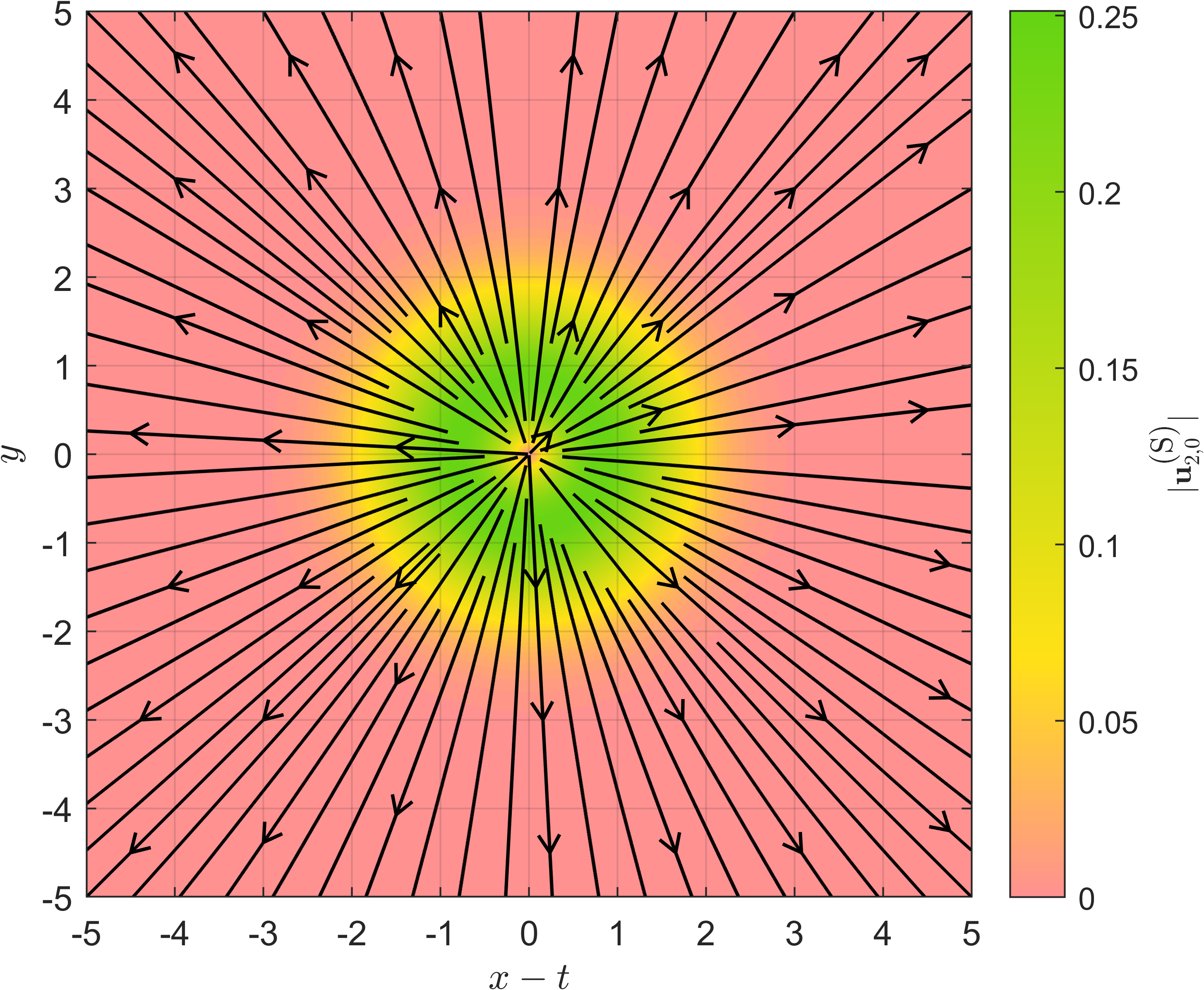}
		\caption{Streamlines for the velocity field $\mathbf{u}_{2,0}^\text{(S)}(x-t,y)$ [as in Eq.~\eqref{eq:u20_decomposn} for the instantaneous flow at order $\alpha^2$] associated with the time-variation of the heat-spot amplitude, for $-5 \leq x-t, y \leq 5$,  with magnitude of the velocity field $ \vert \mathbf{u}_{2,0}^\text{(S)}  \vert$ indicated by colour. The flow decays exponentially in the far field.}
		\label{fig:plot_u20_switch_streamlines}}
\end{figure}
By Eqs.~\eqref{eq:u20_switch_circular} and~\eqref{eq:u_switch}, the velocity field $\mathbf{u}_{2,0}^\text{(S)}$ at order $\alpha^2$ associated with the time-variation of the heat-spot amplitude is given by
\begin{align}
	\mathbf{u}_{2,0}^\text{(S)}(x-t,y) &= \frac{(x-t)\exp(-r^2/2)[1-\exp(-r^2/2)]}{r^2} \mathbf{e}_x + \frac{y\exp(-r^2/2)[1-\exp(-r^2/2)]}{r^2} \mathbf{e}_y\nonumber\\
	&\equiv \frac{\exp(-r^2/2)[1-\exp(-r^2/2)]}{r} \mathbf{e}_r.\label{eq:u20_switch}
\end{align}
We plot the streamlines of this purely radial flow in Fig.~\ref{fig:plot_u20_switch_streamlines}.
This is the leading-order source-like flow $\mathbf{u}_{1,0}^\text{(S)}$ modulated by the heat-spot shape function, so it decays exponentially, not algebraically, in the far field.

By Eqs.~\eqref{eq:u20_translate_general} and~\eqref{eq:p20_translate}, the velocity field $\mathbf{u}_{2,0}^\text{(T)}$ at order $\alpha^2$ associated with the translation of the heat spot is given by
\begin{align}
	\mathbf{u}_{2,0}^\text{(T)}(x-t,y) =  \mathbf{e}_x \bigg \{&-\frac{1}{4r^2} + \frac{(x-t)^2}{2r^4} + \left [-\frac{1}{2r^2} + \frac{(x-t)^2}{r^4} \right ] \exp(-r^2/2) \nonumber\\
	&+ \left [\frac{3}{4r^2} - \frac{3(x-t)^2}{2r^4} - \frac{(x-t)^2}{r^2}\right ] \exp(-r^2) + \frac{1}{4} \E_1(r^2/2) - \frac{1}{4} \E_1(r^2)\bigg \} \nonumber\\
	+ \mathbf{e}_y \bigg \{ & \frac{(x-t)y}{2r^4} + \frac{(x-t)y}{r^4} \exp(-r^2/2) \nonumber\\
	&+ \left [-\frac{3(x-t)y}{2r^4} - \frac{(x-t)y}{r^2}\right ] \exp(-r^2) \bigg \} .\label{eq:u20_translate}
\end{align}
\begin{figure*}[t]
	\subfloat[]
	{\includegraphics[width=0.543\textwidth]{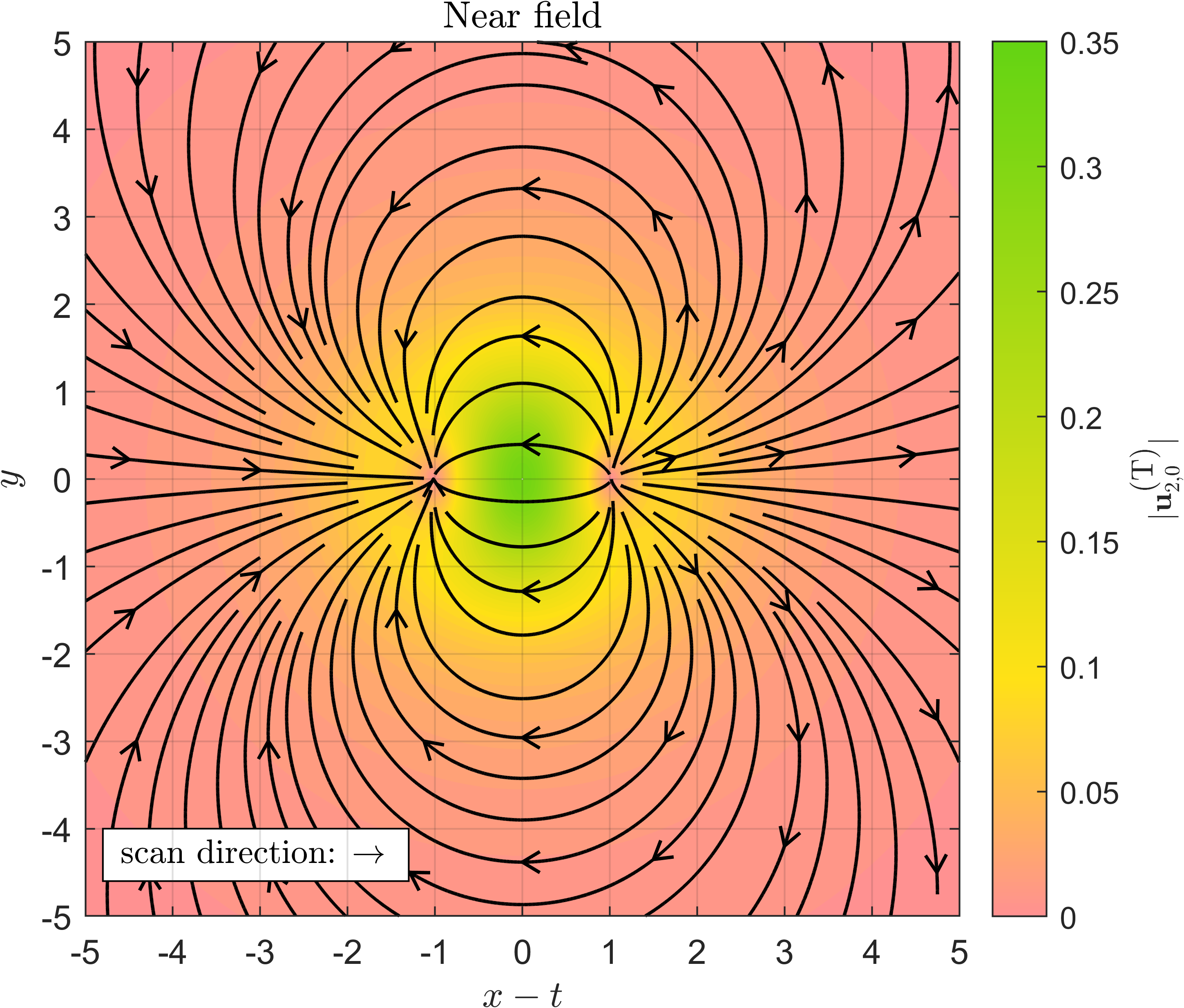}
		\label{fig:plot_u20_translate_streamlines_near}}
	\subfloat[]
	{\includegraphics[width=0.457\textwidth]{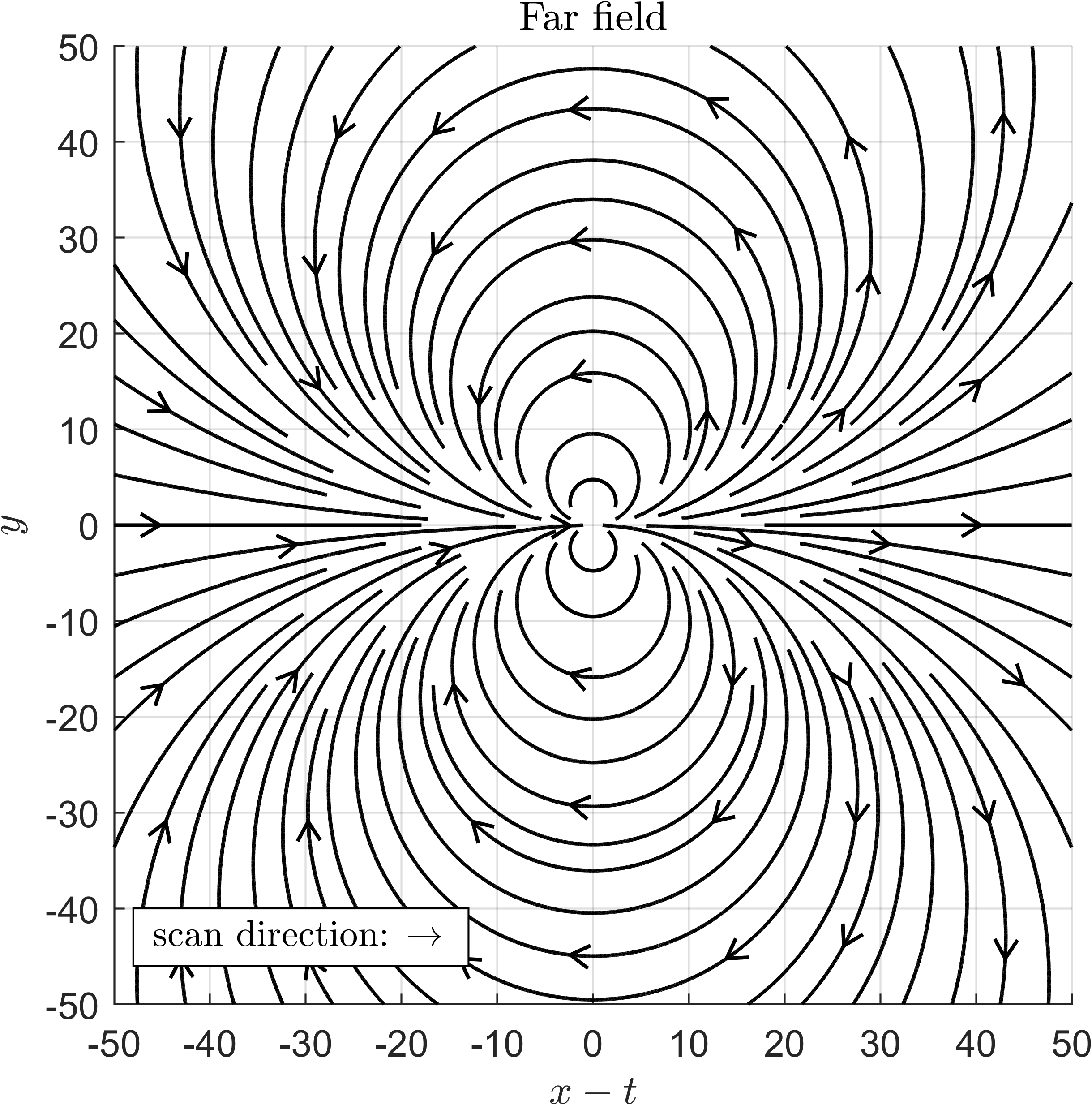}
		\label{fig:plot_u20_translate_streamlines_far}} 
	\caption{Streamlines for the flow field $\mathbf{u}_{2,0}^\text{(T)}(x-t,y)$ [as in Eq.~\eqref{eq:u20_decomposn_general} for the instantaneous flow at order $\alpha^2$] associated with translation of the heat spot in the positive $x$ direction (scan direction).   Left (Fig.~\ref{fig:plot_u20_translate_streamlines_near}): streamlines for $-5 \leq x-t, y \leq 5$, close to the heat spot (near field), with magnitude of the velocity field $ \vert \mathbf{u}_{2,0}^\text{(T)}  \vert$ indicated by colour. Right (Fig.~\ref{fig:plot_u20_translate_streamlines_far}): streamlines for $-50 \leq x-t,y \leq 50$ to illustrate the far-field behaviour.}
	\label{fig:plot_u20_translate_streamlines}
\end{figure*}
\begin{figure}[t]
	{\includegraphics[width=0.6\textwidth]{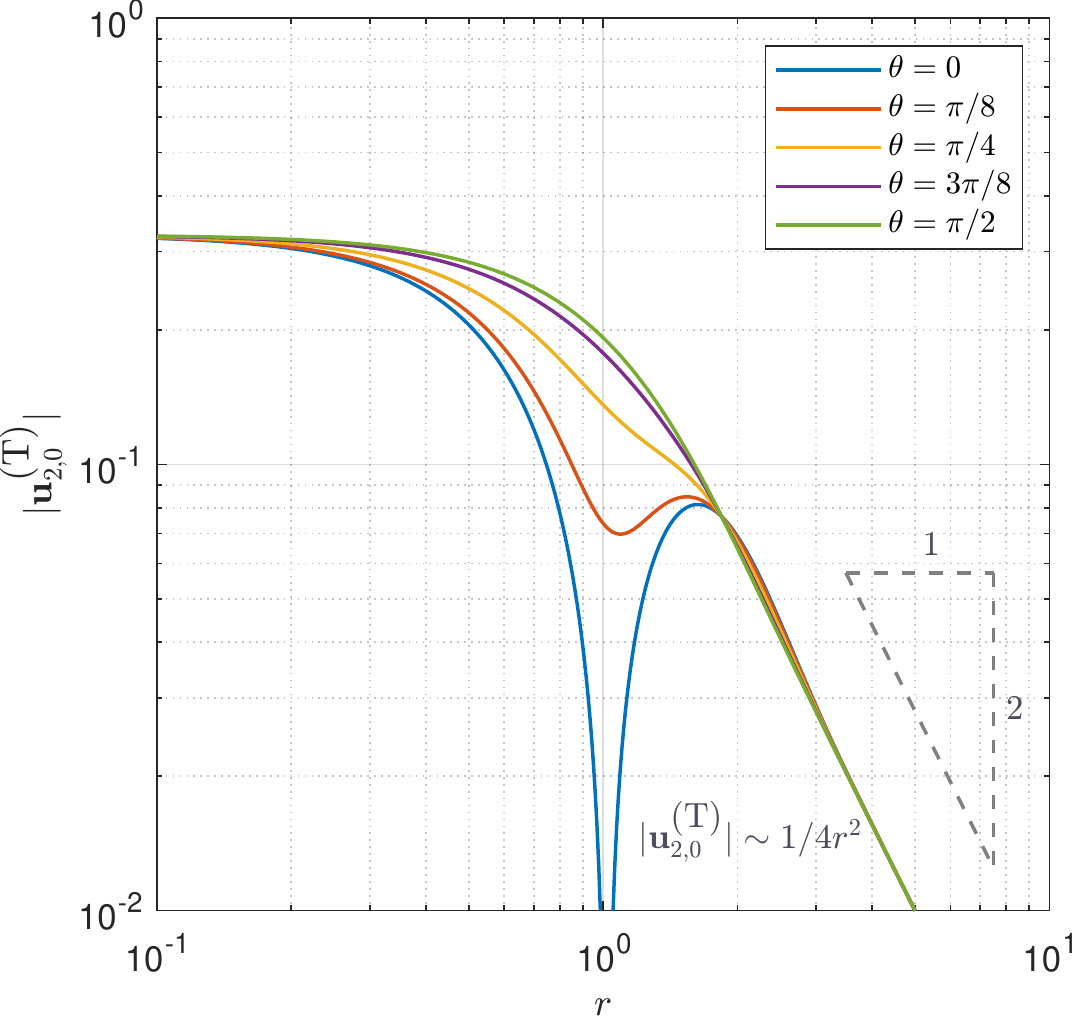}
		\caption{Magnitude of the flow field at order $\alpha^2$ associated with translation of the heat spot, $\vert \mathbf{u}_{2,0}^\text{(T)} \vert$, against the radial distance $r$ to the centre of the heat spot, plotted on a log--log scale, for $\theta = 0$, $\pi/8$, $\pi/4$, $3\pi/8$, and $\pi/2$.
			The far-field behaviour is given by $\vert \mathbf{u}_{2,0}^\text{(T)}   \vert \sim 1/4r^2$.}
		\label{fig:plot_u20_translate_speed_radius}}
\end{figure}
We plot in Fig.~\ref{fig:plot_u20_translate_streamlines} the streamlines of the flow   $\mathbf{u}_{2,0}^\text{(T)}$, with  the near field illustrated in Fig.~\ref{fig:plot_u20_translate_streamlines_near} and the far field in Fig.~\ref{fig:plot_u20_translate_streamlines_far}.
The magnitude $\vert \mathbf{u}_{2,0}^\text{(T)} \vert$ as a function of radius $r$ is shown in Fig.~\ref{fig:plot_u20_translate_speed_radius} on a log--log scale,  along the radial lines $\theta = 0$, $\pi/8$, $\pi/4$, $3\pi/8$, and $\pi/2$.

\subsubsection{Far-field behaviour}

The far-field behaviour of the instantaneous flow $\mathbf{u}_{2,0}$ at order $\alpha^2$ is given by
\begin{align}
	\mathbf{u}_{2,0} \sim -\frac{1}{4} A(t)^2 \left \{\mathbf{e}_x  \left  [\frac{1}{r^2} - \frac{2(x-t)^2}{r^4} \right  ]  + \mathbf{e}_y \left [ -\frac{2(x-t)y}{r^4} \right ]\right \},\label{eq:u20_far-field}
\end{align}
with magnitude decaying as $\vert \mathbf{u}_{2,0} \vert \sim A(t)^2/4r^2$.
This is a  source dipole, the same hydrodynamic singularity as at order $\alpha\beta$ but of opposite sign.
It is provided solely by the contribution $A(t)^2 \mathbf{u}_{2,0}^\text{(T)}(x-t,y)$ associated with heat-spot translation.
This is because the contribution $A(t)A'(t) \mathbf{u}_{2,0}^\text{(S)}(x-t,y)$ associated with the time-variation of the heat-spot amplitude is highly localised to the heat spot, decaying exponentially.
This is in contrast with the leading-order instantaneous flow (at order $\alpha$), where the switching-on of the heat spot gives rise to the algebraic decay (source flow) that dominates in the far field.

\subsubsection{Physical mechanism}\label{sec:physical_mechanism_order_alphasq}

Now that we have solved analytically for the instantaneous flow at order $\alpha^2$, we can interpret our results physically.
We have already explained the contribution $A(t)A'(t)\mathbf{u}_{2,0}^\text{(S)}$ associated with the time-variation of the heat-spot amplitude in Sec.~\ref{sec:u20_switch}.
The mechanism for the contribution $A(t)^2 \mathbf{u}_{2,0}^\text{(T)}$ associated with heat-spot translation, which we address now, is similar to the mechanism at order $\alpha\beta$ in Sec.~\ref{sec:physical_mechanism_order_alphabeta} but relates to density changes instead of viscosity changes, mirroring their mathematical similarities.

As discussed in Sec.~\ref{sec:physical_mechanism_order_alpha} and recapped in Sec.~\ref{sec:physical_mechanism_order_alphabeta}, the instantaneous flow $A(t)\mathbf{u}_{1,0}^\text{(T)}$ at leading order (order $\alpha$) associated with the translation of a heat spot  is leftwards near the heat spot.
The correction to the leading order captured at order $\alpha^2$ is the following.
Heating locally decreases the fluid density from the reference  value.
Therefore, in order to ensure that mass is conserved, the flow speed must increase locally, compensating for the fact that the density is lower than accounted for in the leading-order theory.
Essentially, the potential flow at order $\alpha^2$ reinforces the leading-order flow due to a heat spot.

Specifically, in the perturbation expansion, at order $\alpha$, the density $\rho$ in the term $\nabla \cdot (\rho \mathbf{u})$ (the divergence of the mass flux) in the mass conservation equation is approximated as the reference value $\rho_0$.
The potential flow correction at order $\alpha^2$ is driven by the fact that, in the term $\nabla \cdot (\rho \mathbf{u})$, the density $\rho$ is slightly lower than $\rho_0$, due to heating.
In contrast with this, at order $\alpha$, thermal expansion only forces the flow via the rate of change of density $\partial \rho/\partial t$ in the mass-conservation equation.

We can compare this with the mechanism at order $\alpha\beta$ (Sec.~\ref{sec:order_alpha_beta}).
In the flow at order $\alpha\beta$, a potential flow compensates for the compressibility of the leading-order flow modulated by the temperature profile.
Here, instead, the flow at order $\alpha^2$ is a potential flow that has the same divergence as the leading-order flow modulated by the temperature profile.
The flow at order $\alpha^2$ associated with heat-spot translation is a potential flow that has a source at the front of the heat spot and a sink at the back, just like the corresponding leading-order flow.
It therefore reinforces the source dipole in the far field of the leading-order flow  $\mathbf{u}_{1,0}^\text{(T)}$ associated with heat-spot translation.

\section{Leading-order trajectories of material points during one scan}\label{sec:one_scan}

In Sec.~\ref{sec:inst_flow}, we solved for the instantaneous velocity field induced by a translating heat spot  with arbitrary, time-varying amplitude, in the limit of small thermal expansion coefficient~$\alpha$ and thermal viscosity coefficient~$\beta$.
This was an Eulerian perspective.
In experiments, the heat spot scans repeatedly along a scan path and the resulting flow induces transport of various suspended bodies in the fluid, including proteins inside cells~\cite{mittasch2018non} or tracer beads in controlled experiments in a viscous fluid~\cite{erben2021feedback,weinert2008optically}.
To understand this  transport, we first analyse in this section the leading-order trajectories of material points during one scan, resulting from the leading-order instantaneous flow.
This is a Lagrangian perspective.
We show that at order $\alpha$, the net displacement of any material point is exactly zero, so the leading-order net displacement, which we find in Sec.~\ref{sec:net_displ}, occurs at higher order.

\subsection{Equation of motion for a material point}

\begin{figure}[t]
	{\includegraphics[width=0.65\textwidth]{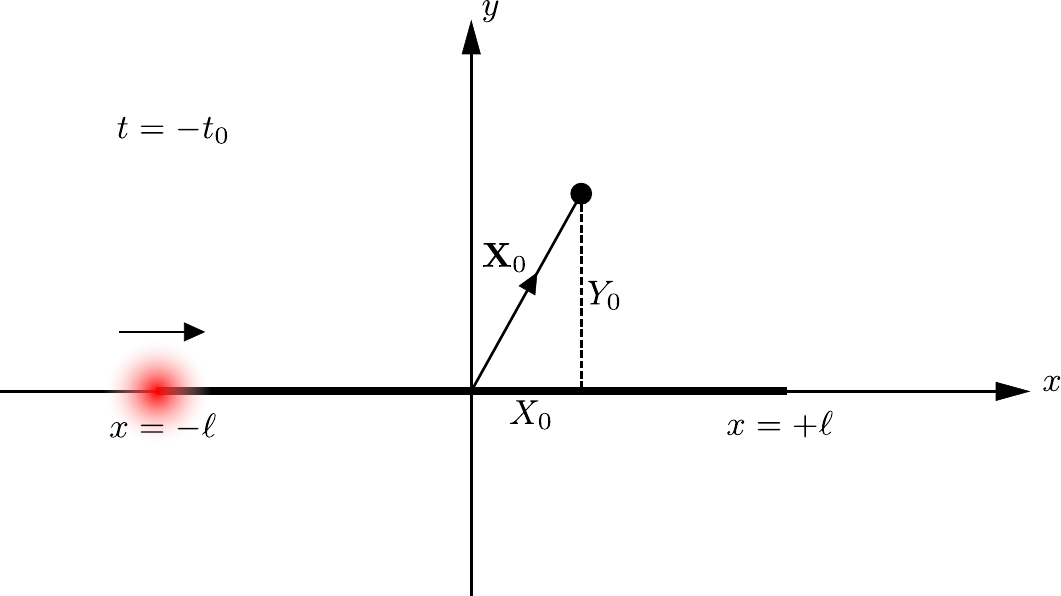}
		\caption{Sketch of a material point initially at position $\mathbf{X}_0$ when the heat spot is at the start of the scan path $x=-\ell$ at time $t=-t_0$.}
		\label{fig:diagram_tracer_position}}
\end{figure}
We begin by solving for the trajectory of a material point during one scan, i.e.,~as the heat spot translates at dimensionless speed $1$ along the scan path from $x=-\ell$ to $x=\ell$, as time progresses from $t=-t_0$ to $t=t_0$, where $t_0=\ell$ (in dimensionless terms).
Consider a material point that has initial position $\mathbf{X}_0 \equiv (X_0,Y_0)$ at time $t=-t_0$; this notation is illustrated in Fig.~\ref{fig:diagram_tracer_position}. 
At time $t$, its position vector relative to the origin is $\mathbf{X}(\mathbf{X}_0;t) \equiv (X(\mathbf{X}_0;t), Y(\mathbf{X}_0;t))$, which we write as $\mathbf{X}(t) \equiv (X(t), Y(t))$ for brevity. 
We aim to solve for this position vector as a function of time.
In the absence of noise, the kinematics of the material point is governed by the ordinary differential equation
\begin{align}
	\frac{d \mathbf{X}}{d t} &= \mathbf{u}(\mathbf{X}(t),t),
\end{align}
i.e.,~the material point is advected by the flow field $\mathbf{u}$ induced by the heat spot.
Integrating both sides and using the initial condition, we find that the equivalent integral equation is given by
\begin{align}
	\mathbf{X}(t) - \mathbf{X}(-t_0) = \mathbf{X}(t) - \mathbf{X}_0 = \int_{-t_0}^t \mathbf{u}(\mathbf{X}(\tilde{t}), \tilde{t}) \, d\tilde{t},\label{eq:X_integral_eq}
\end{align}
where the left-hand side is the displacement of the material point from its initial position at $t=-t_0$, when the heat spot is at the left endpoint, $x=-\ell$, of the scan path.

\subsection{Perturbation expansion}

The position vector $\mathbf{X}(t)$ appears on both sides of Eq.~\eqref{eq:X_integral_eq}.
In order to make further analytical progress, we pose a perturbation expansion for the displacement vector $\Delta \mathbf{X}(t) \equiv (\Delta X(t), \Delta Y(t))\equiv \mathbf{X}(t) - \mathbf{X}_0$ as
\begin{align}
	\Delta \mathbf{X}(t) 
	=   \alpha \Delta\mathbf{X}_{1,0}(t) + \alpha^2 \Delta\mathbf{X}_{2,0}(t) + \alpha\beta \Delta\mathbf{X}_{1,1}(t) + \text{cubic and higher-order terms},
\end{align}
where $\Delta\mathbf{X}_{m,n}(t) \equiv (\Delta X_{m,n} (t), \Delta Y_{m,n}(t))$ is the order $\alpha^m\beta^n$ displacement  of the material point at time $t$ from the position $\mathbf{X}_0$ at $t=-t_0$.
Here we write $\Delta\mathbf{X}(t)$ to mean $\Delta \mathbf{X} (\mathbf{X}_0;t)$, omitting for simplicity the dependence on the initial position from the notation just as we did for the position vector $\mathbf{X}(t)$.
In the above, we   anticipate that the displacement will inherit the structure of the velocity field perturbation expansion, since displacement is the time-integral of velocity.

Using this, we may also expand the velocity field at the position $\mathbf{X}(t)$ of the material point, about the initial position $\mathbf{X}_0$, as
\begin{align}
	\mathbf{u}(\mathbf{X}(t),t)
	=& \mathbf{u}(\mathbf{X}_0 +  \alpha \Delta \mathbf{X}_{1,0}(t) + \text{h.o.t.}, t)\nonumber \\
	=& \alpha\mathbf{u}_{1,0}(\mathbf{X}_0 + \alpha \Delta \mathbf{X}_{1,0}(t) + \text{h.o.t.}, t) \nonumber\\
	&+ \alpha^2\mathbf{u}_{2,0}(\mathbf{X}_0 + O(\alpha), t) + \alpha\beta\mathbf{u}_{1,1}(\mathbf{X}_0 + O(\alpha), t) + \text{cubic and higher-order terms} \nonumber \\
	=& \alpha\mathbf{u}_{1,0}(\mathbf{X}_0,t) + \alpha^2[\mathbf{u}_{2,0}(\mathbf{X}_0,t) + \Delta \mathbf{X}_{1,0}(t) \cdot \nabla \mathbf{u}_{1,0}(\mathbf{X}_0,t)] \nonumber\\
	& + \alpha\beta\mathbf{u}_{1,1}(\mathbf{X}_0,t) 
	+\text{cubic and higher-order terms}.
\end{align}
Substituting this into Eq.~\eqref{eq:X_integral_eq}, we obtain 
\begin{align}
	&\alpha \Delta\mathbf{X}_{1,0}(t) + \alpha^2 \Delta\mathbf{X}_{2,0}(t) + \alpha\beta \Delta\mathbf{X}_{1,1}(t) +\text{cubic and higher-order terms} \nonumber\\
	=&  \alpha\int_{-t_0}^t \mathbf{u}_{1,0}(\mathbf{X}_0,\tilde{t}) \, d\tilde{t} 
	+ \alpha^2\int_{-t_0}^t [\mathbf{u}_{2,0}(\mathbf{X}_0,\tilde{t}) + \Delta \mathbf{X}_{1,0}(\tilde{t}) \cdot \nabla \mathbf{u}_{1,0}(\mathbf{X}_0,\tilde{t})] \, d\tilde{t}
	+  \alpha\beta \int_{-t_0}^t \mathbf{u}_{1,1}(\mathbf{X}_0,\tilde{t}) \,d\tilde{t}.\label{eq:displacement_expanded}
\end{align}
We observe that at order $\alpha$ (leading order) and at order $\alpha\beta$, this displacement is simply the integral over time of the velocity field evaluated at the initial position of the material point. 
In contrast with this, at order $\alpha^2$, there is an additional contribution, associated with the order-$\alpha$ displacement of the material point.

\subsection{Order $\alpha$}

From Eq.~\eqref{eq:displacement_expanded},  the  displacement $\Delta\mathbf{X}_{1,0}(t)$ at order $\alpha$ of a material point is given by
\begin{align}
	\Delta \mathbf{X}_{1,0}(t) &= \int_{-t_0}^t \mathbf{u}_{1,0}(\mathbf{X}_0,\tilde{t}) \, d\tilde{t} ,\label{eq:X10_integral}
\end{align}
where we recall that $\mathbf{u}_{1,0}$ is the instantaneous velocity field at order $\alpha$.
During one scan, the displacement $\Delta\mathbf{X}_{1,0}(t)$ at order $\alpha$  typically gives the leading-order displacement of a material point at time $t$. 
(In Sec.~\ref{sec:net_displ_zero_alpha}, we  show that one full scan of a heat spot results in zero net displacement of the material point at order $\alpha$.)
To evaluate the integral in Eq.~\eqref{eq:X10_integral}, recall from Eq.~\eqref{eq:u10_general_decomposn} that the velocity field $\mathbf{u}_{1,0}$ at order $\alpha$ is an exact time-derivative of a function proportional to the heat-spot amplitude, for the general heat spot in Eq.~\eqref{eq:temp_general}.
Therefore,  for a general heat spot, Eq.~\eqref{eq:X10_integral} becomes
\begin{align}
	\Delta \mathbf{X}_{1,0}(t) &= \int_{-t_0}^t \frac{\partial}{\partial \tilde{t}}  [A(\tilde{t}) \mathbf{u}_{1,0}^\text{(S)}(X_0-\tilde{t},Y_0)] \, d\tilde{t}\nonumber\\
	 &= A(t)\mathbf{u}_{1,0}^\text{(S)}(X_0 - t, Y_0) - A(-t_0)\mathbf{u}_{1,0}^\text{(S)}(X_0 + t_0, Y_0),
\end{align}
by the Fundamental Theorem of Calculus.
If the scan path is finite as in experiments, then the heat-spot amplitude is  zero at the ends of the scan path by definition, so the second term vanishes [$A(-t_0)=0$].
For an infinite scan path ($t_0 \to \infty$) as in related theoretical work~\cite{weinert2008optically}, the second term vanishes regardless of the heat-spot amplitude because the velocity field $\mathbf{u}_{1,0}^\text{(S)}$ decays at infinity. 
In either case, the displacement at order $\alpha$ during one scan simplifies to
\begin{align}
	\Delta \mathbf{X}_{1,0}(t) &= A(t)\mathbf{u}_{1,0}^\text{(S)}(X_0 - t, Y_0).\label{eq:X10_general}
\end{align}

For the Gaussian heat spot [Eq.~\eqref{eq:temp}], the relevant velocity field $\mathbf{u}_{1,0}^\text{(S)}$ is given by Eq.~\eqref{eq:u_switch}.
The displacement  of a material point $\Delta \mathbf{X}_{1,0}(t)$ at order $\alpha$ during one scan is therefore given by
\begin{align}
	\Delta \mathbf{X}_{1,0}(t) &= \left . A(t) \left \{\frac{(x-t)[1-\exp(-r^2/2)]}{r^2} \mathbf{e}_x + \frac{y[1-\exp(-r^2/2)]}{r^2} \mathbf{e}_y \right \}\right \vert_{\mathbf{X}_0} \nonumber\\ 
	&\equiv \left . A(t) \left [ \frac{1-\exp(-r^2/2)}{r} \mathbf{e}_r \right ] \right \vert_{\mathbf{X}_0} ,\label{eq:X10_closed_form}
\end{align}
where we recall that the radius $r$ and radial unit vector $\mathbf{e}_r$ are measured from the centre of the (translating) heat spot.

\subsubsection{Sinusoidal heat-spot amplitude and finite scan path}

\begin{figure*}[t] 
	\subfloat[]
	{\includegraphics[width=0.5\textwidth]{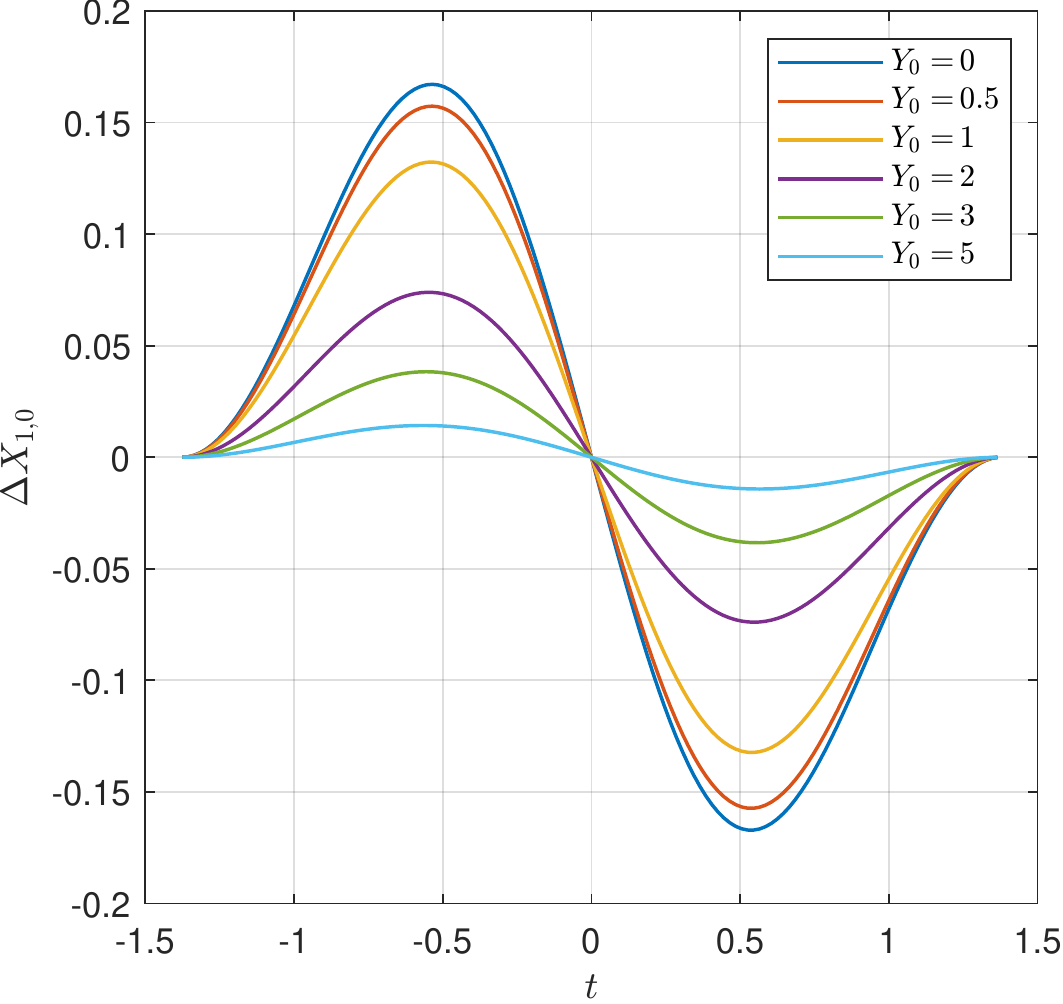}
		\label{fig:plot_DeltaX10_t}} 
	\subfloat[]
	{\includegraphics[width=0.5\textwidth]{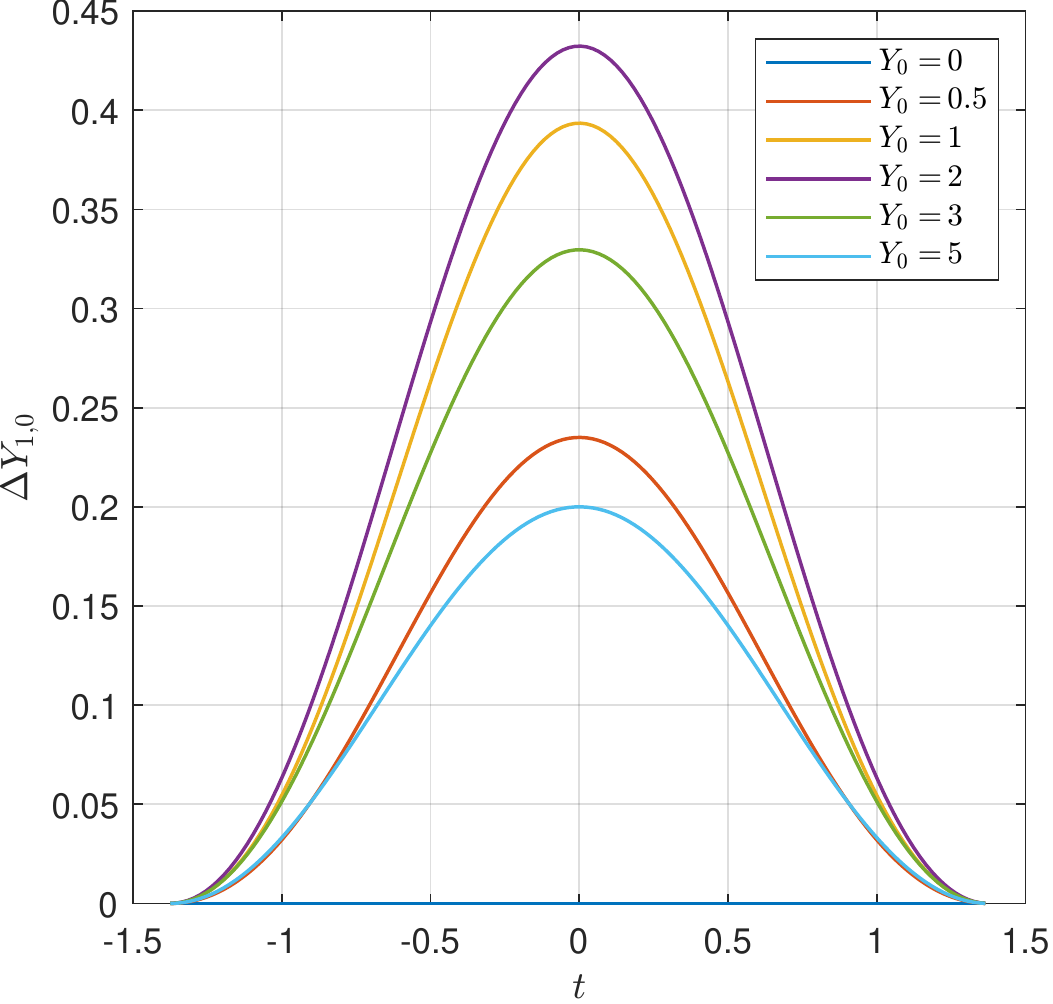}
		\label{fig:plot_DeltaY10_t}}\\
	\subfloat[]
	{\includegraphics[width=0.5\textwidth]{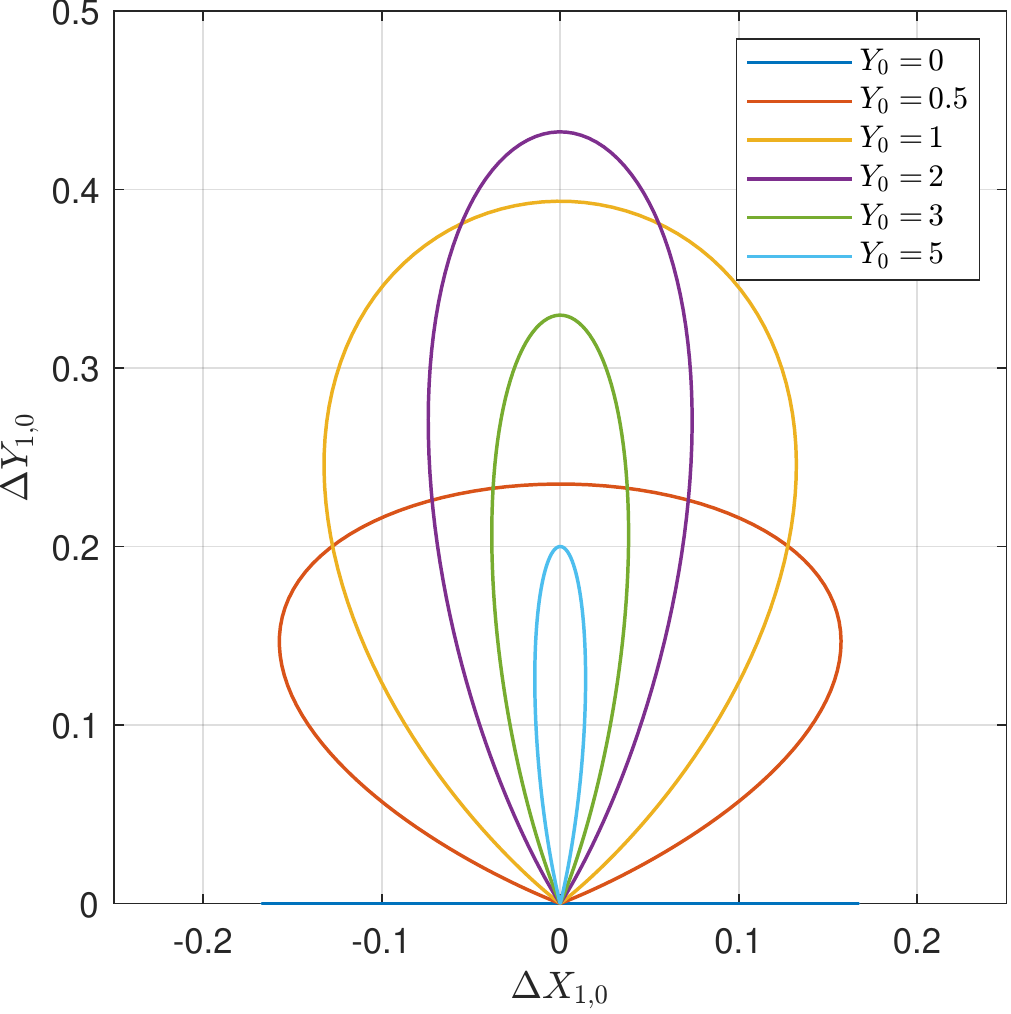}
		\label{fig:plot_DeltaX10_DeltaY10}}
	\caption{Trajectories of material points during one scan at leading order, i.e.,~order $\alpha$. The material points start at position $(0,Y_0)$, on the $y$ axis, at time $t=-t_0$ when the heat spot begins the scan from $x=-\ell$, for selected values  $Y_0=0$, $0.5$, $1$, $2$, $3$, and $5$. 
		Top left (Fig.~\ref{fig:plot_DeltaX10_t}): plot of the $x$ component $\Delta X_{1,0}$ of the  displacement of the material point at order $\alpha$ as a function of time $t$, during one scan, i.e.,~$-t_0 \leq t \leq t_0$.
		Top right (Fig.~\ref{fig:plot_DeltaY10_t}): same as Fig.~\ref{fig:plot_DeltaX10_t} but for the $y$ component $\Delta Y_{1,0}$ of the displacement of the material point at order $\alpha$.
		Bottom (Fig.~\ref{fig:plot_DeltaX10_DeltaY10}): plot of the same  trajectories in the $(\Delta X_{1,0}, \Delta Y_{1,0})$ plane.
		The length of the scan path is chosen to be $2\ell\equiv 2 t_0= 2.75$, to match microfluidic  experiments~\cite{erben2021feedback}.}
	\label{fig:plot_trajectories_onescan_DeltaX10_DeltaY10_t}
\end{figure*}
We illustrate in Fig.~\ref{fig:plot_trajectories_onescan_DeltaX10_DeltaY10_t} the trajectories of material points during one scan. 
We choose the sinusoidal amplitude function from Eq.~\eqref{eq:A_sinusoidal} and we choose the length of the scan path to be $2\ell = 2t_0 = 2.75$ to match experiments~\cite{erben2021feedback}. 
In Fig.~\ref{fig:plot_trajectories_onescan_DeltaX10_DeltaY10_t}, the  initial positions of the material points are $\mathbf{X}_0 = (0,0)$, $(0,0.5)$, $(0,1)$, $(0,2)$,  $(0,3)$, and $(0,5)$.
These lie on the $y$ axis for simplicity.
We plot the $x$ component  $\Delta X_{1,0}(t)$ of displacement  and $y$ component $\Delta Y_{1,0}(t)$ at order $\alpha$ as a function of time  in Figs.~\ref{fig:plot_DeltaX10_t} and~\ref{fig:plot_DeltaY10_t}, respectively.
We plot in Fig.~\ref{fig:plot_DeltaX10_DeltaY10} the corresponding trajectories  in space, $(\Delta X_{1,0}(t),\Delta Y_{1,0}(t))$ for $t=-t_0$ to $t=t_0$.
These trajectories are ``petal-shaped". 
[As an aside, if we instead have a constant-amplitude heat spot translating along an infinitely-long scan path, it can be shown analytically that the trajectory of a material point far from the $x$ axis ($Y_0 \gg 1$) is a circle, due to the translating source-dipole flow far from the heat spot.
The diameter of this circular trajectory scales as $1/Y_0$.] 
At order $\alpha$, each material point in Fig.~\ref{fig:plot_trajectories_onescan_DeltaX10_DeltaY10_t} appears to return precisely to its initial position after one full scan of the heat spot; we  show in the following sections that this is a general result for all material points, not only those on the $y$ axis.

\subsection{Zero net displacement at order $\alpha$ for general heat spot}\label{sec:net_displ_zero_alpha}

The result in Eq.~\eqref{eq:X10_general} gives the displacement at order $\alpha$ of a material point at time $t$ that has position $\mathbf{X}_0$ when the heat spot is at the start of the scan path ($t=-t_0$).
We illustrated in the previous section examples of leading-order trajectories of material points during one scan and saw that material points return to their initial position after one full scan, correct to order $\alpha$.
Previous theoretical work~\cite{weinert2008microscale} for a travelling temperature wave showed that this net displacement of a material point is zero at order $\alpha$ for material points on the scan path.
Here, we extend the result that the net displacement is zero at order $\alpha$ to material points with any initial position, for a general heat spot.

Importantly, note that Eq.~\eqref{eq:X10_general} is valid for the general heat spot in Eq.~\eqref{eq:temp_general}, with arbitrary amplitude function $A(t)$ and arbitrary shape specified by the function $\Theta(x-t,y)$.
In this section, we show that for this general heat spot, the net displacement $\Delta\mathbf{X}_{1,0}(t_0)$ of any material point (i.e.,~the displacement due to one full scan of the heat spot) is exactly zero at order $\alpha$, for finite and for infinite scan paths.
Consequently,  net displacements and hence trajectories of material points due to repeated scanning of the heat spot (Sec.~\ref{sec:net_displ}) are due to quadratic effects, not linear.
This is key for understanding experimentally-observed trajectories of tracers~\cite{erben2021feedback} (Sec.~\ref{sec:comparison_expt}).

\subsubsection{Finite scan path}

First we consider a finite scan path.
This is the relevant case for both biological (FLUCS) experiments~\cite{mittasch2018non} and microfluidic experiments~\cite{erben2021feedback}.
By definition of a finite scan path, the amplitude of the heat spot is zero at the ends of the scan path, so we have $A(t_0)=0$.
By Eq.~\eqref{eq:X10_general}, the net displacement $\Delta \mathbf{X}_{1,0}(t_0)$ at order $\alpha$, of any material point, is therefore given by
\begin{align}
	\Delta \mathbf{X}_{1,0}(t_0) = \mathbf{0},
\end{align}
as claimed.
Mathematically, this net displacement is zero because the velocity field at order $\alpha$ is an exact time-derivative of a function proportional to the heat-spot amplitude [Eq.~\eqref{eq:u10_general_decomposn}] and because the net displacement of a material point is the time-integrated fluid velocity evaluated at the initial position, at this order.
The contribution to the net displacement due to the time-variation of the heat-spot amplitude [related to the velocity field contribution $A'(t)\mathbf{u}_{1,0}^\text{(S)}(x-t,y)$] precisely cancels that due to the translation of the heat spot [related to $A(t)\mathbf{u}_{1,0}^\text{(T)}(x-t,y)$], at order $\alpha$.
This demonstration generalises earlier ideas~\cite{weinert2008microscale} for the case of temperature profiles steady in the comoving frame.

The net displacement of any material point (due to a full scan of the heat spot) is thus always zero at order $\alpha$. 
This holds for any heat-spot shape and amplitude function, not only for the idealised Gaussian profile we imposed to find explicit analytical solutions.
In experiments, the heat spot may become elongated if its speed is sufficiently high, losing the circular symmetry it has for lower speeds.
Crucially, symmetry of the heat spot is not necessary  for the net displacement to be zero at order $\alpha$, according to our theoretical model. Importantly, this means that the leading-order net displacement therefore scales at least quadratically, not linearly, with the heat-spot amplitude.
This is consistent with experiments~\cite{weinert2008optically,mittasch2018non} in which a quadratic scaling with the temperature perturbation was measured.

\begin{figure}[t]
	{\includegraphics[width=0.8\textwidth]{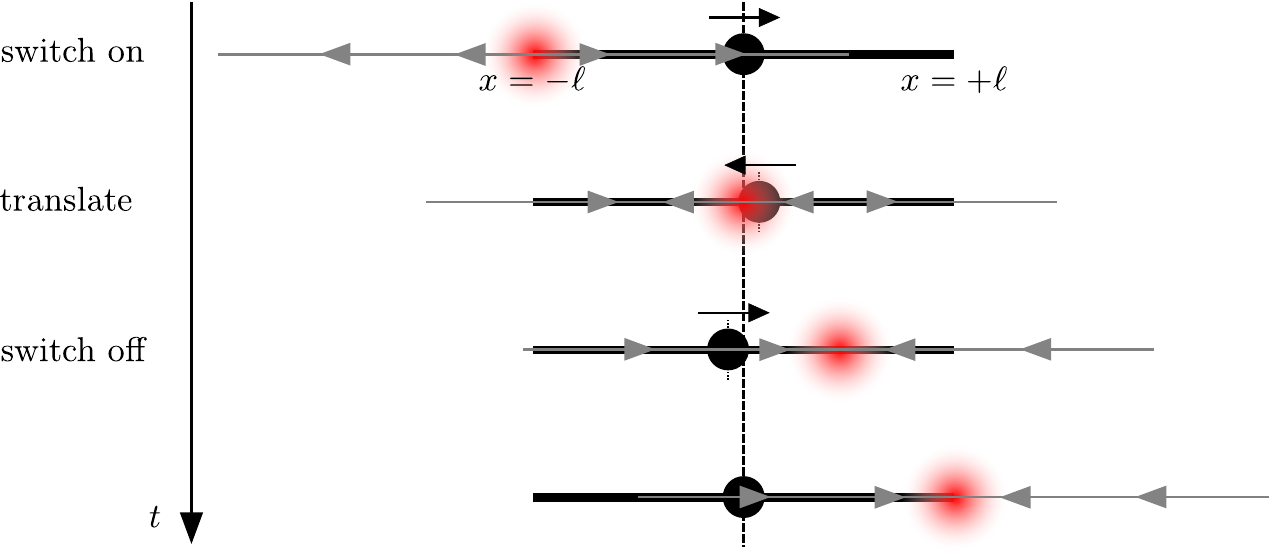} 
		\caption{Cartoon summary of zero net displacement at order $\alpha$ of  a material point, depicted as a black circle. 
			A  heat spot translating along a scan path from $x=-\ell$ to $x=\ell$  induces a velocity field experienced by the material point initially at position vector $\mathbf{X}_0$. (Here, the material point starts at the midpoint of the scan path, $\mathbf{X}_0 = \mathbf{0}$.) 
			This results in a small displacement of the material point at time $t$. The rightward displacement of the material point due to the switching-on and switching-off of the heat spot at the ends of the scan path is cancelled out by the leftward displacement due to the translation of the heat spot, correct to order $\alpha$. 
			}
		\label{fig:cartoon_displacement_alpha}}
\end{figure}

To explain the trajectories of tracers in Fig.~\ref{fig:plot_DeltaX10_DeltaY10} in terms of the leading-order instantaneous flow (Fig.~\ref{fig:plot_u10_streamlines_snapshots}), we illustrate the displacement of a material point due to the heat spot during one scan in Fig.~\ref{fig:cartoon_displacement_alpha}.
For simplicity, the cartoon in Fig.~\ref{fig:cartoon_displacement_alpha} shows the special case where the material point starts at the midpoint of the scan path; it remains on the $x$ axis for all time by symmetry. 
As the heat spot switches on, the resulting source-like flow pushes the material point at the origin rightwards, by a small (order-$\alpha$) distance.
Then, the flow induced by the translation of the heat spot overall pushes the material point leftwards.
This is because the material point is near the heat spot, where the flow is more leftward than in the far field.
Finally, the sink-like flow produced when the heat spot switches off at the end of the scan path pulls the material point rightwards, back to its initial position, correct to order $\alpha$.

To provide intuition on the net displacement of material points in the far field, we observe that the far-field instantaneous flow induced by translation of a constant-amplitude heat spot is a source dipole.
The average velocity of material points far from the scan path inherits the  far-field source dipole from this flow at order $\alpha$, corresponding to a source on the right and a sink on the left (for rightward translation).
Over the course of a scan, this effect is precisely cancelled out by the switching-on of the heat spot at the start of the scan path and the switching-off at the end of the scan path. 
For a finite scan path, the fact that the heat-spot amplitude varies with time is thus essential in our theory.
Over one scan, the switching-on and switching-off give rise to a source dipole in the far field of the average velocity of tracers (average Lagrangian velocity), corresponding to a source on the left and a sink on the right at order $\alpha$, i.e.,~the opposite of that due to translation of the heat spot.

\subsubsection{Infinite scan path}

In order to compare our theory with previous studies~\cite{weinert2008optically,weinert2008microscale}, we consider here the special case of the net displacement at order $\alpha$ of a material point, for an infinite scan path ($t_0\to\infty$). In that case, the heat-spot amplitude $A(t)$ does not necessarily decay as $t \to \pm\infty$, e.g.,~$A(t)=\text{const}$.
However, for a temperature profile that decays at infinity, the velocity field $\mathbf{u}_{1,0}^\text{(S)}$ does decay at infinity.
Using this and Eq.~\eqref{eq:X10_general}, for an infinite scan path, the net displacement $\Delta \mathbf{X}_{1,0}(\infty)$ at order $\alpha$ of any material point is given by
\begin{align}
	\Delta \mathbf{X}_{1,0}(\infty) = \mathbf{0}.
\end{align}

In particular, this holds for the case of constant amplitude $A(t)=A$, considered in earlier   work~\cite{weinert2008optically}.
The material point experiences the velocity field $\alpha A\mathbf{u}_{1,0}^\text{(T)}(x-t,y)$ and so moves rightwards by an order $\alpha$ displacement, then leftwards, and finally rightwards, back to its initial position (correct to order $\alpha$).

\section{Leading-order net displacement of material points due to one scan}\label{sec:net_displ}

We saw in Sec.~\ref{sec:net_displ_zero_alpha} that  at linear order, the net displacement (due to one full scan of the heat spot) of a material point with any initial position is zero, for a  heat spot with arbitrary time-dependent amplitude and arbitrary shape.
We therefore expect  the leading-order net displacement to be quadratic in the heat-spot amplitude, consistent with experiments~\cite{weinert2008optically}.
The model of Ref.~\cite{weinert2008optically} used a constant-amplitude heat spot and examined net displacement of material points on an infinite scan path.
Here we  discuss the leading-order net displacement of material points at any position in the fluid, for a heat spot with arbitrary amplitude and for a scan path of arbitrary length. 
In particular, we illustrate in Fig.~\ref{fig:plot_avg_vel_alphabeta_streamlines} and Fig.~\ref{fig:plot_avg_vel_alphasq_streamlines} the  velocity of tracers averaged over one scan, for a sinusoidal amplitude function, at order $\alpha\beta$ and order $\alpha^2$, respectively. 
We will then apply this in Sec.~\ref{sec:comparison_expt} to recent experimental results on the trajectories and average velocity of tracers due to repeated scanning of the heat spot~\cite{erben2021feedback}.

\subsection{Net displacement at order $\alpha\beta$}

Mathematically, the leading-order net displacement could be a linear combination of net displacements at order $\alpha^2$, order $\alpha\beta$, and order $\beta^2$.
However, we can eliminate order $\beta^2$, as  the instantaneous flow at order $\beta^2$ is zero (Sec.~\ref{sec:order_beta_n}).
Hence, we now consider the net displacement at order $\alpha\beta$; we discuss the contribution at order $\alpha^2$ in the next section.

By Eq.~\eqref{eq:displacement_expanded}, the net displacement $\Delta\mathbf{X}_{1,1}(t_0)$ at order $\alpha\beta$ of a material point with any initial position $\mathbf{X}_0$ is given by
\begin{align}
	\Delta\mathbf{X}_{1,1}(t_0) = \int_{-t_0}^{t_0} \mathbf{u}_{1,1}(\mathbf{X}_0,t) \,dt,\label{eq:net_displ_alphabeta}
\end{align}
where  $\mathbf{u}_{1,1}$ is the instantaneous velocity field at order $\alpha\beta$.
As at order $\alpha$, this is simply the integral of the velocity field evaluated at the initial position of the material point, which approximates the true position of the material point.
For a Gaussian heat spot, the velocity field $\mathbf{u}_{1,1}$ at order $\alpha\beta$ is given explicitly by Eqs.~\eqref{eq:u11_translate} and~\eqref{eq:u11}.

\subsubsection{Sinusoidal heat-spot amplitude and finite scan path}

The net displacement in Eq.~\eqref{eq:net_displ_alphabeta} may then be evaluated numerically once we choose an amplitude function $A(t)$.
To illustrate how the net displacement varies with initial position of the material point, we consider here a sinusoidal heat-spot amplitude function [Eq.~\eqref{eq:A_sinusoidal} as before] and a finite scan path, which is the relevant setup for FLUCS experiments~\cite{mittasch2018non}.

\begin{figure*}[t]
	\subfloat[]
	{\includegraphics[width=0.54\textwidth]{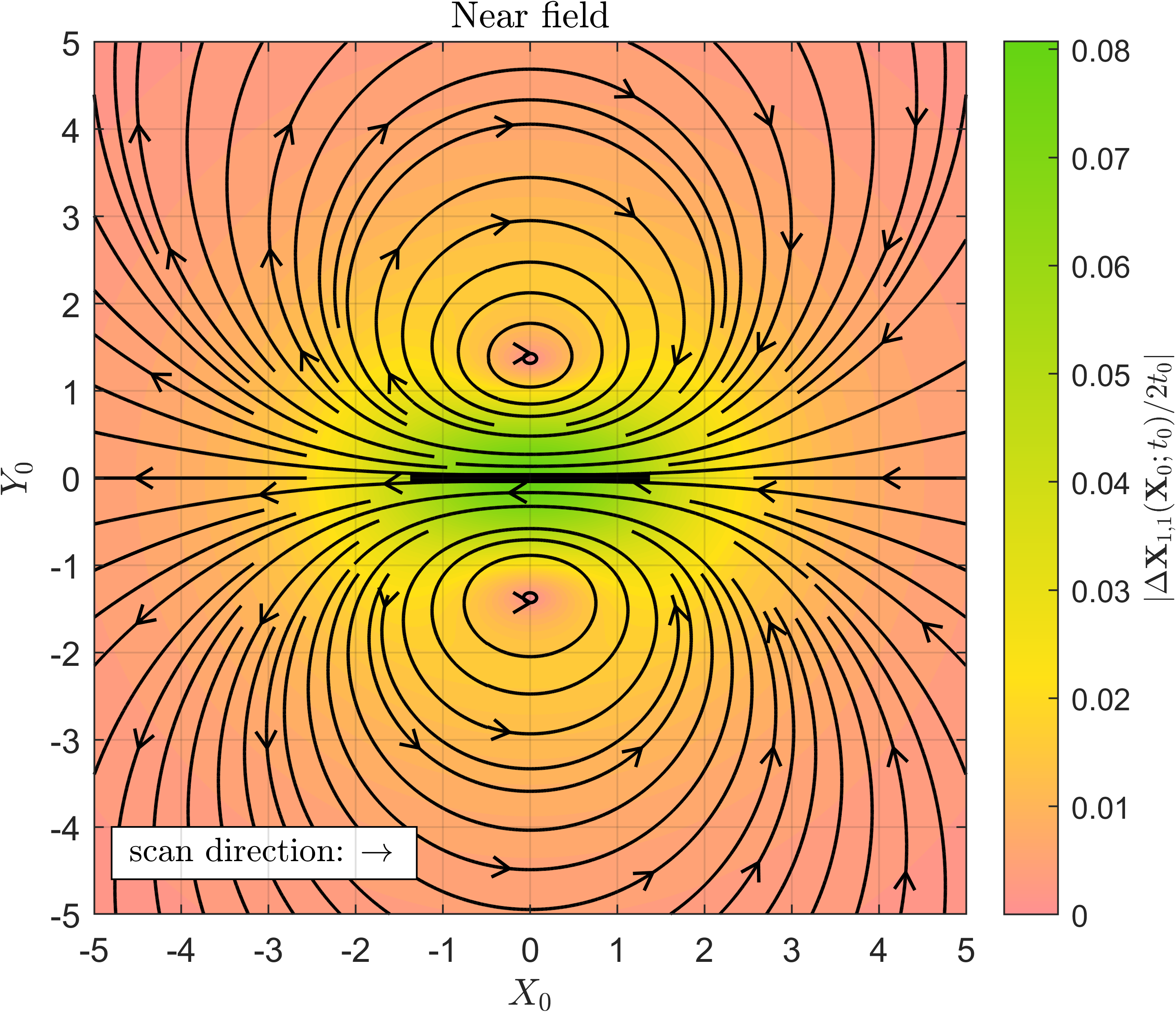}
		\label{fig:plot_avg_vel_alphabeta_streamlines_near}}
	\subfloat[]
	{\includegraphics[width=0.46\textwidth]{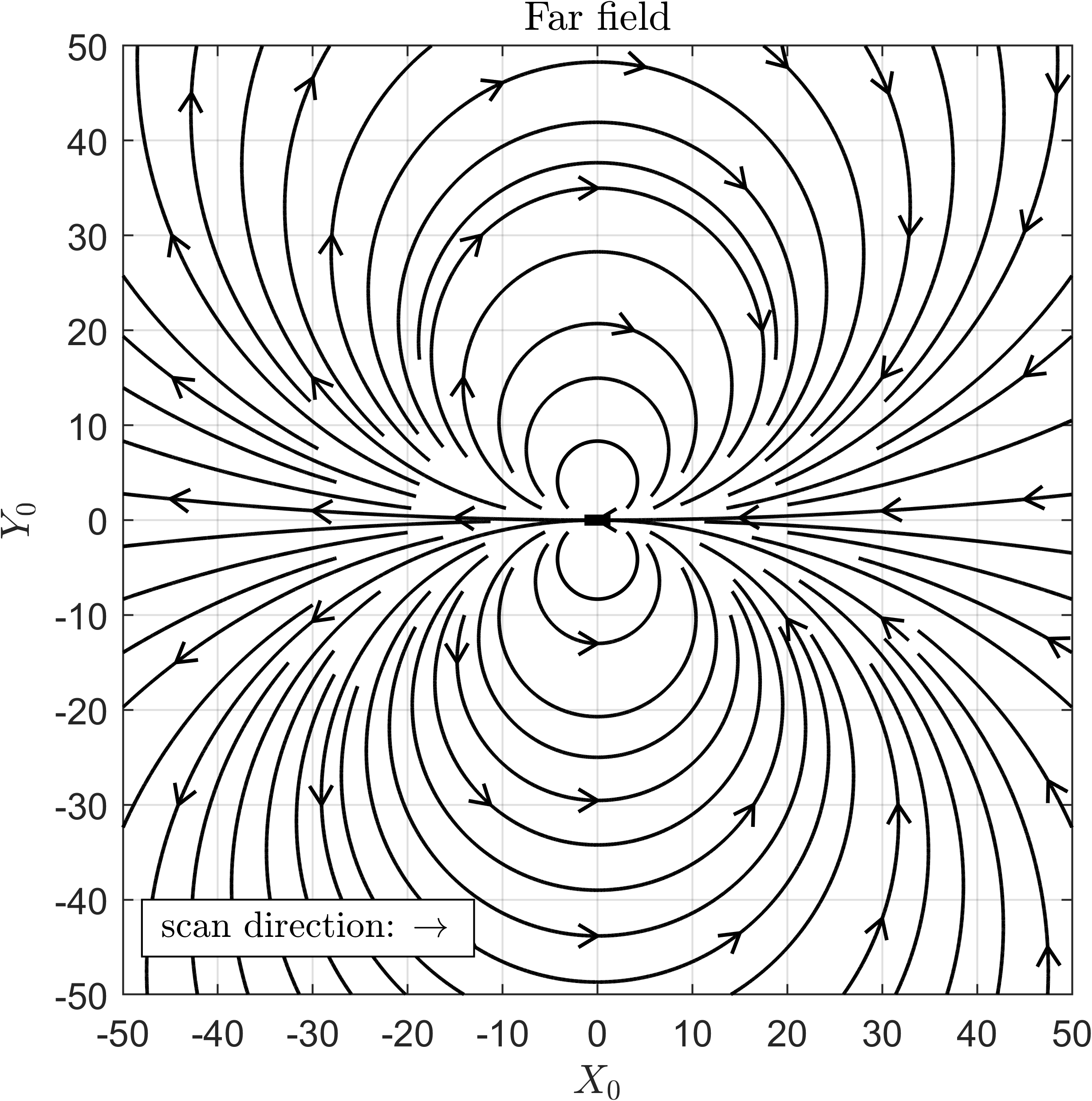}
		\label{fig:plot_avg_vel_alphabeta_streamlines_far}} 
	\caption{Streamlines for the average velocity of a material point at order $\alpha\beta$ over a scan period,  $\Delta \mathbf{X}_{1,1}(\mathbf{X}_0;t_0)/2t_0$. The heat spot translates in the positive $x$ direction (scan direction).  The scan path is indicated with a thick black line segment. 
	Left (Fig.~\ref{fig:plot_avg_vel_alphabeta_streamlines_near}): streamlines for $-5 \leq X_0, Y_0 \leq 5$, close to the scan path (near field), with magnitude of the average velocity of material points $ \vert \Delta\mathbf{X}_{1,1}(\mathbf{X}_0;t_0)/2t_0  \vert$ indicated by colour. 
	Right (Fig.~\ref{fig:plot_avg_vel_alphabeta_streamlines_far}): streamlines for $-50 \leq X_0, Y_0 \leq 50$, which illustrate the far-field behaviour.}
	\label{fig:plot_avg_vel_alphabeta_streamlines}
\end{figure*}
\begin{figure}[t]
	{\includegraphics[width=0.6\textwidth]{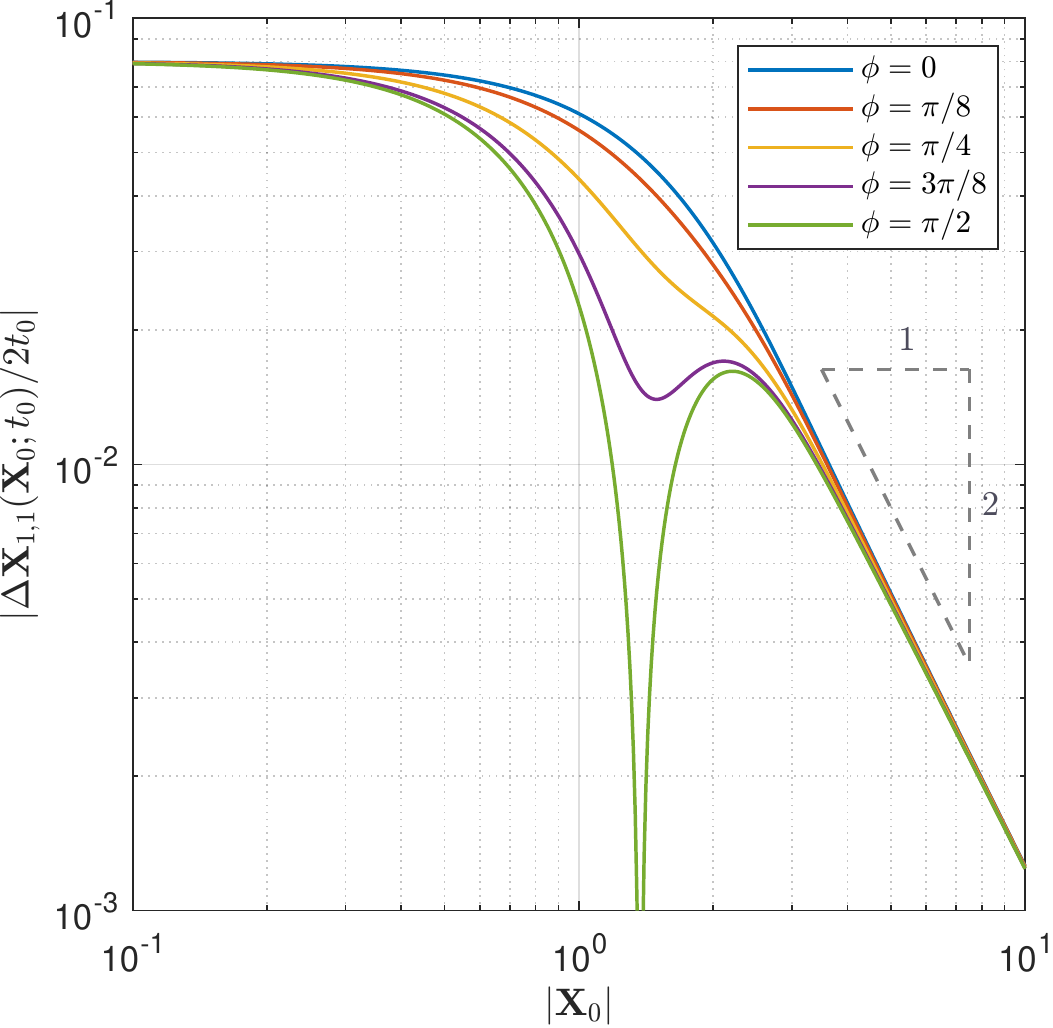}
		\caption{Magnitude of the average Lagrangian velocity $\vert \Delta \mathbf{X}_{1,1}(\mathbf{X}_0;t_0)/2t_0 \vert$ at order $\alpha\beta$ of material points against the  distance $\vert \mathbf{X}_0 \vert$, plotted on a log--log scale, for $\phi = 0$, $\pi/8$, $\pi/4$, $3\pi/8$, and $\pi/2$.
		Here, the initial position of a material point is written as $\mathbf{X}_0 = \vert \mathbf{X}_0 \vert (\cos\phi,\sin\phi)$, i.e.,~in polar coordinates with origin at the midpoint of the scan path.
		In the far field, the magnitude $\vert \Delta\mathbf{X}_{1,1}(\mathbf{X}_0;t_0)/2t_0  \vert$ follows an inverse square law $\sim 1/\vert \mathbf{X}_0 \vert^2$.}
		\label{fig:plot_avg_vel_alphabeta_speed_radius}}
\end{figure}
We plot the streamlines of the velocity $\Delta \mathbf{X}_{1,1}(\mathbf{X}_0;t_0)/2t_0$ of material points at order $\alpha\beta$, averaged over one scan period, in Fig.~\ref{fig:plot_avg_vel_alphabeta_streamlines}, with the near field in Fig.~\ref{fig:plot_avg_vel_alphabeta_streamlines_near} and the far field in Fig.~\ref{fig:plot_avg_vel_alphabeta_streamlines_far}.
To quantify spatial variation, we plot the magnitude of this average Lagrangian velocity, $\vert\Delta \mathbf{X}_{1,1}(\mathbf{X}_0;t_0)/2t_0 \vert$, at order $\alpha\beta$ as a function of the initial distance $\vert \mathbf{X}_0 \vert$ of the material point from the midpoint of the scan path, in Fig.~\ref{fig:plot_avg_vel_alphabeta_speed_radius}.
For this log--log plot, we write the initial position of the material point as $\mathbf{X}_0 = \vert \mathbf{X}_0 \vert (\cos\phi, \sin\phi)$, i.e.,~using polar coordinates with origin at the midpoint of the scan path.

\subsubsection{Far-field behaviour for general heat-spot amplitude and finite scan path}

Recall from Eq.~\eqref{eq:u11_far-field} that in the far field, the instantaneous flow $\mathbf{u}_{1,1}$ at order $\alpha\beta$ is a hydrodynamic source dipole.
Specifically, this applies many heat-spot radii away from the centre of the translating heat spot, $r \gg 1$. 
Substituting this into Eq.~\eqref{eq:net_displ_alphabeta}, the net displacement of a material point that remains many heat-spot radii from the heat spot throughout a scan is given by
\begin{align}
	\Delta\mathbf{X}_{1,1}(\mathbf{X}_0;t_0) \sim \frac{1}{4} \int_{-t_0}^{t_0}  A(t)^2 \left . \left \{\mathbf{e}_x  \left  [\frac{1}{r^2} - \frac{2(x-t)^2}{r^4} \right  ]  + \mathbf{e}_y \left [ -\frac{2(x-t)y}{r^4} \right ]\right \} \right \vert_{\mathbf{X}_0} \, dt,
\end{align}
where we recall that $r^2 = (x-t)^2 + y^2$.

If, in addition to this, the material point is many scan-path lengths away from the heat spot (at all times), i.e.,~$r \gg \ell$, then we can also approximate the spatially-varying factor in the integrand as its value at $t=0$, to leading order.
This gives
\begin{align}
	\Delta\mathbf{X}_{1,1}(\mathbf{X}_0;t_0) \sim \frac{1}{4}\int_{-t_0}^{t_0} A(t)^2 \, dt \left [ \mathbf{e}_x  \left  (\frac{1}{\vert \mathbf{X}_0 \vert^2} - \frac{2X_0^2}{\vert \mathbf{X}_0 \vert^4} \right ) + \mathbf{e}_y \left (- \frac{2X_0 Y_0}{\vert \mathbf{X}_0 \vert^4} \right ) \right ].\label{eq:net_displ_alphabeta_far-field}
\end{align}
Correspondingly, the velocity $\Delta \mathbf{X}_{1,1} (\mathbf{X}_0;t_0) /2t_0$ of the material point initially at $\mathbf{X}_0$, averaged over one scan, is given by
\begin{align}
	\frac{\Delta\mathbf{X}_{1,1}(\mathbf{X}_0;t_0)}{2t_0} 
	\sim \frac{1}{8t_0}
	\int_{-t_0}^{t_0} A(t)^2 \, dt 
	\left [ \mathbf{e}_x  \left  (\frac{1}{\vert \mathbf{X}_0 \vert^2} - \frac{2X_0^2}{\vert \mathbf{X}_0 \vert^4} \right ) + \mathbf{e}_y \left (- \frac{2X_0 Y_0}{\vert \mathbf{X}_0 \vert^4} \right ) \right ],
\end{align}
if the material point is far from the scan path, i.e.,~both many heat-spot radii and many scan-path lengths away. 
We note that this is only possible for scan paths of finite length.
This far-field average Lagrangian velocity field is a source dipole, with strength proportional to the time-average of the square of the heat-spot amplitude.
The direction of circulation corresponds to a source on the left and a sink on the right, inherited from the instantaneous flow at order $\alpha\beta$ (see Sec.~\ref{sec:physical_mechanism_order_alphabeta}).

\subsubsection{Net displacement of material points on the $x$ axis for general heat-spot amplitude and general scan path length}

We now examine the net displacement, or equivalently the average velocity, of material points on the $x$ axis.
For a Gaussian heat spot, the instantaneous velocity field $\mathbf{u}_{1,1}$ [Eq.~\eqref{eq:u11}, Fig.~\ref{fig:plot_u11_streamlines}, and Fig.~\ref{fig:plot_u11_speed_radius}] is in the negative $x$ direction everywhere on the $x$ axis, i.e.,~in the opposite direction to the translation of the heat spot. 
As explained in Sec.~\ref{sec:physical_mechanism_order_alphabeta}, this is because of localised amplification of the leading-order flow due to heat-spot translation in the near field and a source-dipole flow enforcing incompressibility in the far field.
Consequently, the net displacement in Eq.~\eqref{eq:net_displ_alphabeta} is also in the negative $x$ direction for any material point on the $x$ axis and there is strong leftward transport near the scan path.
This is illustrated for a sinusoidal amplitude function in Fig.~\ref{fig:plot_avg_vel_alphabeta_streamlines} and Fig.~\ref{fig:plot_avg_vel_alphabeta_speed_radius}.

\subsubsection{Constant heat-spot amplitude and infinite scan path}

Recall that earlier theoretical work~\cite{weinert2008optically} focused on the net displacement of material points lying on an infinite scan path, for a constant-amplitude heat spot.
To begin comparing our mathematical model with this, we therefore first substitute the heat-spot amplitude $A(t) = A = \text{const}$  and take the limit of infinite scan-path length ($t_0 \to \infty$), but still allow the material point to have arbitrary initial position $\mathbf{X}_0$ (i.e.,~$Y_0$ is not necessarily zero).
For this special case, the net displacement at order $\alpha\beta$ [Eq.~\eqref{eq:net_displ_alphabeta}] may be evaluated analytically as
\begin{align}
	\Delta\mathbf{X}_{1,1}(\infty) 
	&= - A^2 \mathbf{e}_x \{ \sqrt{\pi}[\sqrt{2} \exp(-Y_0^2/2) -  \exp(-Y_0^2)] - \pi Y_0 [\erf(Y_0) - \erf(Y_0/\sqrt{2})] \}.\label{eq:net_displ_alphabeta_infinite}
\end{align}

The above result  [Eq.~\eqref{eq:net_displ_alphabeta_infinite}] allows us to make important physical observations.
First, the net displacement at order $\alpha\beta$ is in the negative $x$ direction for all $Y_0$, i.e.,~for all material points, not only for material points on the scan path.
This holds both for a heat spot ($A>0$) and for a cool spot ($A<0$), since the net displacement is quadratic in the heat-spot amplitude.
There is no net displacement in the $y$ direction, by symmetry of the instantaneous flow.
Secondly, the net displacement at order $\alpha\beta$  is independent of  $X_0$ (the horizontal component of the initial position of the material point).
This is unsurprising, since all $x$ are indistinguishable when the scan path is infinite.

Thirdly, we observe that the net displacement at order $\alpha\beta$ arises purely from the amplification of the leading-order instantaneous flow, not from the potential flow at order $\alpha\beta$.
Recall that the instantaneous flow at order $\alpha\beta$ is a sum of a potential flow $-\nabla p_{1,1}$ and the leading-order flow modulated by the temperature profile $\Delta T \, \mathbf{u}_{1,0}$ [Eq.~\eqref{eq:u11_p11_Theta_u10}].
Since we are considering here the case where the time-derivative of the amplitude is zero, there is no contribution to either of these terms from the switching-on or switching-off of the heat spot, only from the translation of the heat spot.
The instantaneous flow is therefore steady in the comoving frame.
As a result, the $x$ component of the potential flow is the exact time-derivative of the pressure (the potential), which decays at infinity.
The integral of the potential flow over time is therefore zero; that is, the potential flow contributes no net displacement.
A similar result holds for material points on the scan path of a travelling wave temperature profile~\cite{weinert2008microscale}, which has periodic boundary conditions in the $x$ direction instead of decay.

Instead, the nonzero net displacement of the material point arises 
from the amplified leading-order velocity contribution $\Delta T \, \mathbf{u}_{1,0}$ in Eq.~\eqref{eq:u11_p11_Theta_u10}, not only for the Gaussian but for any temperature profile decaying at infinity.
We recall from earlier discussion, summarised in Fig.~\ref{fig:diagram_alphabeta_mechanism_cartoon}, that this contribution is the intuitive effect of the heat spot, amplifying the leftward leading-order flow in the heat spot by decreasing the viscosity there, as described in Ref.~\cite{weinert2008optically}.
This then results in net leftward displacement of the material point.

Therefore, we now see that each of the two terms in the instantaneous velocity field at order $\alpha\beta$, $\mathbf{u}_{1,1} = - \nabla p_{1,1} + \Delta T \, \mathbf{u}_{1,0}$,  is responsible for a key physical feature.
The potential flow, $-\nabla p_{1,1}$, is solely responsible for the hydrodynamic source dipole in the far field of the average Lagrangian velocity  $\Delta \mathbf{X}_{1,1} (\mathbf{X}_0;t_0) /2t_0$, inherited from $\mathbf{u}_{1,1}$ (explained in Sec.~\ref{sec:order_alpha_beta}).
On the other hand, the amplified leading-order flow, $\Delta T \, \mathbf{u}_{1,0}$, is solely responsible for the leftwards leading-order net displacement of a material point for an infinitely long scan path, occurring at order $\alpha\beta$.

\subsubsection{Comparison with earlier theoretical work}

We can now compare our results so far on net displacement with earlier work~\cite{weinert2008optically}.
The authors of Ref.~\cite{weinert2008optically} found that the net displacement of a material point on the infinitely-long scan path, for a Gaussian heat spot with constant amplitude, is given by
\begin{align}
	\Delta \mathbf{X}(\mathbf{0};\infty)_\text{\cite{weinert2008optically}} = - \frac{1}{2}\sqrt{\pi}\alpha\beta  \Delta T_0^2 A^2 a \mathbf{e}_x,\label{eq:net_displ_2008}
\end{align}
where we have converted to the notation of our theory (including using the $z$-averaged velocity instead of the mid-plane velocity).
In our work, we  found in Sec.~\ref{sec:net_displ_zero_alpha} that the net displacement at order $\alpha$ is zero, in agreement with Eq.~\eqref{eq:net_displ_2008}.
We also showed in Sec.~\ref{sec:order_beta_n} that at order $\beta^2$ there is no instantaneous flow, so there is also no net displacement at this order.
For water (used in Ref.~\cite{weinert2008optically}), the value of the thermal expansion coefficient~$\alpha$ is much smaller than the thermal viscosity coefficient~$\beta$; the contribution to net displacement at order $\alpha^2$ in our theory  may therefore be neglected in favour of order $\alpha\beta$.
Hence, within our perturbation-expansion framework, we have shown systematically that the leading-order net displacement of the material point occurs at order $\alpha\beta$, in agreement with Ref.~\cite{weinert2008optically} and Eq.~\eqref{eq:net_displ_2008}.

We substitute the initial position $\mathbf{X}_0 = \mathbf{0}$ into Eq.~\eqref{eq:net_displ_alphabeta_infinite}, which we recall came from integrating our analytical expression for the instantaneous flow field.
Our systematic  theoretical approach therefore predicts the dimensional, leading-order net displacement of this material point to be
\begin{align}
	\Delta \mathbf{X} (\mathbf{0};\infty) = -\sqrt{\pi} (\sqrt{2} - 1)\alpha\beta \Delta T_0^2 A^2 a  \mathbf{e}_x + O(\alpha^2) + \text{cubic and higher-order terms}.\label{eq:net_displ_infinite_tocompare2008}
\end{align}
This is quadratic in the peak temperature change $\Delta T_0 A$ and is always in the negative $x$ direction, i.e.,~in the opposite direction to  translation of the heat spot.
Comparing Eq.~\eqref{eq:net_displ_infinite_tocompare2008} with Eq.~\eqref{eq:net_displ_2008}, we see that our rigorous perturbation  calculation  results in an improved numerical factor ($\sqrt{2}-1\approx 0.41$ vs~$1/2=0.5$); however, all scalings are in agreement.

\subsection{Net displacement at order $\alpha^2$}

We now consider the net displacement at order $\alpha^2$ of a material point. This depends only on thermal expansion, so would exist even if the fluid viscosity were constant.
By Eq.~\eqref{eq:displacement_expanded}, the net displacement $\Delta\mathbf{X}_{2,0}(t_0)$ at order $\alpha^2$ is given by
\begin{align}
	\Delta\mathbf{X}_{2,0}(t_0) = \int_{-t_0}^{t_0} [\mathbf{u}_{2,0}(\mathbf{X}_0,t) + \Delta \mathbf{X}_{1,0}(t) \cdot \nabla \mathbf{u}_{1,0}(\mathbf{X}_0,t)] \, dt,\label{eq:net_displ_alphasq_integral}
\end{align}
where we recall that $\mathbf{u}_{2,0}$ is the instantaneous velocity field at order $\alpha^2$ [Eqs.~\eqref{eq:u20_decomposn},~\eqref{eq:u20_switch}, and~\eqref{eq:u20_translate}], $\Delta\mathbf{X}_{1,0}(t)$ is the displacement at order $\alpha$ at time $t$ of the material point [Eq.~\eqref{eq:X10_closed_form}], $\mathbf{u}_{1,0}$ is the instantaneous velocity field at order $\alpha$ [Eqs.~\eqref{eq:u10_decomposn},~\eqref{eq:u_switch}, and~\eqref{eq:u_translate}], and $\mathbf{X}_0$ is the position of the material point when the heat spot is at the start of the scan path ($t=-t_0$).
The equation numbers correspond to our analytical results for the Gaussian heat spot.

We observe that there are two contributions to this due to the Taylor expansion in Eq.~\eqref{eq:displacement_expanded}.
The first  is the integral of the velocity field at order $\alpha^2$ evaluated at the initial position of the material point.
This is similar to the net displacements at order $\alpha$ and order $\alpha\beta$.
However, in contrast with order $\alpha$ and order $\alpha\beta$, there is a second contribution.
This arises from the order-$\alpha$ displacement at time $t$ of the material point from its initial position, when the material point is experiencing the order-$\alpha$ instantaneous flow induced by the heat spot.

\subsubsection{Sinusoidal heat-spot amplitude and finite scan path}

\begin{figure*}[t]
	\subfloat[]
	{\includegraphics[width=0.54\textwidth]{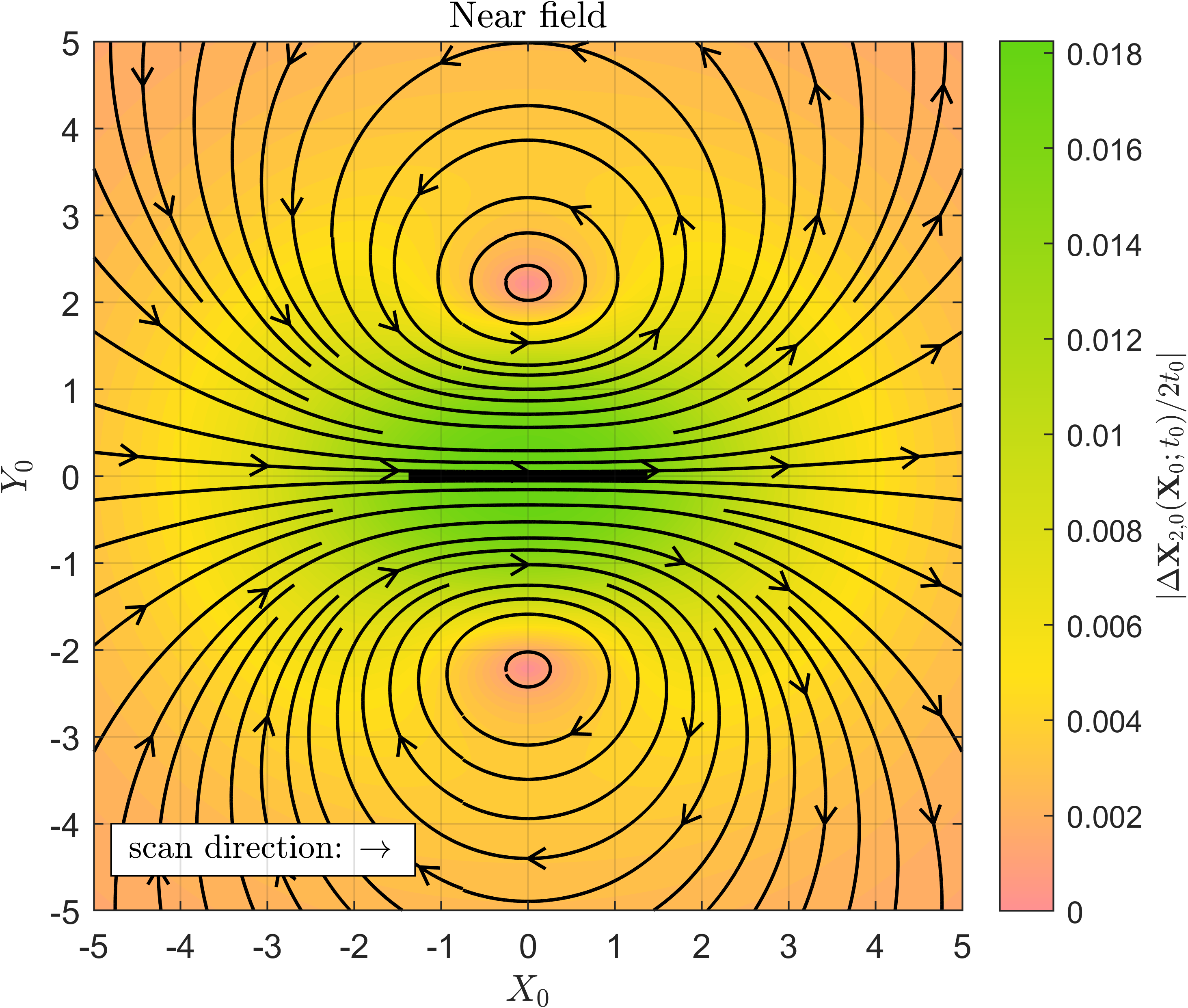}
		\label{fig:plot_avg_vel_alphasq_streamlines_near}}
	\subfloat[]
	{\includegraphics[width=0.46\textwidth]{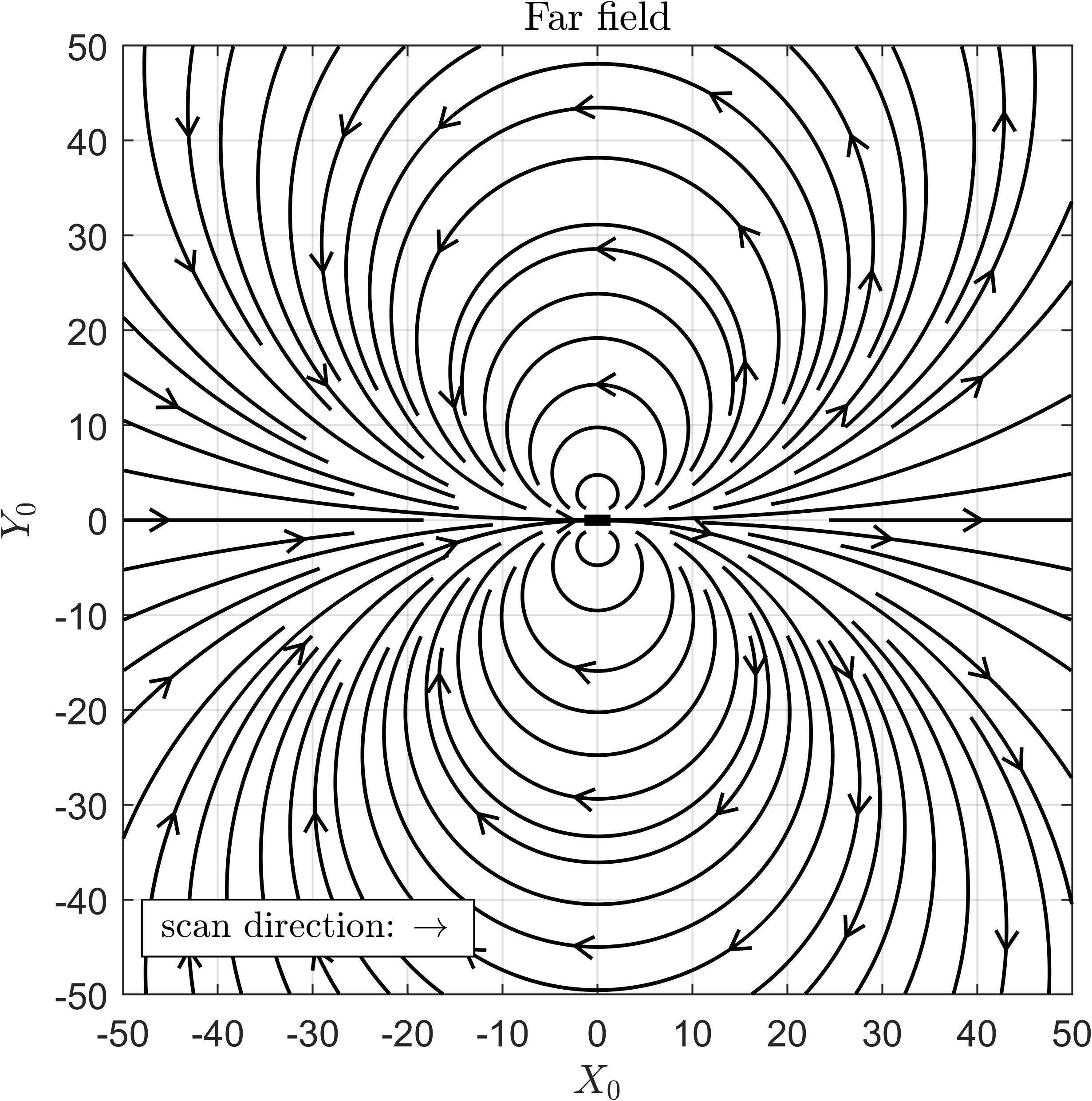}
		\label{fig:plot_avg_vel_alphasq_streamlines_far}} 
	\caption{Streamlines for the average velocity of a material point at order $\alpha^2$ over a scan period, $\Delta \mathbf{X}_{2,0}(\mathbf{X}_0;t_0)/2t_0$. The heat spot translates in the positive $x$ direction (scan direction).  The scan path is indicated with a thick black line segment. 
		Left (Fig.~\ref{fig:plot_avg_vel_alphasq_streamlines_near}): streamlines for $-5 \leq X_0, Y_0 \leq 5$, close to the scan path, with magnitude of the average velocity of material points $ \vert \Delta \mathbf{X}_{2,0}(\mathbf{X}_0;t_0)/2t_0  \vert$ indicated by colour. 
		Right (Fig.~\ref{fig:plot_avg_vel_alphasq_streamlines_far}): streamlines for $-50 \leq X_0, Y_0 \leq 50$ to indicate far-field behaviour.}
	\label{fig:plot_avg_vel_alphasq_streamlines}
\end{figure*}
\begin{figure}[t]
	{\includegraphics[width=0.6\textwidth]{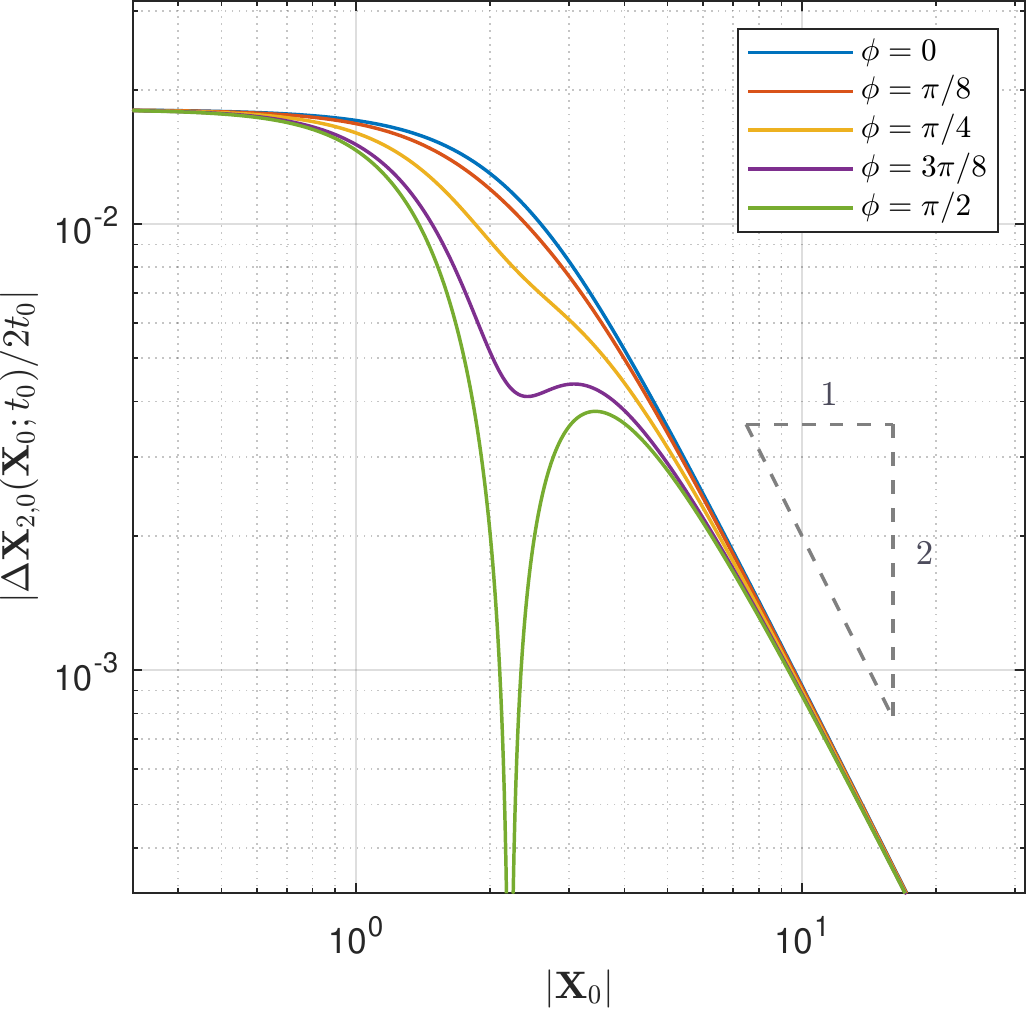}
		\caption{Magnitude of the average velocity $\vert \Delta \mathbf{X}_{2,0}(\mathbf{X}_0;t_0)/2t_0 \vert$ at order $\alpha^2$ of material points against the  distance $\vert \mathbf{X}_0 \vert$, plotted on a log--log scale, for $\phi = 0$, $\pi/8$, $\pi/4$, $3\pi/8$, and $\pi/2$.
			As at order $\alpha\beta$, the initial position of a material point is written as $\mathbf{X}_0 = \vert \mathbf{X}_0 \vert (\cos\phi,\sin\phi)$, i.e.,~in polar coordinates with origin at the midpoint of the scan path.
			In the far field, the magnitude $\vert \Delta \mathbf{X}_{2,0}(\mathbf{X}_0;t_0)/2t_0  \vert$ follows an inverse square law $\sim 1/\vert \mathbf{X}_0 \vert^2$.}
		\label{fig:plot_avg_vel_alphasq_speed_radius}}
\end{figure}
Here we illustrate how the  velocity of material points at order $\alpha^2$ averaged over one scan varies spatially, for the sinusoidal heat-spot amplitude function given by Eq.~\eqref{eq:A_sinusoidal}.
We plot  in Fig.~\ref{fig:plot_avg_vel_alphasq_streamlines} the streamlines of the average Lagrangian velocity $\Delta \mathbf{X}_{2,0}(\mathbf{X}_0;t_0)/2t_0$ at order $\alpha^2$, with the near field in Fig.~\ref{fig:plot_avg_vel_alphasq_streamlines_near} and the far field in Fig.~\ref{fig:plot_avg_vel_alphasq_streamlines_far}.
In Fig.~\ref{fig:plot_avg_vel_alphasq_speed_radius}, we plot on a log--log scale the magnitude of the average Lagrangian velocity, $\vert\Delta \mathbf{X}_{2,0}(\mathbf{X}_0;t_0)/2t_0 \vert$, at order $\alpha^2$ as a function of the initial distance $\vert \mathbf{X}_0 \vert$ of the material point from the midpoint of the scan path.
As before, we write the initial position of the material point as $\mathbf{X}_0 = \vert \mathbf{X}_0 \vert (\cos\phi, \sin\phi)$. 

We observe that the streamlines of the average Lagrangian velocity field at order $\alpha^2$ (Fig.~\ref{fig:plot_avg_vel_alphasq_streamlines}) are qualitatively similar to those at order $\alpha\beta$ (Fig.~\ref{fig:plot_avg_vel_alphabeta_streamlines}), except the direction of circulation is reversed: 
at order $\alpha^2$, near the scan path, the material points have average velocity in the same direction as the translation of the heat spot (rightwards), whereas far from the $x$ axis, the net transport of tracers is instead leftwards and decays in strength. 
Thus, even if viscosity were constant ($\beta=0$), we would predict order-$\alpha^2$ net transport of tracers that varies spatially in a similar way to the net transport due to the interplay between thermal viscosity changes and thermal expansion, but with the opposite direction of circulation.

\subsubsection{Far-field behaviour for general heat-spot amplitude and finite scan path}

For a Gaussian heat spot with arbitrary amplitude, we may find the far-field  velocity of material points at order $\alpha^2$ averaged over one scan. 
First we note how terms in the integrand in Eq.~\eqref{eq:net_displ_alphasq_integral} scale with distance $r$ from the centre of the heat spot, in the limit $r \gg 1$ (many heat-spot radii away).
The instantaneous flow $\mathbf{u}_{2,0}$ at order $\alpha^2$ is a hydrodynamic source dipole in the far field [Eq.~\eqref{eq:u20_far-field}], scaling as $1/r^2$; this comes from the contribution $A(t)^2 \mathbf{u}_{2,0}^\text{(T)}$ associated with heat-spot translation.
The displacement $\Delta\mathbf{X}_{1,0}(t)$ at order $\alpha$ scales as $1/r$.
The flow $\mathbf{u}_{1,0}$ at order $\alpha$ scales as  $1/r$, so its gradient, $\nabla \mathbf{u}_{1,0}$, scales as $1/r^2$.
Therefore, in the far field, the flow $\mathbf{u}_{2,0}$ [specifically the contribution $A(t)^2 \mathbf{u}_{2,0}^\text{(T)}$] at order $\alpha^2$ provides the dominant contribution to the integrand in Eq.~\eqref{eq:net_displ_alphasq_integral}.

Using this and Eq.~\eqref{eq:u20_far-field} gives
\begin{align}
	\Delta\mathbf{X}_{2,0}(\mathbf{X}_0;t_0) \sim -\frac{1}{4}\int_{-t_0}^{t_0}  A(t)^2 \left . \left \{\mathbf{e}_x  \left  [\frac{1}{r^2} - \frac{2(x-t)^2}{r^4} \right  ]  + \mathbf{e}_y \left [ -\frac{2(x-t)y}{r^4} \right ]\right \} \right \vert_{\mathbf{X}_0} \, dt,
\end{align}
provided that $r \gg 1$ throughout the scan period.

If, additionally, the material point is many scan-path lengths away from the heat spot, i.e.,~$r \gg \ell$, then we may also approximate the spatially-varying factor in the integrand with its value at $t=0$.
This is simply the leading-order Taylor-expansion in $t$.
This gives
\begin{align}
	\Delta\mathbf{X}_{2,0}(\mathbf{X}_0;t_0) 
	\sim -\frac{1}{4}\int_{-t_0}^{t_0} A(t)^2 \, dt 
	\left  [ \mathbf{e}_x  \left  (\frac{1}{\vert \mathbf{X}_0 \vert^2} - \frac{2X_0^2}{\vert \mathbf{X}_0 \vert^4} \right ) 
	+ \mathbf{e}_y \left (- \frac{2X_0 Y_0}{\vert \mathbf{X}_0 \vert^4} \right ) \right ].\label{eq:net_displ_alphasq_far-field}
\end{align}
Correspondingly, the velocity $\Delta \mathbf{X}_{2,0} (t_0) /2t_0$ at order $\alpha^2$ of the material point, averaged over one scan, is given by
\begin{align}
	\frac{\Delta\mathbf{X}_{2,0}(\mathbf{X}_0;t_0)}{2t_0} \sim -\frac{1}{8t_0}\int_{-t_0}^{t_0} A(t)^2 \, dt \left  [ \mathbf{e}_x  \left  (\frac{1}{\vert \mathbf{X}_0 \vert^2} - \frac{2X_0^2}{\vert \mathbf{X}_0 \vert^4} \right ) + \mathbf{e}_y \left (- \frac{2X_0 Y_0}{\vert \mathbf{X}_0 \vert^4} \right ) \right ],
\end{align}
for material points far from the scan path (many heat-spot radii and many scan-path lengths away).
Similarly to   order $\alpha\beta$ (i.e.,~the other quadratic order), this far-field average Lagrangian velocity is a hydrodynamic source dipole, with strength proportional to the time-average of the square of the heat-spot amplitude.
However, here at order $\alpha^2$, the direction of far-field flow  corresponds to a sink on the left and a source on the right.
This is therefore opposite to order $\alpha\beta$, consistent with the physical mechanism for the instantaneous flow $A(t)^2 \mathbf{u}_{2,0}^\text{(T)}$ explained in Sec.~\ref{sec:physical_mechanism_order_alphasq}.

\subsection{Net displacement in the far field}

We computed above  the net displacement of a material point correct to quadratic order and derived analytical expressions for the net displacements at order $\alpha\beta$ and order $\alpha^2$ of material points far from the scan path, for a Gaussian heat spot with arbitrary amplitude. 
Combining Eqs.~\eqref{eq:net_displ_alphabeta_far-field} and~\eqref{eq:net_displ_alphasq_far-field}, the dimensional net displacement of a material point far from the scan path is therefore given by
\begin{align}
	\Delta\mathbf{X}(\mathbf{X}_0;t_0) 
	\sim \frac{1}{4} (\alpha\beta-\alpha^2)\Delta T_0^2 U \int_{-t_0}^{t_0} A(t)^2 \, dt 
	\left  [ \mathbf{e}_x  \left  (\frac{a^2}{\vert \mathbf{X}_0 \vert^2} - \frac{2a^2 X_0^2}{\vert \mathbf{X}_0 \vert^4} \right ) + \mathbf{e}_y \left (- \frac{2 a^2 X_0 Y_0}{\vert \mathbf{X}_0 \vert^4} \right ) \right ],
\end{align}
correct to quadratic order.
The corresponding average velocity (over a scan period) of the material point with initial position $\mathbf{X}_0$ is given by
\begin{align}
	\frac{\Delta\mathbf{X}(\mathbf{X}_0;t_0)}{2t_0} 
	\sim \frac{1}{8t_0} (\alpha\beta-\alpha^2)\Delta T_0^2 U \int_{-t_0}^{t_0} A(t)^2 \, dt 
	\left  [ \mathbf{e}_x  \left  (\frac{a^2}{\vert \mathbf{X}_0 \vert^2} - \frac{2 a^2 X_0^2}{\vert \mathbf{X}_0 \vert^4} \right ) + \mathbf{e}_y \left (- \frac{2 a^2 X_0 Y_0}{\vert \mathbf{X}_0 \vert^4} \right ) \right ].\label{eq:avg_vel_far-field_quadratic}
\end{align}
This is a hydrodynamic source dipole, with strength quadratic in the heat-spot amplitude.
We have thus shown that the exponentially-decaying heat spot induces motion of tracers with average velocity decaying algebraically in the far field (i.e.,~slower than the forcing). 
The direction of net displacement of material points in the far field depends critically on the relative size of the thermal expansion coefficient~$\alpha$ and the thermal viscosity coefficient~$\beta$, via the prefactor of $(\alpha \beta - \alpha^2)$. 
As a reminder, this prefactor is expected to be positive for water or glycerol-water solution, i.e.,~the order-$\alpha\beta$ behaviour dominates.

\section{Comparison with experiments}\label{sec:comparison_expt}

\subsection{Trajectories of tracers over many scan periods}

In Sec.~\ref{sec:inst_flow}, we introduced our mathematical model for the microfluidic experiments reported in Ref.~\cite{erben2021feedback}.
We  solved explicitly for the instantaneous flow induced by a translating Gaussian heat spot with arbitrary amplitude and then examined in Sec.~\ref{sec:one_scan} the leading-order trajectories of material points during one scan period, occurring at order $\alpha$.
In particular, we showed that the  net displacement of any material point after a full scan period is zero at order $\alpha$.
The leading-order net displacement instead occurs at higher order and is a quadratic effect (Sec.~\ref{sec:net_displ}), which in fact holds for a general heat spot. In experiments~\cite{erben2021feedback,mittasch2018non,weinert2008optically}, repeated scanning of the heat spot along a finite scan path, i.e.,~localised forcing, creates a large-scale fluid flow that transports material points over large distances, over the course of many scan periods.
To understand this net transport due to repeated scanning, we use our results in Sec.~\ref{sec:net_displ} on the net displacement and average velocity of material points, due to a full scan.
In this section, we  quantitatively compare these  theoretical results with experimental data~\cite{erben2021feedback}.
Specifically, we  focus on  Figures~2d and 2f from Ref.~\cite{erben2021feedback}.

In the controlled experiments in Ref.~\cite{erben2021feedback}, the laser spot scanned at frequency $1\text{--}3~\si{kHz}$ and the scan path had length $11~\si{\micro\metre}$.
The laser spot heated the fluid locally, causing a maximum temperature perturbation of a few kelvins; the characteristic radius of the temperature perturbation was $a = 4$--$4.5~\si{\micro\metre}$.
The resulting trajectories of tracer beads in the fluid were recorded over a period of $100~\si{\second}$. 
The flow recordings were made with an exposure time of at least $10$ times the scan period.
As a result, the trajectories show the displacement of tracers  over many scan periods, instead of the displacement during a single scan period.

In experiments, the motion of the tracer beads was tracked in the mid-plane, i.e.,~halfway between the parallel plates, at $z=h/2$ in terms of notation in Fig.~\ref{fig:diagram_setup}.
Recall that in our theory, we dealt with the  $z$-averaged velocity field $\overline{\mathbf{u}}_\text{H}$, which we wrote as $\mathbf{u}$ to simplify notation.
To convert to the mid-plane velocity field, we recall that within the lubrication approximation, the velocity in the mid-plane $z=h/2$ is given by 
\begin{align}
	\left .\mathbf{u}_\text{H}\right \vert_{z=h/2} = \frac{3}{2} \overline{\mathbf{u}}_\text{H},
\end{align}
since the flow has quadratic dependence on $z$ between the two no-slip surfaces. 
We model the tracer beads as material points for simplicity and discuss this assumption later.

To produce our theoretical plots, we choose the sinusoidal heat-spot amplitude given by Eq.~\eqref{eq:A_sinusoidal}, based on earlier experiments and modelling~\cite{mittasch2018non}.
We use parameter values within the experimental range:
characteristic heat-spot radius $a= 4~\si{\micro\metre}$, scan-path length $2\ell = 11~\si{\micro\metre}$, scan frequency $f = 3.2~\si{\kilo\hertz}$, and characteristic  temperature change $\Delta T_0 = 8~\si{\kelvin}$. 
The fluid in the experiment was 50\% v/v glycerol-water solution.
For this, we estimate the dimensional thermal expansion coefficient as $\alpha = 5\times 10^{-4}~\si{\kelvin}^{-1}$ and the thermal viscosity coefficient as $\beta = 4\times 10^{-2}~\si{\kelvin}^{-1}$~\cite{rumble2017crc}. 
The dimensionless values are then $\alpha =  4\times 10^{-3}$ and $\beta =  0.32$.
Since we have $\beta \gg \alpha$, we may neglect order $\alpha^2$ contributions in our theory relative to order $\alpha\beta$.

\begin{figure*}[t] 
	\subfloat[]
	{\includegraphics[width=0.5\textwidth]{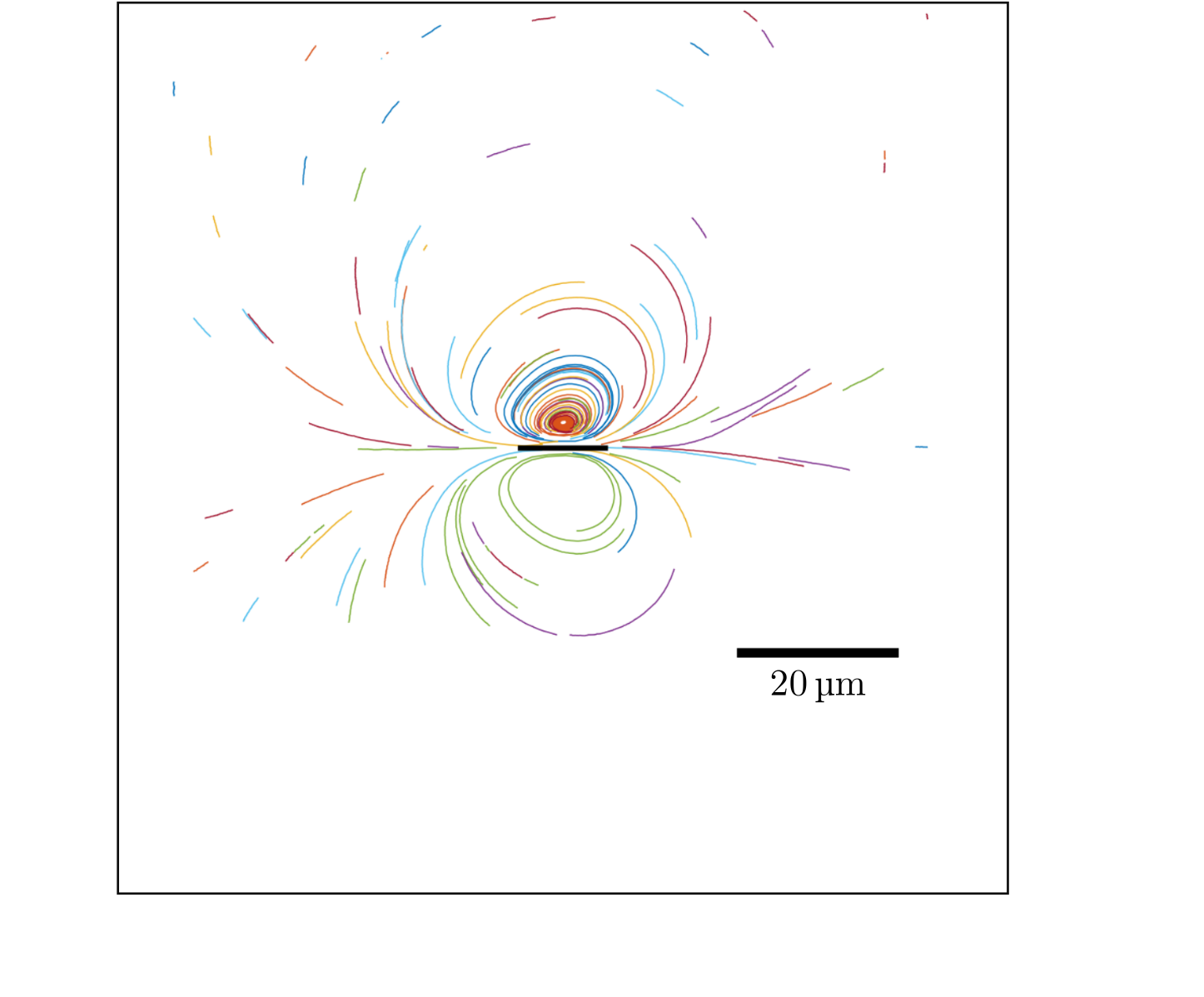}
		\label{fig:plot_OpticsExpress2d}} 
	\subfloat[]
	{\includegraphics[width=0.5\textwidth]{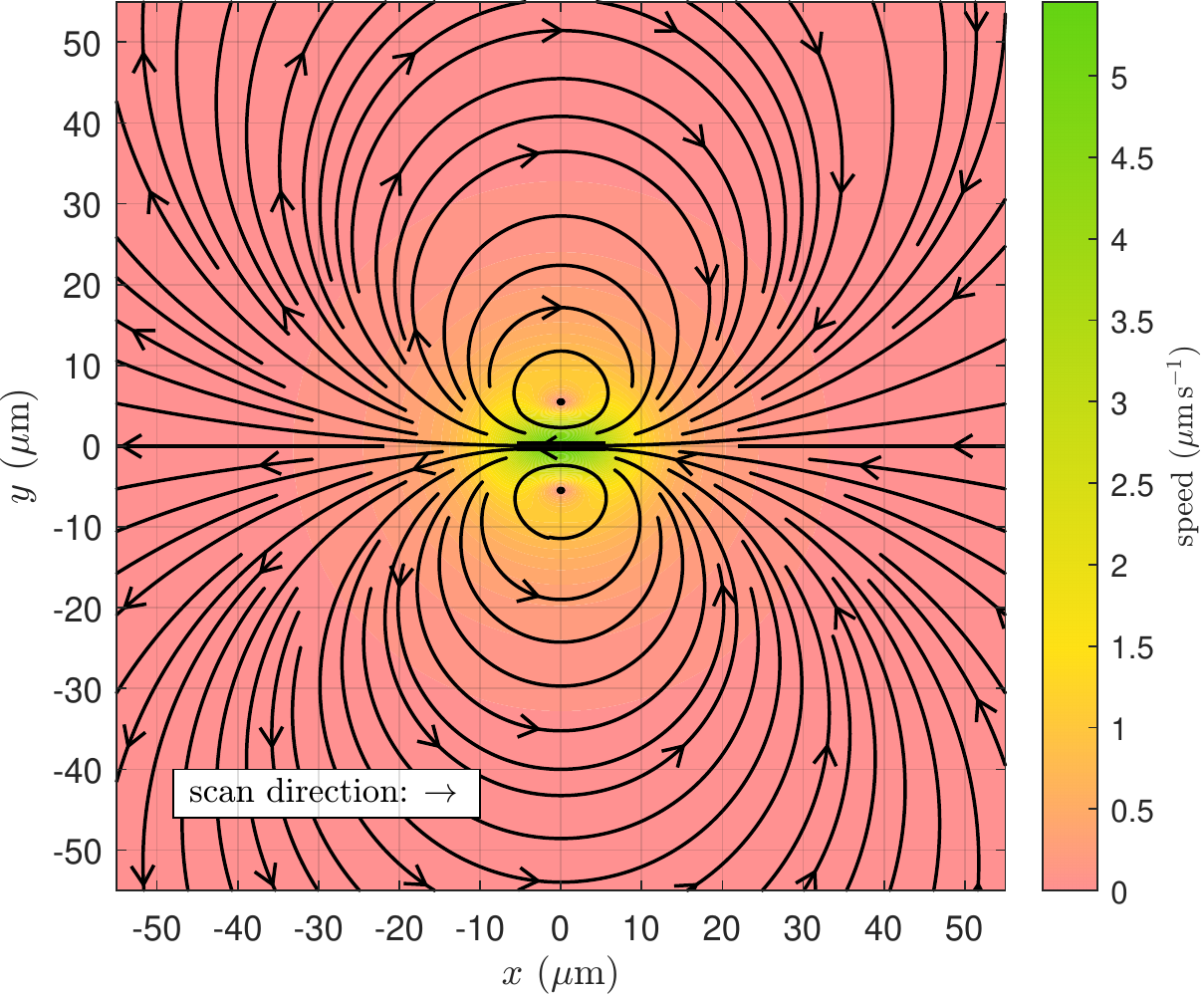}
		\label{fig:plot_diml_avg_vel_alphabeta_streamlines}}
	\caption{Comparison between experimental  and theoretical results for trajectories of tracer beads over many scans of the heat spot. Left (Fig.~\ref{fig:plot_OpticsExpress2d}): experimentally-found trajectories of tracer beads in the mid-plane (halfway between the parallel plates), from data gathered over a time period of $100~\si{\second}$,	adapted with permission from Figure 2d of Ref.~\cite{erben2021feedback} \copyright The Optical Society. 
	Right (Fig.~\ref{fig:plot_diml_avg_vel_alphabeta_streamlines}):  theoretical leading-order trajectories of material points in the mid-plane ($z=h/2$) over many scans, with fully dimensional units; the  average  speed of material points is indicated with colour.
	During each scan, the heat spot translates in the positive $x$ direction in our theory (scan direction).  
	In both panels, the scan path is indicated with a thick black line segment.}
	\label{fig:plot_OpticsExpress2d_compare}
\end{figure*}
We begin with a  qualitative comparison of the experimental and theoretical trajectories of tracer beads over many scans.
In Figure 2d of Ref.~\cite{erben2021feedback}, the experimentally-found trajectories of individual tracer beads are plotted. 
We  compare in Fig.~\ref{fig:plot_OpticsExpress2d_compare} these experimental data (Fig.~\ref{fig:plot_OpticsExpress2d}) with our dimensionalised theory (Fig.~\ref{fig:plot_diml_avg_vel_alphabeta_streamlines}).
Recall that our theory is valid in the limit of small (dimensionless) thermal expansion coefficient~$\alpha$ and thermal viscosity coefficient~$\beta$, with the leading-order trajectories of material points over many scans  given by the streamlines of the leading-order average velocity of the material points (order $\alpha\beta$). 
We plot these streamlines in Fig.~\ref{fig:plot_diml_avg_vel_alphabeta_streamlines}; this is a dimensionalised version of Fig.~\ref{fig:plot_avg_vel_alphabeta_streamlines}.

\begin{figure*}[t] 
	\subfloat[]
	{\includegraphics[width=0.45\textwidth]{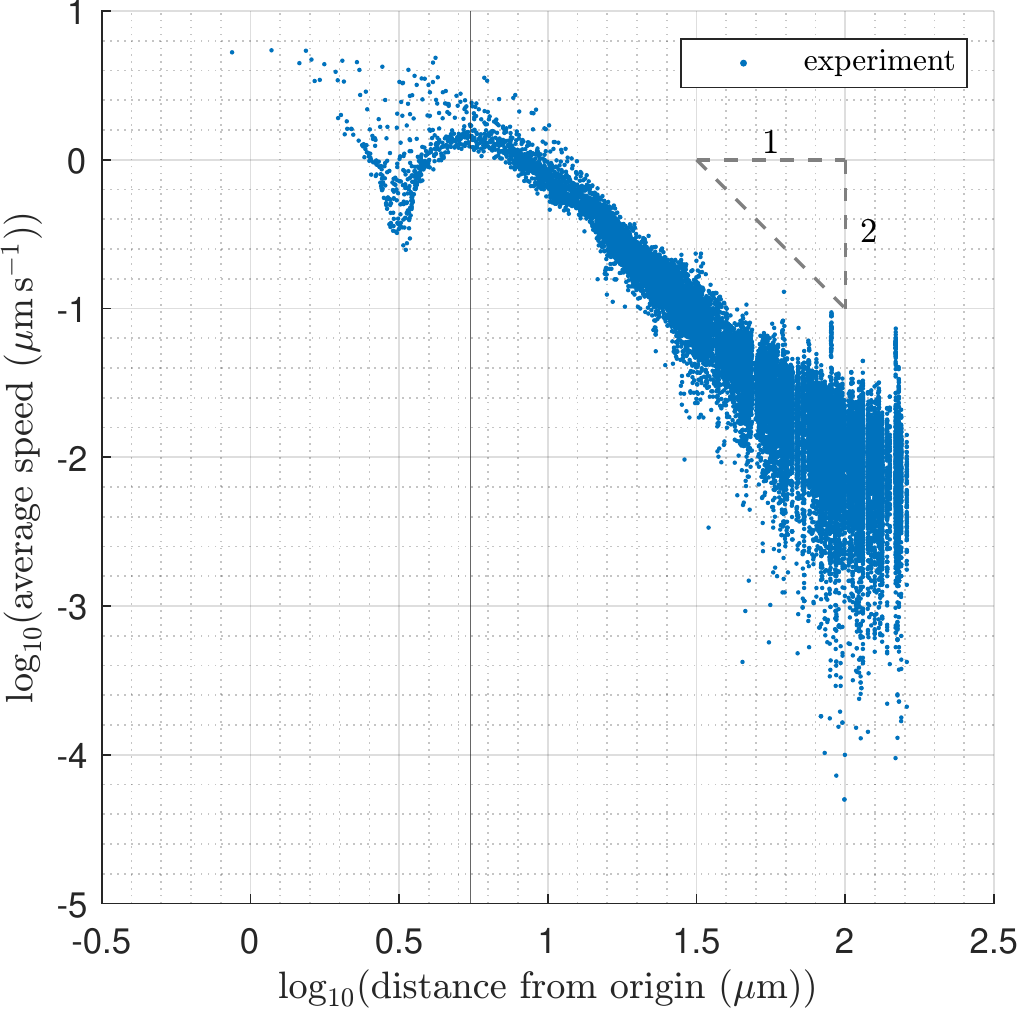}
		\label{fig:plot_OpticsExpress2f_speed_radius}} 
	\subfloat[]
	{\includegraphics[width=0.45\textwidth]{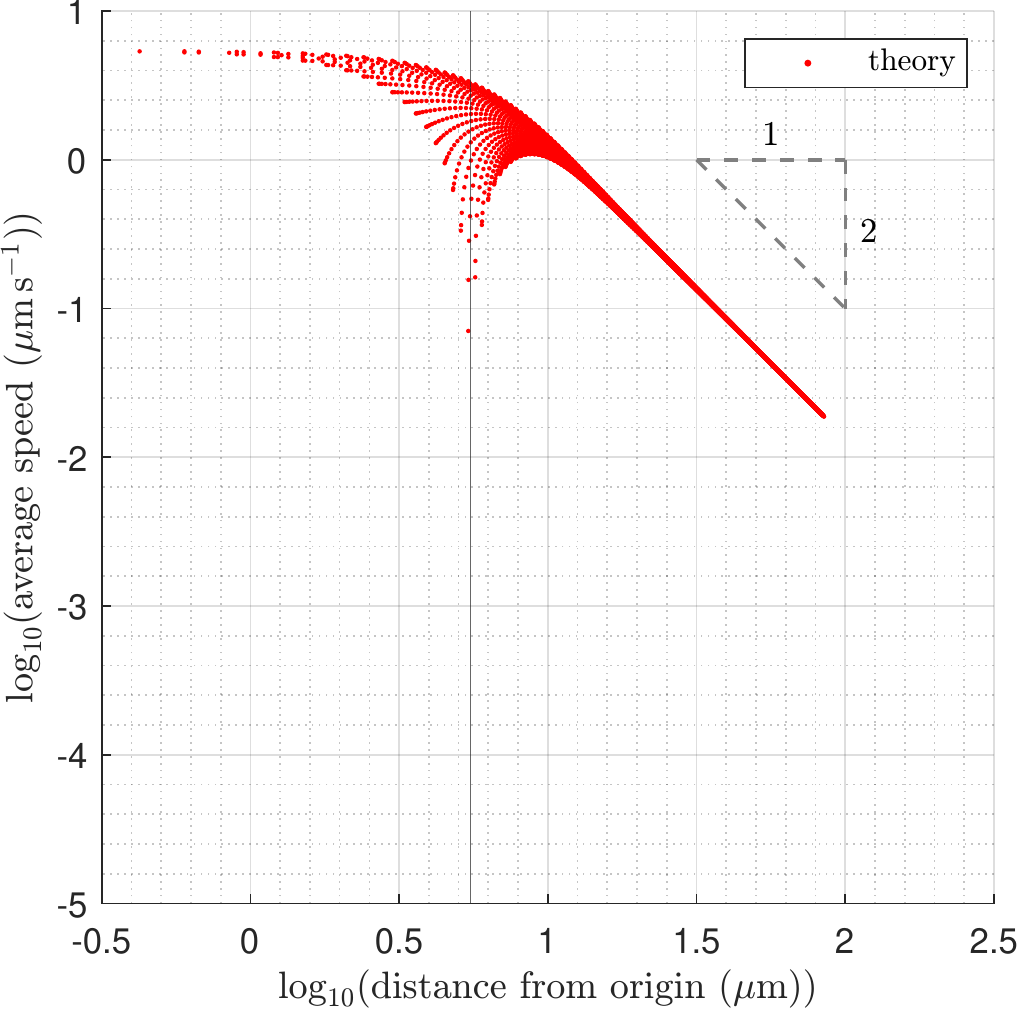}
		\label{fig:plot_diml_scatter_avg_speed_radius}} \\
	\subfloat[]
	{\includegraphics[width=0.45\textwidth]{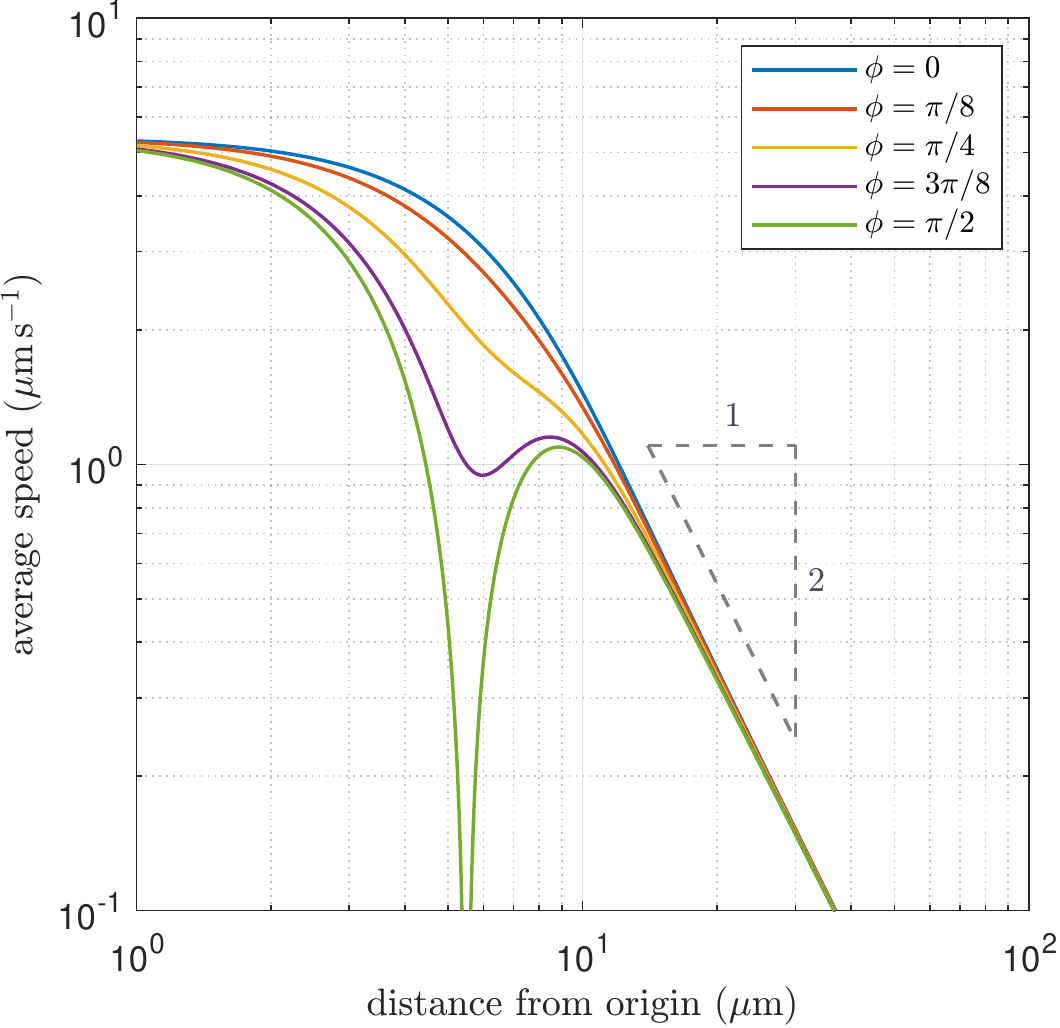}
		\label{fig:plot_diml_avg_speed_radius}}
	\caption{Comparison between experimental data and theory for average speed of tracer beads. 
		Top left (Fig.~\ref{fig:plot_OpticsExpress2f_speed_radius}): experimental data~\cite{erben2021feedback}. 
		Log--log scatter plot of speed of tracer beads in the mid-plane averaged over $0.5~\si{\second}$ vs distance of tracer bead from the origin (the midpoint of the scan path). 
		Data from Figure 2f in Ref.~\cite{erben2021feedback} replotted. 
		Top right (Fig.~\ref{fig:plot_diml_scatter_avg_speed_radius}): theoretical leading-order prediction for Fig.~\ref{fig:plot_OpticsExpress2f_speed_radius}, now dimensional. 
		To generate the data points, the tracers are placed at initial positions lying on a square grid with spacing of~$0.3~\si{\micro\metre}$. 
		In Figs.~\ref{fig:plot_OpticsExpress2f_speed_radius} and~\ref{fig:plot_diml_scatter_avg_speed_radius}, the black vertical line marks the distance from the origin corresponding to the end of the scan path, i.e.,~$\ell=5.5~\si{\micro\metre}$. 
		Bottom (Fig.~\ref{fig:plot_diml_avg_speed_radius}): theoretical prediction. Log--log plot of leading-order speed of a material point averaged over one scan vs distance of the material point from the origin, along radial lines at fixed angles $\phi=0$, $\pi/8$, $\pi/4$, $3\pi/8$, and $\pi/2$ to the $x$ axis. 
		Parameters are the same as for Fig.~\ref{fig:plot_diml_scatter_avg_speed_radius}.}
	\label{fig:plot_OpticsExpress2f_compare}
\end{figure*}

Next, to compare theory with experiment quantitatively, we illustrate in Fig.~\ref{fig:plot_OpticsExpress2f_compare} how the speed of tracer beads varies spatially.
In Fig.~\ref{fig:plot_OpticsExpress2f_speed_radius}, we replot the data in Figure 2f from Ref.~\cite{erben2021feedback}.
This is  a log--log scatter plot of the speed of tracer beads in the mid-plane, averaged over 
$0.5~\si{\second}$, against the distance of the bead from the midpoint of the scan path (the origin in our theoretical model).
We display in Fig.~\ref{fig:plot_diml_scatter_avg_speed_radius} an equivalent theoretical scatter plot for material points, with initial positions lying on a square grid with spacing of~$0.3~\si{\micro\metre}$. 
Finally, in Fig.~\ref{fig:plot_diml_avg_speed_radius}, we plot  the leading-order speed of a material point in the mid-plane, averaged over one scan, against the distance of the material point from the origin (log--log scale), at fixed angles $\phi=0$, $\pi/8$, $\pi/4$, $3\pi/8$, and $\pi/2$ to the $x$ axis.
This will allow us to better understand the first two, directly comparable plots, in Figs.~\ref{fig:plot_OpticsExpress2f_speed_radius} and~\ref{fig:plot_diml_scatter_avg_speed_radius}.

\subsection{Predictions of theoretical model}\label{sec:predictions}

From Figs.~\ref{fig:plot_OpticsExpress2d_compare} and~\ref{fig:plot_OpticsExpress2f_compare}, we see that our theoretical model can produce qualitative and quantitative predictions in agreement with the experimental data, from the near field to the far field.

First, our theory accurately predicts the far-field decay of the average speed of tracers as the inverse squared distance.
Crucially, we showed rigorously that in the far field, even though the leading-order instantaneous velocity field is a source flow  (decaying as $1/r$), the leading-order average velocity of tracers is a source dipole (decaying as $1/r^2$)  because the leading-order instantaneous flow always averages to zero over one scan period.
As a result, the leading-order average velocity of tracers inherits the $1/r^2$ decay from higher-order contributions to the instantaneous flow induced by the heat spot.

Secondly, our theory rigorously shows that the average velocity of tracers throughout space scales as $\alpha\beta \Delta T_0^2 U$  at leading order, for a general heat spot and finite scan path, for a fluid with $\beta \gg \alpha$.
This again follows from our result that the net displacement of a material point is zero at order $\alpha$; the leading-order average velocity is therefore quadratic in the heat-spot amplitude.
The linear scaling with $\alpha$ is consistent with experiments in which $\alpha$ is negative (for water below 4 \degree C)~\cite{weinert2008optically} instead of positive; the observed net transport of tracers reversed in direction.
The quadratic scaling with heat-spot amplitude is also supported by experimental data~\cite{weinert2008optically,mittasch2018non} and earlier theory for tracers lying on the scan path~\cite{weinert2008optically}.

Thirdly, our hydrodynamic theory successfully reproduces the closed-loop trajectories of tracers over many scans seen in experiments and approximately predicts the position of the middle of the closed loops, where the average speed of  tracer beads dips.
The direction of circulation of tracers, relative to the scan direction, matches that reported in experiments~\cite{weinert2008optically}.
Importantly, from  Eq.~\eqref{eq:avg_vel_far-field_quadratic} for the far-field average Lagrangian velocity, this direction is set by the relative sizes of the thermal expansion coefficient~$\alpha$ and the thermal viscosity coefficient~$\beta$.
Here, we have $\beta \gg \alpha$ for glycerol-water solution; consistent with this, our theory correctly predicts that on the scan path, the average velocity of tracers  is in the opposite direction to the translation of the heat spot.

Exploring this near-field behaviour further, on the $y$ axis at the middle of a closed loop, the average speed of a material point over one scan is zero, as we see in Fig.~\ref{fig:plot_diml_avg_speed_radius}.
In contrast with this, in the experimental data (Fig.~\ref{fig:plot_OpticsExpress2f_speed_radius}) and the theoretical equivalent (Fig.~\ref{fig:plot_diml_scatter_avg_speed_radius}), data points do not reach zero average speed in the near field.
By comparing the two theoretical plots in Fig.~\ref{fig:plot_diml_scatter_avg_speed_radius} and Fig.~\ref{fig:plot_diml_avg_speed_radius}, we see that this can be 
accounted for by the nonzero spacing between the tracer beads in experiments.
In our fully deterministic, hydrodynamic theory, spacing between material points that contribute to the scatter plots can ensure that none of them have average speed of precisely zero. 

\subsection{Scaling arguments for neglected physical effects}\label{sec:neglect}

Recall that in our model, we neglected both gravity and inertia, due to the small length scales involved. 
We now verify  using scaling arguments that this is justified, \textit{a posteriori}. 

As a reminder on the effect of gravity, the horizontal density gradients produce horizontal gradients in hydrostatic pressure, driving a gravity current~\cite{simpson1982gravity,yariv2004flow}. From the momentum equations, this gravity-driven flow scales as ${ \rho_0 \alpha g h^3 \Delta T_0 }/{2\eta_0 a}$. 
On the other hand, the flow driven by thermal expansion, which we focus on in our article, scales as $\alpha \Delta T_0 U$. The ratio of the gravity-driven flow to the thermal-expansion-driven flow  therefore scales as ${\rho_0 g h^3}/{2\eta_0 a U}$, which is approximately 0.03 for the experiments in Ref.~\cite{erben2021feedback}. 
This is indeed small, so gravity can be neglected.

Similarly, we can check the size of inertial terms relative to viscous terms in the momentum equation, based on our solution to the inertialess problem. 
The inertial term scales as
\begin{align}
	\rho \frac{\partial \mathbf{u}_\text{H} }{\partial t} \sim \rho_0 \alpha \Delta T_0 \left \{2af^2, Uf, \frac{U^2}{2a}  \right \},
\end{align}
where the three options exist due to the two contributions to the flow (switching on and translation), i.e.,~due to the two different time scales in the problem (scan period and advective time scale).
The viscous term scales as 
\begin{align}
	\eta \frac{\partial^2 \mathbf{u}_\text{H}}{\partial z^2}  
	\sim \frac{\eta_0 \alpha \Delta T_0}{h^2} \{  2af, U \}.
\end{align}
Then the ratio of the inertial term to the viscous term scales as (using $U=2\ell f$)
\begin{align}
	\frac{\text{inertial term}}{\text{viscous term}} 
	&\sim \frac{\rho_0 h^2 f}{ \eta_0  }  \left \{1, \frac{a}{\ell}, \frac{\ell}{a}, \frac{\ell^2}{a^2 }  \right \} .
\end{align}
The maximum value of this is approximately 0.1 for the experiments in Ref.~\cite{erben2021feedback}, which is indeed small. 

\subsection{Limitations of theoretical model}\label{sec:limitations}
 
We observe in Figs.~\ref{fig:plot_OpticsExpress2d_compare} and~\ref{fig:plot_OpticsExpress2f_compare} two key differences between the experimental and theoretical results, which we now discuss. 
First, in the experimental data (Fig.~\ref{fig:plot_OpticsExpress2f_speed_radius}), the dip in the average speed of tracer beads is a distance of less than half the scan-path length (i.e.,~less than $\ell$) from the midpoint of the scan path, whereas  this dip is slightly further away in the theoretical equivalent (Fig.~\ref{fig:plot_diml_scatter_avg_speed_radius}). 
Secondly, in the far field, the experimental data shows a much greater range of speeds at any given distance from the origin than the theoretical model predicts. 

\subsubsection{Lubrication limit}

What are the  limitations in  our model that could contribute to these discrepancies? 
First, we used the lubrication limit in order to simplify the momentum equations.
This corresponds to the limit where the characteristic horizontal length scale is much larger than the vertical; specifically, where the heat-spot diameter is much larger than the separation of the parallel plates.
However, in the experiments in which the spatial decay of average speed of tracers was quantified~\cite{erben2021feedback}, the characteristic heat-spot diameter was $2a = 8$--$9~\si{\micro\metre}$ and the separation of the parallel plates was $h = 15~\si{\micro\metre}$, i.e.,~the characteristic heat-spot diameter and plate separation were comparable. 
In the near field, this order-1 aspect ratio (instead of a thin-film geometry) is expected to become important, which could therefore help explain the slight difference between the predicted and experimentally-found position of the dip in the average speed of tracers.

\subsubsection{Diffusion of tracer beads}

Secondly, our theoretical model is fully deterministic; we have not included any noise.
We modelled the tracer beads as material points, simply advected by the flow induced by the heat spot.
In the experiments~\cite{erben2021feedback}, the tracer beads had a radius of $b=0.25~\si{\micro\metre}$, which is much smaller than the length scale on which the flow field varies, the characteristic radius of the heat spot $a = 4~\si{\micro\metre}$. 
What is the  contribution of thermal noise in comparison with the average (deterministic) velocity of the tracers?
A diffusive scaling for the speed of a tracer is given by $D/b$, where $D$ is the diffusion constant associated with the tracer bead. 
By the Stokes--Einstein relationship, this diffusion constant scales as
\begin{align}
	D \approx \frac{k_\text{B} T_0}{6 \pi \eta_0 b},
\end{align}
where $k_B$ is Boltzmann's constant, $T_0$ is the reference temperature, and $\eta_0$ is the reference viscosity of the fluid.
The diffusive scaling for speed is therefore 
\begin{align}
	\frac{D}{b} \approx \frac{k_\text{B} T_0}{6 \pi \eta_0 b^2}.
\end{align}
Using $k_\text{B} \approx 10^{-23}~\si{\joule}~\si{\kelvin}^{-1}$, $T_0 \approx 300~\si{\kelvin}$, and $\eta_0 \approx 10^{-2}~\si{\pascal}~\si{\second}$, we find that $D/b$ is on the order of $0.1~\si{\micro\metre}~\si{\second}^{-1}$.

How does this compare with the deterministic results? 
The average velocity of tracers varies spatially, as we saw in Fig.~\ref{fig:plot_OpticsExpress2f_speed_radius}.
In the near field, the average speed of a tracer is on the order of $1~\si{\micro\metre}~\si{s}^{-1}$, which is much greater than the diffusive speed.
We may therefore typically neglect the effect of noise in the near field. 
As a result, in the near field, our deterministic theory successfully predicts the shape of the envelope in the log--log plot of speed vs radius in Fig.~\ref{fig:plot_OpticsExpress2f_compare}.

At the middle of the closed-loop trajectories (Fig.~\ref{fig:plot_diml_avg_vel_alphabeta_streamlines}), our deterministic theory predicts that the average speed of tracers is zero (Fig.~\ref{fig:plot_diml_avg_speed_radius}). 
On the other hand, the experimental and theoretical scatter plots (Figs.~\ref{fig:plot_OpticsExpress2f_speed_radius} and~\ref{fig:plot_diml_scatter_avg_speed_radius}, respectively) both only show a dip in the average speed.
As explained earlier, the spacing between the tracer beads that contribute to the scatter plots can  account for this in our deterministic theory; noise can also contribute to this feature.

In the far field, our deterministic theory predicts that the average speed of tracers decays, following an inverse-square law.
Noise therefore becomes important for tracers far from the scan path. 
This may help explain the scatter seen in the far field for the experimental log--log plot of speed vs radius (Fig.~\ref{fig:plot_OpticsExpress2f_speed_radius}), which is not seen in the deterministic hydrodynamic theory (Fig.~\ref{fig:plot_diml_scatter_avg_speed_radius}).
Notably, this scatter is prominent where the average speed is smaller than around  $0.1~\si{\micro\metre}~\si{s}^{-1}$, precisely the diffusive scale for the average speed of the tracer bead predicted here.
In Ref.~\cite{erben2021feedback}, the transport induced by the repeatedly-scanning heat spot was used in order to manipulate beads and 
a feedback loop was needed in order to deal with the stochasticity of the particle positions.
Our hydrodynamic theory predicts quantitatively where this stochasticity becomes important in comparison with the deterministic flow.

\subsubsection{Temperature profile}

In our hydrodynamic theory, the two-dimensional, Gaussian temperature profile we impose to solve our equations analytically is an idealisation.
It is possible that the temperature perturbation in the microfluidic experiments~\cite{erben2021feedback} did not have circular symmetry, depending on the scan frequency used.
The sinusoidal amplitude function is also idealised, chosen based on earlier numerical work~\cite{mittasch2018non}.
Importantly, however, our theoretical result that the average velocity of material points is quadratic in the temperature perturbation holds for a heat spot of any shape with arbitrary time-dependent amplitude. 

Note also that our theory dealt with temperature profiles independent of $z$, as we assumed that the separation of the parallel plates is small compared with the heat-spot diameter.
In experiments, the temperature profile was  three-dimensional, while the horizontal and vertical length scales were of similar magnitude.
A laser beam has a waist; the width of the laser beam varies in the $z$ direction, though this variation is indeed slower than the exponential decay in the horizontal directions.
The temperature perturbation correspondingly also  has a waist, which is not included in our theory.

\subsubsection{Energy balance}

Finally, in our theory, we prescribe  the temperature field and then  solve  the momentum and mass equations.
In reality, the temperature profile satisfies an energy balance equation~\cite{yariv2004flow}, where
the fluid is heated due to the laser spot, and the heat   diffuses through the fluid and is advected by the flow, which in turn depends on the temperature field.
This couples the heat and fluid-flow problems.
Scaling suggests that advection of heat ($\mathbf{u}\cdot\nabla T$) is a much smaller effect than the rate of change of temperature at a fixed position due to scanning ($\partial T / \partial t$).
However, comparing this rate of change of temperature with diffusion of heat gives a thermal P\'eclet number scaling as
\begin{align}
	\text{Pe}_\text{thermal} \sim \frac{\rho c_p \frac{\partial T}{\partial t}}{\nabla \cdot (k \nabla T)} \sim \frac{(2a)^2 \rho_0 c_p f  }{k } \left \{ 1, \frac{\ell}{a} \right \},
\end{align}
where $c_p$ is the specific heat capacity and $k$ is the thermal conductivity of the fluid; again, the two possible scalings arise from the two time scales appearing in the temperature profile.
This P\'eclet number is small provided the scan frequency is sufficiently small, but could be  approximately 1 for some of the experiments in Ref.~\cite{erben2021feedback}.
Hence, future theoretical models could therefore solve for the temperature via its transport equation instead of prescribing the temperature field.

\section{Summary and perspective}\label{sec:conclusion}

In summary, motivated by recent experimental advances in artificial cytoplasmic streaming, we presented in this article an analytical, theoretical model for both the fluid flow and the transport induced by a scanning heat spot.
In Sec.~\ref{sec:inst_flow}, we solved analytically for the instantaneous  flow field of viscous fluid between two parallel plates, driven by small, prescribed temperature changes, i.e.,~in the limit of small dimensionless thermal expansion coefficient~$\alpha$ and thermal viscosity coefficient~$\beta$.
Our model allows the heat spot to have time-varying amplitude and the scan path to be of arbitrary length. 
We showed mathematically that the flow is driven by thermal expansion.
The leading-order instantaneous flow field is a potential flow solely due to thermal expansion of the fluid and is independent of the thermal viscosity coefficient, in agreement with earlier studies~\cite{weinert2008microscale}.
Specifically, it is proportional to the characteristic temperature change and occurs at order $\alpha$.
We demonstrated that provided the heat-spot amplitude varies with time, the far-field instantaneous flow is a source (or sink) flow, due to the switching-on (or switching-off) of the heat spot.
This generically dominates over the hydrodynamic source dipole in the far field due to the translation of the heat spot. 
However, the flow associated with the translation of the heat spot forms the basis of our explanations of net transport due to thermoviscous flow.

We next solved for the instantaneous flow at quadratic order, i.e.,~at order $\alpha\beta$ and order $\alpha^2$, as this in fact gives rise to the leading-order net displacement of material points.
These are both quadratic in the characteristic temperature perturbation. 
The first correction to the leading-order flow due to thermal viscosity changes, at order $\alpha\beta$, is made up of two contributions, the leading-order flow associated with translation of the heat spot, modulated by the temperature profile, and another potential flow.
The leading-order flow modulated by the temperature profile is the intuitive effect of thermal viscosity changes, as described previously~\cite{weinert2008microscale,weinert2008optically}.
We then saw mathematically that the potential flow enforces incompressibility of the flow at order $\alpha\beta$ and is solely responsible for the far field of the flow at order $\alpha\beta$, which is a hydrodynamic source dipole.

The other flow at quadratic order, i.e.,~at order $\alpha^2$, is compressible.
In the far field, this flow is also a hydrodynamic source dipole, in the opposite direction to that at order $\alpha\beta$.
	
We next took a Lagrangian perspective in Sec.~\ref{sec:one_scan} and examined the trajectories of material points during one scan. 
We solved analytically for the leading-order trajectory of an individual material point as a function of time.
The displacement of a material point typically occurs at order $\alpha$.
In experiments~\cite{erben2021feedback}, the heat spot scans repeatedly along a finite scan path and the transport induced is visualised using tracers.
Because the exposure time of recordings is much longer than a scan period, we see the net displacement of these tracers.
However, we showed that the net displacement of a material point (due to a full scan) at order $\alpha$ is zero, for any heat-spot amplitude and shape and for any scan path length, generalising earlier work~\cite{weinert2008microscale}.

We then discussed the net displacement of material points at higher order in Sec.~\ref{sec:net_displ}.
The leading-order net displacement of a material point occurs at quadratic order and so is quadratic in the heat-spot amplitude.
This can be at order $\alpha\beta$ or at order $\alpha^2$, but for the fluid used in experiments~\cite{weinert2008optically,erben2021feedback}, the order-$\alpha\beta$ contribution dominates.
At order $\alpha\beta$, the net displacement results from  amplification of flow due to reduced viscosity in the locally heated fluid.
This gives rise to strong leftward transport near the scan path, as explained in Ref.~\cite{weinert2008optically}.
Using our analytical solutions, we further showed that the far-field average velocity of the tracers is a source dipole, with rightward transport far from the $x$ axis, inherited from the instantaneous flow at the same order.
For other fluids, the contribution at order $\alpha^2$ to the average Lagrangian velocity may also be important; this is also a source dipole in the far field.

In Sec.~\ref{sec:comparison_expt}, we revisited the net displacement of material points, to compare with the trajectories of material points over many scans from experiments~\cite{erben2021feedback}.
In particular, our analytical expression for the flow field allowed us to understand the net displacement of material points throughout space, not only those on the scan path.
We found that our model could provide quantitative agreement with the experimental results for realistic parameter values, explaining the experimentally-found $1/r^2$ far-field spatial decay of the average speed of tracers as a source dipole. 
 
We also discussed limitations of our model in the context of the particular experiments of Ref.~\cite{erben2021feedback}, most significantly, the role of thermal noise, the lubrication approximation, and the prescribed temperature profile. 
These simplifications could of course be revisited in future work, allowing us to obtain further insight into thermoviscous fluid flows.

Our analysis also suggests potential future experiments to validate the model. 
For example, existing imaging has focused on the net  transport induced by repeated scanning of the laser, but new experiments with a lower laser scanning frequency may allow the instantaneous fluid flow to be visualised and compared with the theory. 
The choice of properties of the laser scanning, such as amplitude function or the  shape of the heat spot, could also be investigated and optimised for specific applications.

In addition to the fundamental understanding of the   experiments in Ref.~\cite{erben2021feedback}, our work provides a rigorous theoretical platform that will facilitate design of future FLUCS experiments at lower computational expense. 
Indeed, in this study we derived not only how all transport quantities scale with the parameters of the problem but also all mathematical prefactors. 
Our approach can therefore predict the detailed nature of the thermoviscous fluid flows in both space and time. 

Even in the geometrical setup considered in this article, our  work may  be extended and adapted to the case where the laser scan paths follow  arbitrary two-dimensional curves, instead of the straight line segments assumed here.  
We could also consider a localised heating that involves the simultaneous scanning of multiple heat spots along multiple scan paths, enabling the generation of new patterns of fluid flow and transport. 
This may in turn allow finer control of particles at the micrometre scale, with potential  applications in thermal trapping, microrobotics  and medicine~\cite{mast2013escalation,osterman2015thermooptical,morasch2016heat,kreysing2019probing}. 
Beyond the parallel-plate setup addressed here, future  theoretical modelling could examine the nature of similarly-generated thermoviscous flow in different geometries, and probe in particular the role of confinement in flow generation. 

Furthermore, while we assumed that the flow of glycerol-water solution had a Newtonian behaviour in our model of microfluidic experiments~\cite{erben2021feedback}, we expect that in FLUCS experiments in cell biology~\cite{mittasch2018non}, the flow of cytoplasm would instead have complex rheological behaviour. 
In particular, given that the forcing from the laser is periodic in time, qualitatively different flows to the Newtonian case could be  induced provided that scanning occurs at a rate faster than the relevant relaxation for the fluid. 
The typical scanning rate used in FLUCS experiments is on the order of $1~\si{\kilo\hertz}$, whereas typical relaxation times in cellular flows can be   on the order of tenths of seconds~\cite{bausch1999measurement}, so viscoelasticity is expected to play a role in many applications of FLUCS.

\appendix

\section{Scaling argument to derive the momentum equations in the lubrication limit}\label{sec:lubrication_limit_scaling}

	We present in this Appendix the derivation of the momentum equations in the lubrication limit, Eqs.~\eqref{eq:lubrication_x}--\eqref{eq:lubrication_z}, via a scaling argument, as referred to in Sec.~\ref{sec:setup}.
	Recall that in Sec.~\ref{sec:setup}, we introduced the two assumptions that the vertical separation of the plates is much less than the characteristic heat-spot diameter, $h \ll 2a$, and that the temperature profile is independent of $z$, $\frac{\partial \Delta T}{\partial z} = 0$. 
	The horizontal coordinates  $x$ and $y$ scale as the characteristic heat-spot radius  $a$ (omitting numerical factors); the vertical coordinate $z$ scales as the plate separation $h$. 
	Partial derivatives with respect to horizontal coordinates thus scale as $\partial/\partial x, \partial/\partial y \sim 1/a$, whereas the partial derivative with respect to $z$ scales as $ \partial/\partial z \sim 1/h$. 
	We write $u$ for the horizontal velocity scale and $w$ for the vertical velocity scale. 
	The shear viscosity, bulk viscosity, and density of the fluid scale as their reference values $\eta_0$, $\kappa_0$, and $\rho_0$, respectively, at the reference temperature $T_0$.

	In order to deduce the relative scalings of the horizontal and vertical fluid velocities, we first consider the mass conservation equation. This may be written as
	\begin{align}
		\frac{\partial \rho}{\partial t} + \nabla_\text{H}\cdot(\rho\mathbf{u}_\text{H}) + \frac{\partial (\rho w)}{\partial z}=0,\label{eq:mass_consn_3D_H}
	\end{align}
	where $\mathbf{u}_\text{H} \equiv (u,v)$ is the horizontal velocity field and the horizontal gradient is $\nabla_\text{H} \equiv (\partial/\partial x, \partial / \partial y)$. 
	Focusing on the relative scalings of the second and third terms in this equation, there are three cases to consider: (i) $u/a \sim w/h$, (ii) $u/a \gg w/h$, and (iii) $u/a \ll w/h$. 
It can be shown that  both case (ii) and case (iii) are inconsistent, leading to contradictions. 
In case (ii), the term $\frac{\partial (\rho w)}{\partial z}$ in Eq.~\eqref{eq:mass_consn_3D_H} is neglected in favour of the other two terms.  
Although a scaling argument yields the same momentum equations in case (ii) as in case (i), the resulting parabolic horizontal velocity profile cannot satisfy the leading-order mass conservation equation under the assumption of case (ii). 
		In case (iii), the horizontal contribution, $\nabla_\text{H}\cdot(\rho\mathbf{u}_\text{H})$, is neglected in Eq.~\eqref{eq:mass_consn_3D_H} in favour of the other two terms. 
The mass conservation equation in this case therefore may be directly integrated with respect to $z$ (noting that the density field is independent of $z$) to find the vertical velocity field, but the result then cannot satisfy the no-slip boundary condition on both of the parallel plates; thus, this case is also inconsistent.

	The only consistent scaling, relating the horizontal and vertical velocities, is therefore given by
	\begin{align}
		\frac{u}{a} \sim \frac{w}{h}.
	\end{align}
In this paper, we solve for the horizontal velocity field. 
From this, the vertical velocity field $w$ can be computed by integrating the mass conservation equation with respect to $z$; the scaling above may therefore be verified  to be consistent.

	We next consider the momentum equation. 
	As commented below Eqs.~\eqref{eq:lubrication_x}--\eqref{eq:lubrication_z} in Sec.~\ref{sec:setup}, we neglect  inertia and gravity due to the small length scales and thin-film geometry involved; we verify this \textit{a posteriori} in Sec.~\ref{sec:neglect} with a scaling argument.
	From Eq.~\eqref{eq:Cauchy_momentum} and Eq.~\eqref{eq:stress_tensor_Newt}, the momentum equation, in suffix notation ($i=1,2,3$), therefore simplifies to 
	\begin{align}
		- \frac{\partial p}{\partial x_i} 
		+ \kappa \frac{\partial }{\partial x_i} \frac{\partial u_j}{\partial x_j}  
		+ \frac{\partial \kappa}{\partial x_i} \frac{\partial u_j}{\partial x_j} 
		+ \eta \frac{\partial }{\partial x_j} \frac{\partial u_i}{\partial x_j} 
		+ \frac{1}{3} \eta \frac{\partial }{\partial x_i} \frac{\partial u_j}{\partial x_j}
		+ \frac{\partial \eta}{\partial x_j}  \left (\frac{\partial u_i}{\partial x_j} +\frac{\partial u_j}{\partial x_i} \right )
		- \frac{2}{3} \frac{\partial \eta}{\partial x_i} \frac{\partial u_j}{\partial x_j} 
		= 0.
	\end{align}

	We observe  that since the temperature profile is assumed to be independent of $z$ and the bulk and shear viscosities are functions of temperature, the bulk and shear viscosities are also independent of $z$; their $z$-derivatives, $\frac{\partial \kappa}{\partial z}$ and $\frac{\partial \eta}{\partial z}$, respectively, are therefore zero. 
	We now  write down the scalings of all the terms in the horizontal momentum equation ($i=1,2$) as
	\begin{align}
		\frac{\partial p}{\partial x_i}  &\sim \frac{p}{a};\\
		\kappa \frac{\partial }{\partial x_i} \frac{\partial u_j}{\partial x_j}, \frac{\partial \kappa}{\partial x_i} \frac{\partial u_j}{\partial x_j}     
		\sim \frac{\kappa_0 }{a} \max \left \{\frac{u}{a},\frac{w}{h}\right \}
		&\sim \frac{\kappa_0 u}{a^2}, \text{ using $\frac{u}{a} \sim \frac{w}{h}$}; \\
		\eta \frac{\partial }{\partial x_j} \frac{\partial u_i}{\partial x_j} 
		\sim \eta_0 u \max\left \{\frac{1}{a^2},\frac{1}{h^2} \right \} 
		\sim \eta \frac{\partial^2 u_i}{\partial z^2} 
		&\sim \frac{\eta_0 u}{h^2}, \text{ using $h \ll a$};\\
		\eta \frac{\partial }{\partial x_i} \frac{\partial u_j}{\partial x_j}, \frac{\partial \eta}{\partial x_i} \frac{\partial u_j}{\partial x_j} 
		\sim \frac{\eta_0 }{a} \max \left \{\frac{u}{a},\frac{w}{h}\right \}
		&\sim \frac{\eta_0 u}{a^2}, \text{ using $\frac{u}{a} \sim \frac{w}{h}$};\\
		\frac{\partial \eta}{\partial x_j}  \frac{\partial u_i}{\partial x_j},
		\frac{\partial \eta}{\partial x_j}  \frac{\partial u_j}{\partial x_i} 
		&\sim \frac{\eta_0 u}{a^2}, \text{ using $\frac{\partial \eta}{\partial z} = 0$}.
	\end{align}
	Comparing these scalings, we have $\frac{\eta_0 u}{a^2} \ll \frac{\eta_0 u}{h^2}$, since $h \ll a$. 
	We also assume that $\kappa_0$ and $\eta_0$ are of similar magnitude, so we  have $\frac{\kappa_0 u}{a^2} \ll \frac{\eta_0 u}{h^2}$.
	In the lubrication limit, the leading-order balance in the horizontal momentum equation, in terms of scalings, is thus given by
	\begin{align}
		\frac{p}{a} \sim \frac{\eta_0 u}{h^2},\label{eq:pressure_scale}
	\end{align}
	from which we deduce the pressure scale, identical to that for the classical, incompressible case.

	Next, we write down the scalings of all the terms in the vertical momentum equation as
	\begin{align}
		\frac{\partial p}{\partial z} \sim \frac{p}{h}
		&\sim \frac{\eta_0 a u}{h^3}, \text{ using Eq.~\eqref{eq:pressure_scale}};\\
		\kappa \frac{\partial }{\partial z}\frac{\partial u_j}{\partial x_j}  
		\sim \frac{\kappa_0 }{h} \max \left \{\frac{u}{a},\frac{w}{h}\right \}
		&\sim \frac{\kappa_0 u}{ah}, \text{ using $\frac{u}{a} \sim \frac{w}{h}$};\\
		\frac{\partial \kappa}{\partial z} \frac{\partial u_j}{\partial x_j}  
		&\sim 0, \text{ using $\frac{\partial \kappa}{\partial z}  = 0$}; \\
		\eta \frac{\partial }{\partial x_j} \frac{\partial w}{\partial x_j} 
		\sim \eta_0 w \max\left \{\frac{1}{a^2},\frac{1}{h^2} \right \} 
		\sim \eta   \frac{\partial^2 w}{\partial z^2}
		\sim \frac{\eta_0 w}{h^2} 
		&\sim \frac{\eta_0 u}{ah}, 
		\text{ using $h \ll a$ and $\frac{u}{a} \sim \frac{w}{h}$};\\
		\eta \frac{\partial }{\partial z} \frac{\partial u_j}{\partial x_j} 
		\sim \frac{\eta_0 }{h} \max \left \{\frac{u}{a},\frac{w}{h}\right \}
		&\sim \frac{\eta_0 u}{ah}, \text{ using $\frac{u}{a} \sim \frac{w}{h}$};\\
		\frac{\partial \eta}{\partial x_j} \frac{\partial w}{\partial x_j}  
		\sim \frac{\eta_0 w}{a^2}
		&\sim \frac{\eta_0 uh}{a^3}, \text{ using $\frac{\partial \eta}{\partial z}  = 0$ and $\frac{u}{a} \sim \frac{w}{h}$}; 	\\
		\frac{\partial \eta}{\partial x_j} \frac{\partial u_j}{\partial z}  
		&\sim \frac{\eta_0 u}{ah}, \text{ using $\frac{\partial \eta}{\partial z}  = 0$}; \\	
		\frac{\partial \eta}{\partial z} \frac{\partial u_j}{\partial x_j} 
		&\sim 0, \text{ using $\frac{\partial \eta}{\partial z} = 0$}.
	\end{align}
	We see that the scaling for the vertical pressure gradient is at least a factor of $a^2/h^2 \gg 1$ times as large as the scalings of the other terms in the vertical momentum equation; the vertical pressure gradient is unbalanced, the same as in standard lubrication theory for incompressible flow.
	
	The  momentum equations in the lubrication limit are therefore finally given by
	\begin{align}
		-\frac{\partial p}{\partial x} + \eta\frac{\partial^2 u}{\partial z^2} &=0,\\
		-\frac{\partial p}{\partial y} + \eta \frac{\partial^2 v}{\partial z^2} &=0,\\
		\frac{\partial p}{\partial z}&=0,
	\end{align}
as stated in Eqs.~\eqref{eq:lubrication_x}--\eqref{eq:lubrication_z}.

\section{Derivation of solution for the instantaneous flow at order $\beta^n$}\label{sec:derivation_order_beta_n}

In this Appendix, we derive the result, stated in Sec.~\ref{sec:order_beta_n}, that there is no flow at order $\beta^n$ for all $n \geq 0$, for any prescribed temperature profile $\Delta T$ that decays at infinity. We proceed by induction.

For the base case, at order $\beta^0$ of Eq.~\eqref{eq:pressure_expanded} we simply have Laplace's equation
\begin{align}
	\nabla^2 p_{0,0} = 0,
\end{align}
for the pressure $p_{0,0}$, with the boundary condition that the pressure gradient decays at infinity. 
The solution to this is unique, up to an additive constant that we choose to be zero without loss of generality, and given by 
\begin{align}
	p_{0,0} = 0.
\end{align}
From the velocity perturbation expansion Eq.~\eqref{eq:vel_pert_exp}, we have the corresponding flow field
\begin{align}
	\mathbf{u}_{0,0} &= -\nabla p_{0,0}\nonumber\\
	&=\mathbf{0}.
\end{align}
That is, at order $1$, there is no flow.
This is to be expected physically since in the limit of small thermal expansion coefficient~$\alpha$ and thermal viscosity coefficient~$\beta$, the coupling of the fluid flow to the temperature change is weak, so volume changes and flows driven by the heating are also weak.

For the inductive step, we assume  that  $p_{0,n-k}=0$ for all $k$ such that $1 \leq k \leq n$, where $n \geq 1$.
We will show that this implies that $p_{0,n}=0$ (where we again choose the arbitrary constant to be zero without loss of generality), which completes the proof by induction.
Order $\beta^n$ of Eq.~\eqref{eq:pressure_expanded} reads
\begin{align}
	\nabla^2 p_{0,n} + \nabla \cdot \left [ \sum_{k=1}^{n} (\Delta T)^k \nabla p_{0,n-k} \right ] = 0.
\end{align}
By the induction hypothesis, the Poisson equation   becomes Laplace's equation,
\begin{align}
	\nabla^2 p_{0,n} = 0,
\end{align}
so, again using the boundary conditions at infinity, the solution at order $\beta^n$ is
\begin{align}
	p_{0,n} &=0,
\end{align}
as claimed.
This result completes the proof by induction and corresponds to the velocity field given by
\begin{align}
	\mathbf{u}_{0,n} &= -\sum_{k=0}^{n} (\Delta T)^k \nabla p_{0,n-k}\nonumber\\
	&=\mathbf{0},
\end{align}
i.e.,~no flow at order $\beta^n$.

\begin{acknowledgments}
	We thank Francesco Boselli and  John Lister for helpful discussions.
	We gratefully acknowledge funding from the  Engineering and Physical Sciences Research Council (studentship to W.L.), from the European Research Council (Grant ``GHOSTs" \# 8953619 to M.K.), and from the European Research Council under the European Union’s Horizon 2020 research and innovation programme (Grant Agreement No. 682754 to E.L.).
\end{acknowledgments}

\bibliography{FLUCS_bibliography.bib}

\end{document}